\documentclass[11pt]{article}
\usepackage{epsf}
\newcommand{\s}{\mbox{\boldmath$\sigma$}}
\newcommand{\os}{\mbox{$\left[\:\!\mbox{\boldmath$\sigma$}\right]$}}
\begin{document}
\thispagestyle{empty}
\begin{titlepage}
\title{Phase Transition in Lattice Surface Systems with Gonihedric Action}
\author{R. Pietig\thanks{email: pietig@hybrid.tphys.uni-heidelberg.de}
        and F.J. Wegner\thanks{email: wegner@hybrid.tphys.uni-heidelberg.de} \\
         Insitut f\"ur Theoretische Physik\\
         Ruprecht-Karls-Universit\"at Heidelberg\\
         Philosophenweg 19, D-69120 Heidelberg, Germany}
\date{}
\maketitle
\begin{abstract}
We prove the existence of an ordered low temperature phase in a model of
soft-self-avoiding closed random surfaces on a cubic lattice by a suitable
extension of Peierls contour method. The statistical weight of each surface
configuration depends only on the mean extrinsic curvature and on an 
interaction term arising when two surfaces touch each other along some contour.
The model was introduced by F.J.\ Wegner and G.K.\ Savvidy as a lattice version
of the gonihedric string, which is an action for triangulated random surfaces.
\end{abstract}
\end{titlepage}
\section{Introduction}
The gonihedric string was introduced by Savvidy et.\ al.\ 
\cite{savvidy1,savvidy2,savvidy3}
as a new action for random surfaces. For triangulated surfaces, the action
reads 
\begin{equation}
\label{gonihedric}
S=\frac{1}{2}\sum_{<ij>}|\vec{X}_{i}-\vec{X}_{j}|\,\theta(\alpha_{ij}),
\end{equation}
where
\begin{equation}
\theta(\alpha_{ij})=|\pi-\alpha_{ij}|
\end{equation}
and $\alpha_{ij}$ is the angle between the neighbouring triangles with common
link $<ij>$. This string model can be considered as a natural extension of the
Feynman integral over paths to an integral over  surfaces in the sense, that
both amplitudes coincide in cases, when the surface degenerates into a single
particle world line. The simulation of (\ref{gonihedric}) shows flat surfaces
\cite{baillie}, although some problems arise from the failure to suppress the
wanderings of vertices \cite{durhuus}. \par
One possibility to regularize the gonihedric action is to formulate the
model on the euclidean lattice, which has been done by Wegner and 
Savvidy. There are two essentially distinct cases which 
correspond to non-self-avoiding surfaces \cite{wegner2} and to 
soft-self-avoiding ones \cite{wegner1}.
In the former case self-inter"-sections of the surface do not produce any
additional energy, while in case of soft-self-avoiding surfaces, the energy
contribution of a link, where self-intersection occurs was defined by
$\theta(\frac{\pi}{2})$ times the number of pairs of plaquettes which meet
under a right angle. It was shown, that in both cases a $Z_{2}$ Ising model
can be constructed, which reproduce the same surface dynamics and an 
extension to $(d-n)$-dimensional hypersurfaces on a $d$-dimensional lattice 
was also given. \par
In this paper we consider the case of $(d-1)$-dimensional soft-self-avoiding
surfaces on a $d$-dimensional cubic lattice. The statistical weight of each 
surface configuration is given by $E=l_{2}+4\,l_{4}$, where $l_{2}$ is the
number of links, where two $(d-1)$-dimensional plaquettes meet under a right 
angle (i.e.\ the mean extrinsic curvature) and $l_{4}$ is the number of
plaquettes, where four plaquettes meet. This additional term arises, when two 
surfaces touch each other and is responsible for the soft-self-avoidance. 
The equivalent 
spin systems contain just ferromagnetic nearest neighbour and antiferromagnetic
next nearest neighbour couplings. If one wants to allow arbitrary 
self-intersection coupling  $k>0$, i.e.\ $E=l_{2}+4k\,l_{4}$, the corresponding 
spin-hamiltonian 
contains also a plaquette term. The two dimensional model for $k=1$ does not 
seem to show a phase phase transition \cite{landau} \cite{savvidy4}, whereas in 
three dimensions a second order phase transition occurs at 
$\beta_{c}\approx 0.44$ which is close two the critical temperature of the two 
dimensional Ising model. These results were obtained by numerical simulations. 
The three dimensional gonihedric Ising system was also discussed for arbitrary 
$k>0$ in \cite{johnston} using mean field methods and simulations. The authors 
find a second order phase transition for $k>0$ and a qualitatively different 
behaviour for the case $k=0$. The critical temperature for $k>0$  was shown to
increase with increasing $k$. \par 
In this work we prove the existence of an ordered low temperature phase 
for $k>0$ in $d\geq 3$ dimensions using a suitable extension of Peierls contour
method \cite{peierls}. After recapitulating the model, we adapt Peierls 
argument to the case $k=1$. An upper bound for $\beta_{c}$ for the three 
dimensional model is found, which is in agreement with \cite{savvidy4,johnston}.
Finally we show, how to generalize the argument for arbitrary $k>0$.
\section{The Model}
Consider a $d$-dimensional euclidean lattice, with lattice points 
${\bf r}\in Z^{d}$.
We define a $(d-n)$-dimensional hyperplaquette 
$\Omega_{\alpha_{1},\ldots,\alpha_{n}}({\bf r})$ (all $\alpha_{i}$ different) by
\begin{eqnarray}
&& \Omega_{\alpha_{1},\ldots,\alpha_{n}}({\bf r})  \nonumber\\
&=& \left\{  {\bf x}\in R^{d} 
\mid x_{\alpha_{i}}=r_{\alpha_{i}},\: r_{\alpha}\leq x_{\alpha}\leq r_{\alpha}+1
\:\:\mbox{for all}\:\:  \alpha\not=\alpha_{i} \right\}
\end{eqnarray}
For convenience we call a $(d-1)$-dimensional hyperplaquette simply a
plaquette, a $(d-2)$-dimensional hyperplaquette a link and a 
$(d-3)$-dimensional hyperplaquette a vertex. \par
A closed surface on the lattice is a collection of plaquettes, where at each 
link an even number of plaquettes meet. In \cite{wegner1} a hamiltonian for 
closed surfaces $M$ on the lattice is defined in the following way: \par
Attach plaquette variables $U_{P}$ to each plaquette $P$ of the lattice and 
define
\begin{equation}
U_{P} = \left\{ 
\begin{array}{lll}
-1 & \mbox{if} & P\in M  \\
+1 & \mbox{if} & P\not\in M
\end{array} \right. 
\end{equation}
The energy of $M$ is given as a sum over all links
\begin{equation}
H=\sum_{\mbox{\scriptsize all links}}H_{\mbox{\scriptsize link}},
\end{equation}
where $H_{\mbox{\scriptsize link}}$ contributes $4J$ if four plaquettes meet at
the corresponding link, $J$ if two plaquettes meet perpendicular and zero in
all other cases. Note that $4$ is always the maximum number of plaquettes,
which can meet at a link, independent of the dimension of the lattice. In terms
of the link variables this can be written as
\begin{equation}
\label{hlink}
H_{\mbox{\scriptsize link}}=\frac{1}{4}J(2-U_{1}-U_{-1})(2-U_{2}-U_{-2}).
\end{equation}
\begin{picture}(10,4.5)
\epsfysize=3cm
\put(5.5,1){\mbox{\epsfbox{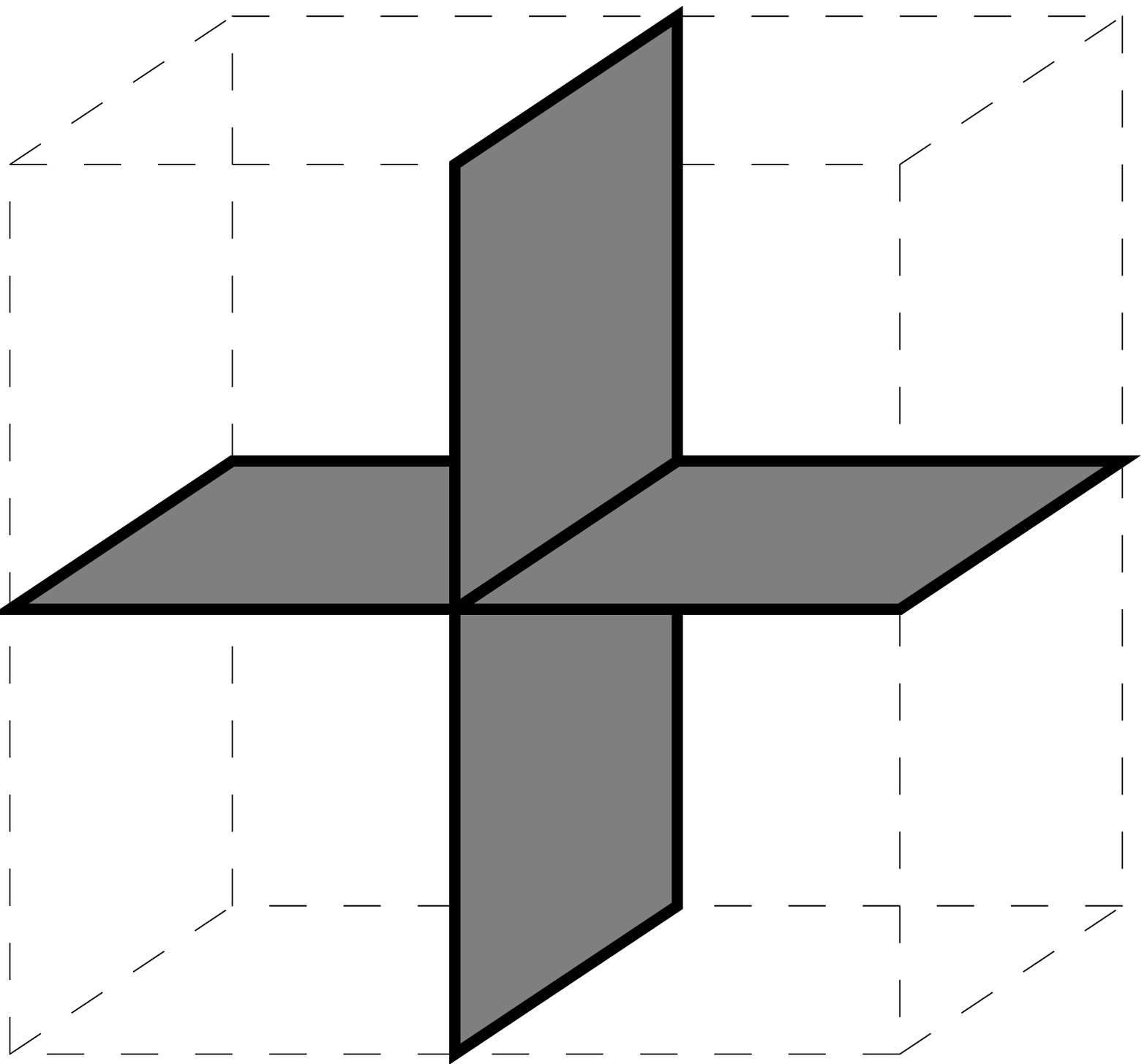}}}
\put(6.8,0.5){\mbox{$U_{-2}$}}
\put(6.8,4.3){\mbox{$U_{2}$}}
\put(4.7,2.5){\mbox{$U_{-1}$}}
\put(9,2.5){\mbox{$U_{1}$}}
\end{picture}\\
$U_{1},U_{-1},U_{2},U_{-2}$ can not be chosen independently since the surface
is closed. This condition requires
\begin{equation}
U_{1}U_{-1}U_{2}U_{-2}=1.
\end{equation}
To resolve this constraint, spin variables $\sigma=\pm 1$ were attached to the
sites of the dual lattice.\\[3mm]
\begin{picture}(10,4.5)
\epsfysize=3cm
\put(5.5,1){\mbox{\epsfbox{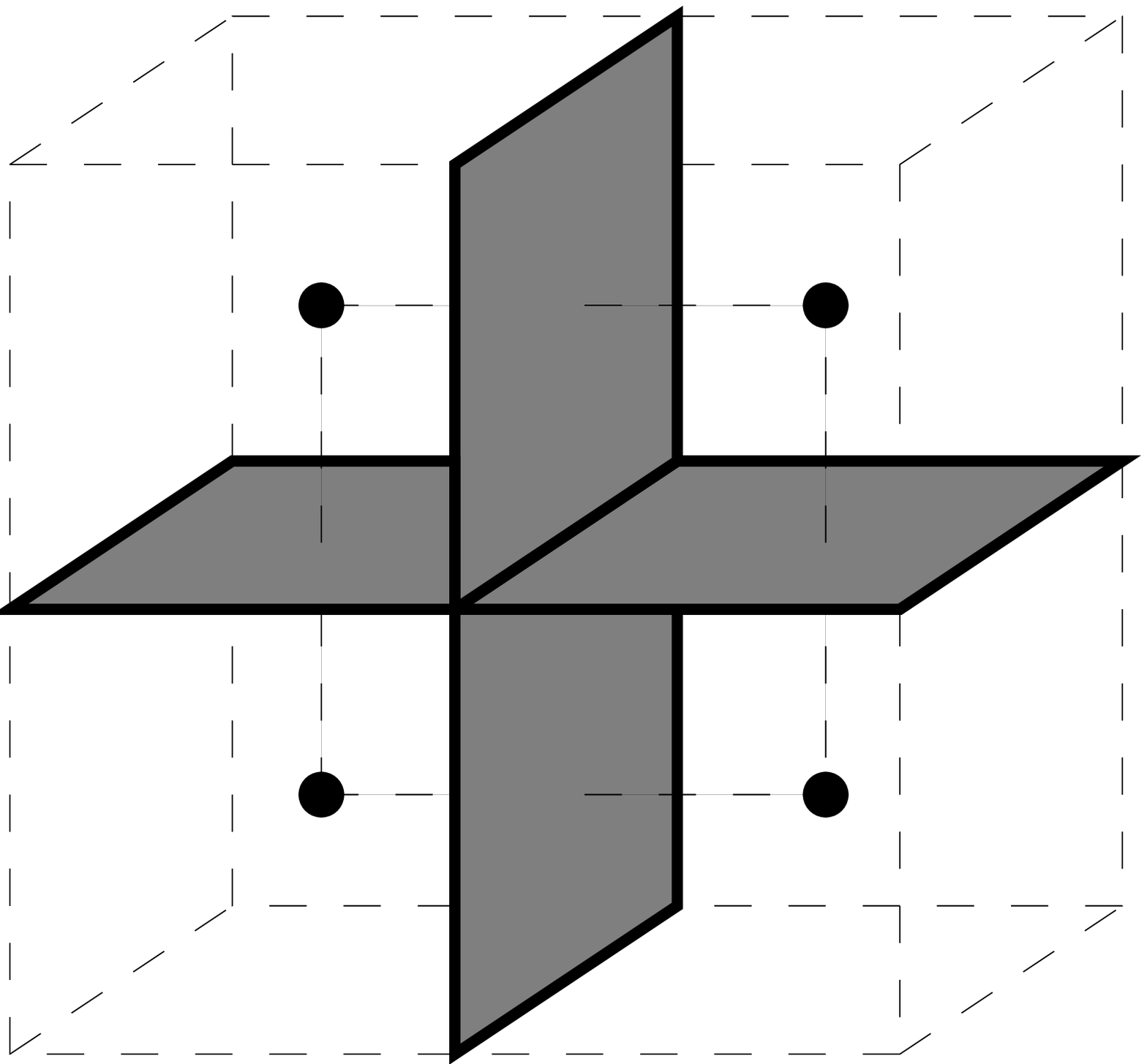}}}
\put(6.8,0.5){\mbox{$U_{-2}$}}
\put(6.8,4.3){\mbox{$U_{2}$}}
\put(4.7,2.5){\mbox{$U_{-1}$}}
\put(9,2.5){\mbox{$U_{1}$}}
\put(4.7,3.5){\mbox{$\sigma_{-2}$}}
\put(9,1.6){\mbox{$\sigma_{2}$}}
\put(4.7,1.6){\mbox{$\sigma_{-1}$}}
\put(9,3.5){\mbox{$\sigma_{1}$}}
\end{picture}\\
The plaquette variables can then be represented as
\begin{equation}
\label{plaquette_rep}
U_{1}=\sigma_{1}\sigma_{2},\quad
U_{-1}=\sigma_{-1}\sigma_{-2},\quad
U_{2}=\sigma_{1}\sigma_{-2},\quad
U_{-2}=\sigma_{-1}\sigma_{2},\quad
\end{equation}
and the full hamiltonian becomes equivalent to 
\begin{eqnarray}
\label{hamiltonian}
H & = & J\sum_{{\bf r}}\Bigg[\frac{d}{2}(d-1)
-(d-1)\sum_{\alpha}\sigma({\bf r})\sigma({\bf r}-{\bf e}_{\alpha}) \nonumber \\
\label{spin-h}
&& +\frac{1}{2}\sum_{\alpha<\beta}
\Big(\sigma({\bf r}-{\bf e}_{\alpha})\sigma({\bf r}-{\bf e}_{\beta})
+\sigma({\bf r})\sigma({\bf r}-{\bf e}_{\alpha}-{\bf e}_{\beta})\Big)  \Bigg].
\end{eqnarray}
In this expression ${\bf r}$ runs over all sites of the dual lattice, where the 
spins are located and ${\bf e}_{\alpha},{\bf e}_{\beta}$ denote unit vectors 
parallel to the $d$ cubic axes.  One sees that the equivalent spin system 
(\ref{spin-h}) contains ferromagnetic nearest neighbour and antiferromagnetic 
next nearest neighbour couplings, where the ratio between them is fixed and 
given by
\begin{equation}
\frac{J_{\mbox{\scriptsize ferromagnet}}}
{J_{\mbox{\scriptsize antiferromagnet}}}
=2(d-1).
\end{equation}
For arbitrary self-intersection coupling $k>0$, the energy contribution of a 
link, where four plaquettes meet is equal to $4Jk$. In this case equation 
(\ref{hlink}) must be replaced by
\begin{eqnarray}
H_{\mbox{\scriptsize link}} & = & \frac{1}{4}Jk(2-U_{1}-U_{-1})(2-U_{2}-U_{-2})
\nonumber \\
&& +\frac{1}{4}J(1-k)(2-U_{1}U_{-1}-U_{2}U_{-2})
\end{eqnarray}
and the corresponding spin-hamiltonian becomes
\begin{eqnarray}
H & = & J\sum_{{\bf r}}\Bigg[\frac{d}{2}(d-1)\frac{k+1}{2}
-(d-1)k\sum_{\alpha}\sigma({\bf r})\sigma({\bf r}-{\bf e}_{\alpha}) 
\nonumber \\
&& +\frac{1}{2}k\sum_{\alpha<\beta}
\Big(\sigma({\bf r}-{\bf e}_{\alpha})\sigma({\bf r}-{\bf e}_{\beta})
+\sigma({\bf r})\sigma({\bf r}-{\bf e}_{\alpha}-{\bf e}_{\beta})\Big)
\nonumber \\
&& -\frac{1-k}{2}\sum_{\alpha<\beta}
\Big(\sigma({\bf r})\sigma({\bf r}-{\bf e}_{\alpha})
\sigma({\bf r}-{\bf e}_{\alpha}-{\bf e}_{\beta})
\sigma({\bf r}-{\bf e}_{\beta})\Big)  \Bigg].
\end{eqnarray}
This hamiltonian also contains a plaquette term as long as $k\not=1$. \par
Because in (\ref{plaquette_rep}) the plaquette variables were represented as
products of nearest neighbour spins, we can restore the original surface $M$
from the spin configuration by simply choosing all plaquettes between spins of 
opposite sign. In this sense, the surface $M$ can be considered as a domain
wall, which separates spins of positive from spins of negative sign.
In contrast to the ordinary ferromagnetic Ising model, the energy of a spin 
configuration is not given by the total number of plaquettes of $M$, i.e.\
the surface,  but by $E=(l_{2}+4k\,l_{4})J$, where $l_{2}$ is the number
of links contained in $M$, where two plaquettes meet perpendicular and $l_{4}$
is the number of links, where 4 plaquettes meet. Because of this, one can always
insert plain domain walls by swaping a whole ($(d-1)$-dimensional) spin layer
without changing the total energy, provided the inserted walls do not cross
any existing surface. In particular the ground state of the finite lattice
system containing $N^{d}$ spins ($d$-dimensional cubus) is 
$(d\,(2^{N}-2)+2)$-fold degenerate. 
\section{Existence of a phase transition for ${\bf d\geq 3}$}
In this section we will prove, that the model defined in the previous
section shows a phase transition, provided $d\geq 3$. This will be done with a
method which is due to Peierls \cite{peierls}. The argument proceeds in
two steps. (i) We show that the spontaneous magnetization $\hat{M}_{N}$ for a 
finite system with special boundary conditions is bounded from below, if the
temperature is lower than a critical temperature $T_{c}$, i.e.\
\begin{equation}
\label{magnetization}
\hat{M}_{N}\geq \alpha > 0 \qquad\mbox{for $T<T_{c}$}.
\end{equation}
The hat indicates the special boundary conditions and the index $N$
refers to the finite system containing $N$ spins. $\alpha$ is independent of
$N$, but sensitive to the special boundary conditions we impose. (ii) The free
energy $\hat{f}_{N}(T,H)$ is concave as a function of the magnetic field $H$,
i.e.\
\begin{equation} 
\label{concav}
-\frac{\hat{f}_{N}(T,H)-\hat{f}_{N}(T,0)}{H}\geq \hat{M}_{N}(H=0)\geq \alpha >0,
\end{equation}
for $T<T_{c}$ and $H\geq 0$.
The thermodynamic limit $\lim_{N\to\infty}\hat{f}_{N}(T,H)$ exists in the sense 
of van Hove and is independent of the boundary conditions. The limiting free 
energy is concave as well. This implies that equation (\ref{concav}) remains
valid in the thermodynamic limit. Since the magnetization is antisymmetric in 
$H$, it follows that $f(T,H)$ is not analytic in $H=0$ if $T<T_{c}$. \par
To get an idea how equation (\ref{magnetization}) can be shown, consider a 
3-dimensional finite volume with $N$ spins and fix the spins at the boundary to
+1. The ground state of this system is given by the spin configuration, in 
which all spins are positive, since then no domain walls are present and hence 
the total energy is zero. If we now swap a little `island' of spins inside the 
volume to -1, we need an amount of energy $E$, which is essential proportional
to the total number of edge elements of the domain wall, which was established
by swaping the spins. Therefore, the probability of occurrence of a domain wall
with $l$ edge elements is of order $\nu(l)e^{-\beta J l}$, where $\nu(l)$ is
the total number of possible edge configurations with $l$ edge elements.
Now for $d=3$, the total number of negative spins inside a domain wall is 
bounded by $(\frac{l}{12})^{3}$ (volume of a cube), hence the mean number of 
negative spins will be bounded by
\begin{equation}
\label{p1}
\left<N_{-}\right>\leq C\sum_{l}\left(\frac{l}{12}\right)^{3}\nu(l)\,
e^{-\beta Jl},
\end{equation}
where $C$ is some constant. If $\nu(l)$ does not increase too rapidly for
increasing $l$, the right hand side of (\ref{p1}) will be smaller than
$\frac{N}{2}$ for large enough $\beta>\beta_{c}$. Thus the spontaneous 
magnetisation will be strictly non zero, if $\beta>\beta_{c}$. The 
argument can easily be extended to $d\geq 3$, if $(\frac{l}{12})^{3}$ is
replaced by $(\frac{l}{2d(d-1)})^{\frac{d}{d-2}}$. It does not work in two
dimensions, since in this case, the edge elements are vertices and the total
number of negative spins inside a domain wall with a given number of vertices
can be arbitrary large. Indeed, in two dimensions, the model does not seem to 
show a phase transition \cite{landau,savvidy4}. 
We now proceed with the details for the case $k=1$. \par
Consider a finite volume of the dual lattice containing $N$ spins in its
interior $V\setminus\partial V$ and fix the spins at the boundary:
\begin{equation}
\sigma({\bf r})=+1 \quad\mbox{if}\quad {\bf r}\in\partial V
\end{equation}
If a certain configuration $\s=(\sigma_{1},\ldots,\sigma_{N})$
of the $N$ spins is given we denote by $M\os$ the corresponding 
closed surface on the original lattice which separates regions of negative from
regions of positive spins. We call
a link contained in $M\os$ an edge element, if one of the
following two conditions is fulfilled:
\begin{itemize}
\item Four plaquettes of $M\os$ meet at the given link.
\item Two plaquettes of $M\os$ meet perpendicular at the given link.
\end{itemize}
The set $C^{\mbox{\scriptsize tot}}$ of all edge elements of $M\os$ can be
divided into connected edge diagrams $C_{1},\ldots,C_{n}$ in a unique way
\begin{equation}
C^{\mbox{\scriptsize tot}}=C_{1}\cup C_{2}\cup\ldots\cup C_{n}
\end{equation}
The energy $E\os$ of the spin configuration can then easily be expressed
in terms of connected edge diagrams since $M\os$ is the original interacting
surface and therefore only the edge elements of $M\os$ contribute to $E\os$.
We define the energy contribution $E\left[\:\!C\right]$ of a connected edge 
diagram $C$ as
\begin{equation}
E\left[\:\!C\right]=J(l_{2}+4\,l_{4}),
\end{equation}
where $l_{2}$ denotes the number of edge
elements of $C$, where two plaquettes meet and $l_{4}$ the number of edge
elements, where four plaquettes meet. The total energy of a spin configuration
can then be written as
\begin{equation}
E\os=\sum_{C}E\left[\:\!C\right].
\end{equation}
Clearly, the number of all connected edge diagrams with a given energy 
contribution $E$ that fit into the finite volume $V$ is finite. We denote
this number by $g(E)$. Next we attach variables $\chi_{i}^{E}\os, i=1,\ldots,
g(E)$ to all these connected edge diagrams $C_{i}^{E}$. $\chi_{i}^{E}\os$ 
assume the value
+1 if the corresponding edge diagram appears in $\s$ and 0 otherwise. \par
The partition function $Z_{N}$ for the $N$ spins is given by
\begin{equation}
Z_{N}=\sum_{\s}e^{-\beta E[\s]}
\end{equation}
We are interested in the thermodynamical expectation value 
$\left<\chi_{i}^{E}\right>_{N}$, that is the probability of appearance of
$C_{i}^{E}$:
\begin{equation}
\label{mean1}
\left<\chi_{i}^{E}\right>_{N}=\frac{1}{Z_{N}}\sum_{\s}\chi_{i}^{E}\os
e^{-\beta E[\s]}
\end{equation}
In this expression only spin configurations which contain $C_{i}^{E}$
contribute. Numbering them by $\s_{1},\ldots,\s_{k}$, (\ref{mean1}) reads:
\begin{equation}
\label{mean2}
\left<\chi_{i}^{E}\right>_{N}=\frac{1}{Z_{N}}\sum_{j=1}^{k}
e^{-\beta E[\:\!\s_{j}]}
\end{equation}
Consider a particular spin configuration $\s_{j}$ in which $C_{i}^{E}$
appears. The connected edge diagram $C_{i}^{E}$ belongs to a domain wall, which
encloses a certain number of spins. By reversing all these spins,
the domain wall disappears and the resulting configuration $\s_{j}^{*}$ will
not contain $C_{i}^{E}$. Therefore
\begin{equation}
E[\:\!\s_{j}] = E[\:\!\s_{j}^{*}]+E[C_{i}^{E}].
\end{equation}
From this we obtain the following estimate:
\begin{equation}
Z_{N}\geq \sum_{j=1}^{k}e^{-\beta E[\:\!\s_{j}^{*}]}
\geq e^{\beta E[C_{i}^{E}]}\sum_{j=1}^{k}e^{-\beta E[\:\!\s_{j}]}.
\end{equation}
Together with (\ref{mean2}) this implies:
\begin{equation}
\label{mean3}
\left<\chi_{i}^{E}\right>_{N} \leq e^{-\beta E[C_{i}^{E}]}.
\end{equation}
We can now find an upper bound for the mean number of negative spins inside
$V$ as follows: \par
Consider a given configuration $\s$ and denote its connected edge diagrams
by $C_{1},\ldots,C_{n}$. Each diagram belongs to a domain wall, which contains
a number $n_{j}$ of negative spins. If $l_{j}\leq E[C_{j}]/J$ denotes the 
number of edge elements in $C_{j}$, $n_{j}$ is bounded by
\begin{equation}
n_{j} \leq \left(\frac{l_{j}}{2d(d-1)}\right)^{\frac{d}{d-2}}
\leq \left(\frac{E[C_{j}]}{2d(d-1)J}\right)^{\frac{d}{d-2}}.
\end{equation}
This relation becomes an equation, if the domain wall is a simple cube.
The total number of negative spins $N_{-}\os$ is therefore bounded by
\begin{equation}
N_{-}\os \leq 
\sum_{j=1}^{n}\left(\frac{E[C_{j}]}{2d(d-1)J}\right)^{\frac{d}{d-2}}.
\end{equation}
Using the variables $\chi_{i}^{E}\os$, we can write 
\begin{equation}
N_{-}\os \leq 
\sum_{E}\sum_{i=1}^{g(E)}\chi_{i}^{E}\os
\left(\frac{E}{2d(d-1)J}\right)^{\frac{d}{d-2}},
\end{equation}
where the first summation is over all possible energies $E$, a connected edge
diagram can assume. From (\ref{mean3}) we get an upper bound for the
thermodynamical expectation value of $N_{-}\os$:
\begin{equation}
\label{nminus}
\left<N_{-}\right>_{N} \leq \sum_{E}
\left(\frac{E}{2d(d-1)J}\right)^{\frac{d}{d-2}}
g(E)\,e^{-\beta E}
\end{equation}
Note that the right hand side of (\ref{nminus}) is not defined for $d=2$.
To proceed with the argument, we need an upper bound for $g(E)$. 
Consider the following construction method, which can be 
used to build up every possible diagram:
\begin{itemize}
\item[1.] First we number all the vertices ($(d-3)$-dimensional 
hyperplaquettes) of the lattice and keep this numbering once for all fixed.
\item[2.] Next we choose one of the $N$ lattice points and attach 3 edge 
elements, pointing in three given directions. For example if 
$p=(x_{1},\ldots,x_{d})$ denotes the chosen point, we can attach the edge
elements \\
$c_{1}=(x_{1},x_{2},\lambda_{3},\ldots,\lambda_{d})$, \\
$c_{2}=(x_{1},\lambda_{2},x_{3},\lambda_{4},\ldots,\lambda_{d})$, \\
$c_{3}=(\lambda_{1},x_{2},x_{3},\lambda_{4},\ldots,\lambda_{d})$, \\
$x_{i}\leq \lambda_{i}\leq x_{i}+1$.
This can be done since such
an edge configuration must occur at least once in any possible 
connected edge diagram. We think of this vertex as a corner of the closed
surface. This fixes the 8 spins surrounding the vertex $p$.
\item[3.] Imagine that a connected subdiagram exists 
already. Now we choose the vertex with the lowest number, which occurs in the
construction so far, i.e.\ where at least one edge element ends and where the 
configuration of the edge elements
surrounding this vertex has not been specified yet. Next we do this 
specification by choosing the values of not more than four of the eight spins, 
surrounding the vertex, since we know that at least one edge element ends
at this vertex already.
\item[4.] Finally by repeating construction step 3., all possible connected
edge diagrams can be constructed.
\end{itemize}
To find an upper bound for $g(E)$, we estimate the maximum number of possible
outcomes of the construction procedure just given. If the considered finite
lattice volume contains $N$ spins, we have $N$ choices for construction step
2. Each time we perform construction step 3, we have different numbers of 
choices how to specify the remaining spins. This depends on how many edge 
elements and what kind of edge elements are already ending at the vertex under
consideration.
The maximum number of choices arises, if just one edge element, surrounded by 
two or four plaquettes, is already present at the vertex, leaving four spins to
specify, i.e.\ 16 possibilities. These possibilities are shown in figure 1, 
together with the corresponding energy contribution of the specified 
vertex. Edge elements where two plaquettes meet are indicated by a simple edge.
Edge elements, where four plaquettes meet are indicated as double 
edges. The left column shows all cases, where one simple edge coming 
from the left was already present and correspondingly
the right column shows all cases, where one
double edge coming from the left was already present.
From this picture we read of the maximum numbers of choices $n(E)$, how 
to specify the spins, such that a vertex of energy $E$ arises:\\
\begin{center}
\begin{tabular}{c|c}
\parbox{3.2cm}{maximum number \\ of choice $n(E)$} & $x^{E/J}$ \\ \hline
1 & $x^2$ \\ 
4 & $x^3$ \\
2 & $x^4$ \\
2 & $x^6$ \\
4 & $x^7$ \\
3 & $x^8$ \\
4 & $x^{10}$ \\
4 & $x^{15}$ \\
1 & $x^{24}$ 
\end{tabular}\\
\end{center}
The function $G_{n}(x)=N\,x^3\,f^{n}(x)$, where $f(x)$ is defined by
\begin{eqnarray}
f(x) & = & \sum_{E}n(E)\,x^{E/J} \nonumber\\
& = & x^2+4\,x^3+2\,x^4+2\,x^6+4\,x^7+3\,x^8+4\,x^{10}+4\,x^{15}+x^{24}
\end{eqnarray}
can now be interpreted as a generating function which counts the number of all
connected edge diagrams with energy $E$ at least once, which can be constructed
by repeating construction step 3 $n$-times, i.e.\ the coefficient of 
$x^{2(d-2)\,E/J}$ in an expansion of $G_{n}(x)$ is an upper bound for the 
number of those diagrams. Since all edge elements were counted $2(d-2)$ times, 
we have to calculate the coefficient of $x^{2(d-2)\,E/J}$ and not simply 
$x^{E/J}$. Summing $G_{n}(x)$ over all possible numbers of steps
\begin{equation}
\label{generator}
G(x)=\sum_{n=0}^{\infty}G_{n}(x)=
N\,\frac{x^3}{1-f(x)},
\end{equation}
we get a generation function, which counts all connected edge diagrams
at least once. Therefore
\begin{equation}
\label{bound1}
g(E)\leq \left.\frac{1}{(2(d-2)E/J)!}\:
\frac{\partial^{2(d-2)E/J}}{\partial\,x^{2(d-2)E/J}}\:G(x)\right|_{x=0}
\end{equation}
provides an upper bound for $g(E)$. We can write $G(x)$ as
\begin{equation}
G(x)=N\,\sum_{k=1}^{24}\frac{a_{k}}{x-x_{k}}
\end{equation}
Putting this expression in equation (\ref{bound1}), we obtain
\begin{equation}
\label{bound2}
g(E)\leq N\,\sum_{k=1}^{24}\left(-\frac{a_{k}}{x_{k}}\right)\,
\left(\frac{1}{x_{k}^{2(d-2)}}\right)^{E/J}
\end{equation} 
If we denote the smallest $|x_{k}|$ by $x_{min}$, which clearly dominates
the increase of $g(E)$ for large $E$, we can further approximate this expression
as follows:
\begin{equation}
g(E)\leq N\,\left(\sum_{k=1}^{24}\left|\frac{a_{k}}{x_{k}}\right|\right)\,
\left(\frac{1}{x_{min}^{2(d-2)}}\right)^{E/J} =: N\,a\,c^{(d-2)E/J},
\end{equation}
where
\begin{eqnarray}
a & \leq & 0.625 \nonumber\\
c & \leq & 3.882 \nonumber
\end{eqnarray}
Substituting this result into equation (\ref{nminus}), we find an upper bound
for the density of negative spins:
\begin{equation}
\frac{\left<N_{-}\right>_{N}}{N}\leq\sum_{E}\left(
\frac{E}{2d(d-2)J}\right)^{\frac{d}{d-2}}
a\,c^{(d-2)E/J}\,e^{-\beta E}
\end{equation}
This inequality remains valid in the thermodynamical limit. The sum on 
the right hand side converges, if $\beta$ is larger than $\frac{d-2}{J}$ln(c)
and approaches zero, if $\beta$ goes to infinity. Therefore for 
$\beta>\beta_{c}$, it will be smaller than $\frac{1}{2}$, which implies a non 
zero spontaneous magnetisation. This proves the existence of a 
phase transition. \par
Equation (\ref{nminus}) together with an upper bound for $g(E)$ given in 
(\ref{bound2}) also provides a lower bound for the critical temperature $T_{c}$.
We find
\begin{eqnarray}
T_{c}\geq 0.67\,J \qquad\mbox{for $d=3$} \\
T_{c}\geq 0.36\,J \qquad\mbox{for $d=4$}
\end{eqnarray}
The value for $d=3$ is in agreement with the critical temperature found in 
\cite{savvidy4,johnston}, which was $T_{c}\approx 2.28\,J$ \par
The method used above to prove the existence of an ordered low temperature
phase for $k=1$ can easily be generalized to arbitrary $k=\frac{p}{q}>0$, where
$p,q\in N$. We have
\begin{eqnarray}
\frac{E}{J} & = & l_{2}+4k\,l_{4} \geq l=l_{2}+l_{4} 
\qquad\mbox{if}\quad k\geq \frac{1}{4} \\
\frac{E}{4kJ} & = & \frac{l_{2}}{4k}+l_{4} \geq l=l_{2}+l_{4} 
\qquad\mbox{if}\quad k< \frac{1}{4} 
\end{eqnarray}
Therefore equation (\ref{nminus}) remains unchanged if $k\geq \frac{1}{4}$,
whereas in case $k<\frac{1}{4}$ we have to replace it by
\begin{equation}
\left<N_{-}\right>_{N} \leq \sum_{E}
\left(\frac{E}{2d(d-1)4kJ}\right)^{\frac{d}{d-2}}
g(E)\,e^{-\beta E}
\end{equation}
An upper bound for $g(E)$ can be found by expanding the generating function
\begin{equation}
G(x)=N\,\frac{x^3}{1-f(x)},
\end{equation}
\begin{eqnarray}
f(x) & = & x^2+4x^3+2x^4+2x^6+4x^{3+4\frac{p}{q}}+2x^{4+4\frac{p}{q}}
\nonumber\\
&& +x^{8\frac{p}{q}}+4x^{2+8\frac{p}{q}}+4x^{3+12\frac{p}{q}}+x^{24\frac{p}{q}}
\end{eqnarray}
in powers of $x^{\frac{1}{q}}$, i.e.
\begin{equation}
G(x)=N\sum_{j}\frac{a_{j}}{x^{\frac{1}{q}}-x_{j}}=\sum_{n=0}^{\infty}
\left(N\sum_{j}\Big(-\frac{a_{j}}{x_{j}}\Big)
\Big(\frac{1}{x_{j}}\Big)^{n}\right)\:x^{\frac{n}{q}}
\end{equation}
From this we obtain
\begin{equation}
g(E)\leq N\,\sum_{j}\left(-\frac{a_{j}}{x_{j}}\right)\,
\left(\frac{1}{x_{j}^{2(d-2)}}\right)^{qE/J}
\end{equation}
The estimate for $g(E)$ becomes smaller for increasing $k$ and diverges for
$k\to 0$. For $k\to\infty$ the behaviour of $g(E)$ for large $E$ is dominated
by the smallest zero of the function $1-x^2-4x^3-2x^4-2x^6$, which gives
\begin{equation}
g(E)\leq const\,(3.74)^{(d-2)E/J} \qquad\mbox{for}\quad k\to\infty
\end{equation} 
\section{Summary}
We have proven the existence of an ordered low temperature phase for the spin
systems defined in section 2, which can be considered as the lattice
regularization of the gonihedric string. Each spin configuration corresponds to
a particular random surface which is simply the domain wall separating regions 
of spin with opposite sign. This observation was crucial, since it allowed us
to apply Peierls contour method. However the naive application of the argument
would yield an inequality of the following form in $d=3$ dimensions:
\begin{equation}
\label{naiv}
\left<N_{-}\right>_{N} \leq \sum_{b}
\left(\frac{b}{6}\right)^{\frac{3}{2}}
\nu(b)\,e^{-\beta E_{\scriptstyle min}(b)},
\end{equation}
where $\left<N_{-}\right>_{N}$ is the expectation value of the number of
negative spins, calculated for the finite system with boundary condition
$\sigma({\bf r})=+1, {\bf r}\in\partial V$. The factor 
$(b/6)^{\frac{3}{2}}$ is the maximum number of negative
spins that a closed surface with $b$ plaquettes can enclose and $\nu(b)$
denotes the number of closed surfaces with $b$ plaquettes. This number is
known to grow exponentially, in fact
\begin{equation}
\nu(b)\leq N\,3^{b-1}.
\end{equation}
For $E_{\scriptstyle min}(b)$ we have to put the minimum energy of a closed
domain wall with $b$ plaquettes which is
\begin{equation}
E_{\scriptstyle min}(b) = 12\sqrt{\frac{b}{6}} \qquad (d=3)
\end{equation}
This shows, that the right hand side of (\ref{naiv}) actually diverges for
every $\beta$ due to the exponential growth of $\nu(b)$. We therefore had to
modify the argument as discussed in section 3. This was possible since the
energy of a closed surface is essentially proportional to the number $l$ of 
edges and the number of negative spins inside is bounded by 
$(\frac{l}{12})^{3}$. Instead of (\ref{naiv}) we arrived at
\begin{equation}
\left<N_{-}\right>_{N} \leq \sum_{E}
\left(\frac{E}{2d(d-1)4kJ}\right)^{\frac{d}{d-2}}
g(E)\,e^{-\beta E}
\end{equation}
for $0<k<\frac{1}{4}$ and
\begin{equation}
\left<N_{-}\right>_{N} \leq \sum_{E}
\left(\frac{E}{2d(d-1)J}\right)^{\frac{d}{d-2}}
g(E)\,e^{-\beta E}
\end{equation}
for $k\geq \frac{1}{4}$. Finally we proved that the number of connected
edge diagrams with given energy $g(E)$ does not grow faster than exponentially
for any $k>0$. \par
As pointed out before, the argument does not work in two dimensions, because
in this case the ($0$-dimensional) edges are not connected and therefore do not
give a restriction for the number of negative spins.
\section{Acknowledgement}
The authors would like to thank G.K.\ Savvidy for helpful discussions.
\newpage\noindent
\begin{picture}(15,2)
\put(0,2){\line(1,0){15}}
\put(0,0){\line(1,0){15}}
\put(7.5,0){\line(0,1){2}}
\put(0,0){\line(0,1){2}}
\put(15,0){\line(0,1){2}}
\put(0.3,1.6){\mbox{left and right}}
\put(0.6,1.2){\mbox{spin layer}}
\put(3.5,1.6){\mbox{visual}}
\put(3.4,1.2){\mbox{picture}}
\put(5.5,1.3){\mbox{$x^{\mbox{\small energy/J}}$}}
\put(7.8,1.6){\mbox{left and right}}
\put(8.1,1.2){\mbox{spin layer}}
\put(11,1.6){\mbox{visual}}
\put(10.9,1.2){\mbox{picture}}
\put(13,1.3){\mbox{$x^{\mbox{\small energy/J}}$}}
\epsfxsize=0.3cm
\put(1.2,0.4){\mbox{\epsfbox{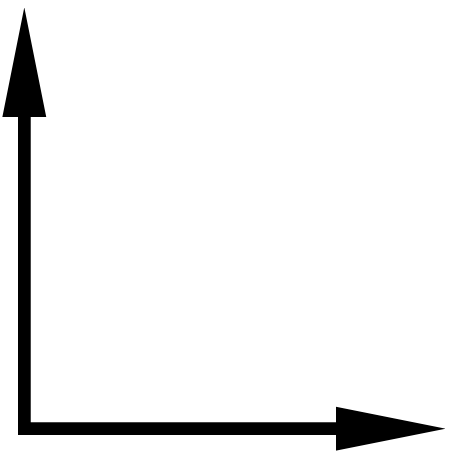}}}
\put(1,0.7){\mbox{\tiny z}}
\put(1.55,0.3){\mbox{\tiny y}}
\epsfxsize=0.3cm
\put(3.8,0.4){\mbox{\epsfbox{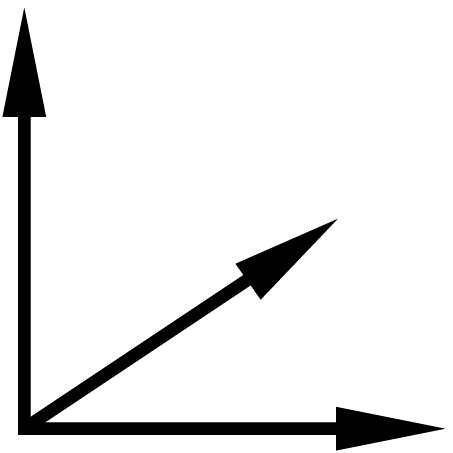}}}
\put(3.6,0.7){\mbox{\tiny z}}
\put(4.15,0.3){\mbox{\tiny x}}
\put(4.1,0.6){\mbox{\tiny y}}
\epsfxsize=0.3cm
\put(8.7,0.4){\mbox{\epsfbox{arrow2.eps}}}
\put(8.5,0.7){\mbox{\tiny z}}
\put(9.05,0.3){\mbox{\tiny y}}
\epsfxsize=0.3cm
\put(11.3,0.4){\mbox{\epsfbox{arrow3.eps}}}
\put(11.1,0.7){\mbox{\tiny z}}
\put(11.65,0.3){\mbox{\tiny x}}
\put(11.6,0.6){\mbox{\tiny y}}
\end{picture}\\
\begin{picture}(15,2)
\put(7.5,0){\line(0,1){2}}
\epsfxsize=0.7cm
\put(0.5,0.5){\mbox{\epsfbox{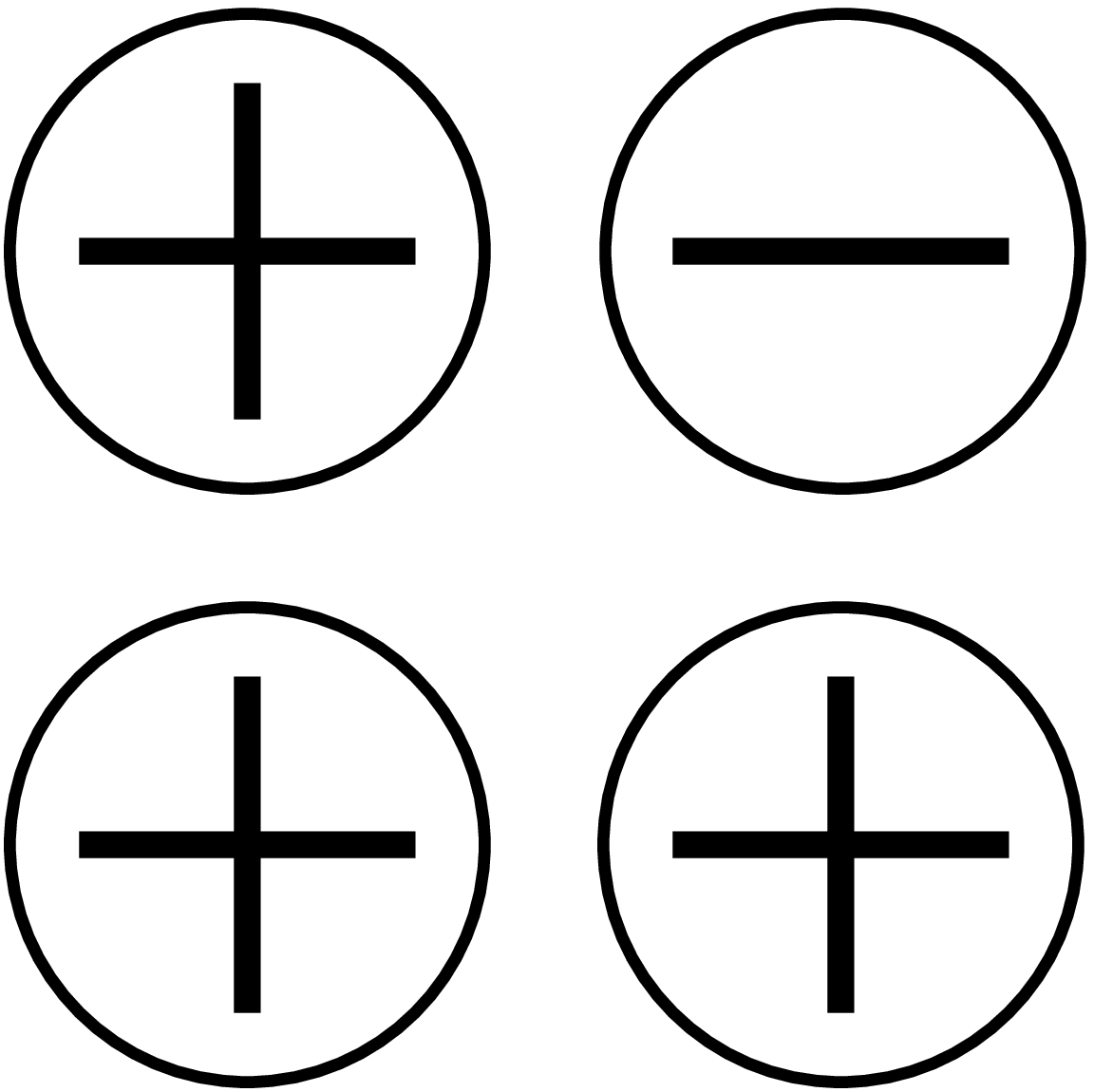}}}
\epsfxsize=0.7cm
\put(1.5,0.5){\mbox{\epsfbox{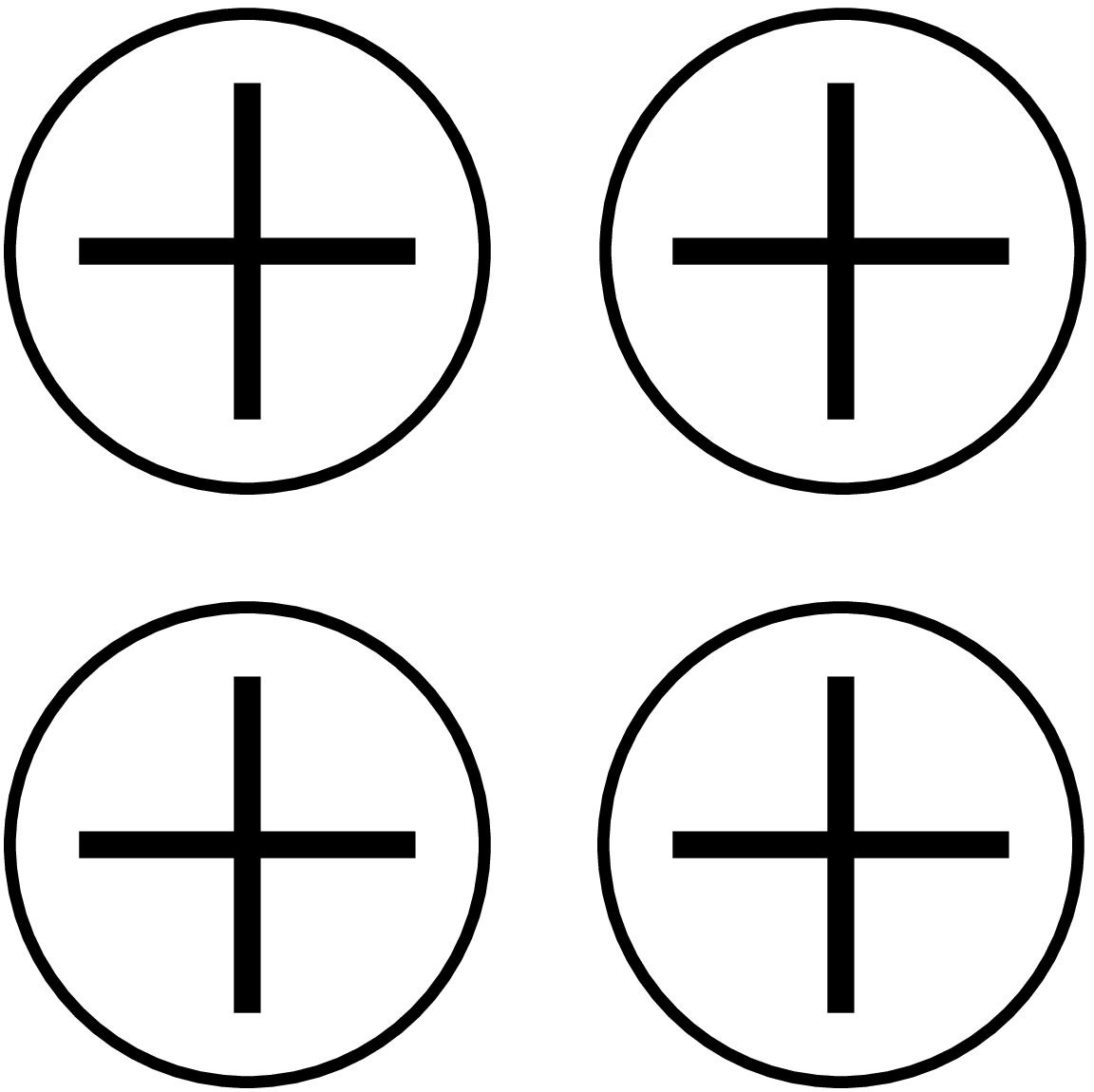}}}
\epsfxsize=1.8cm
\put(3.1,0.1){\mbox{\epsfbox{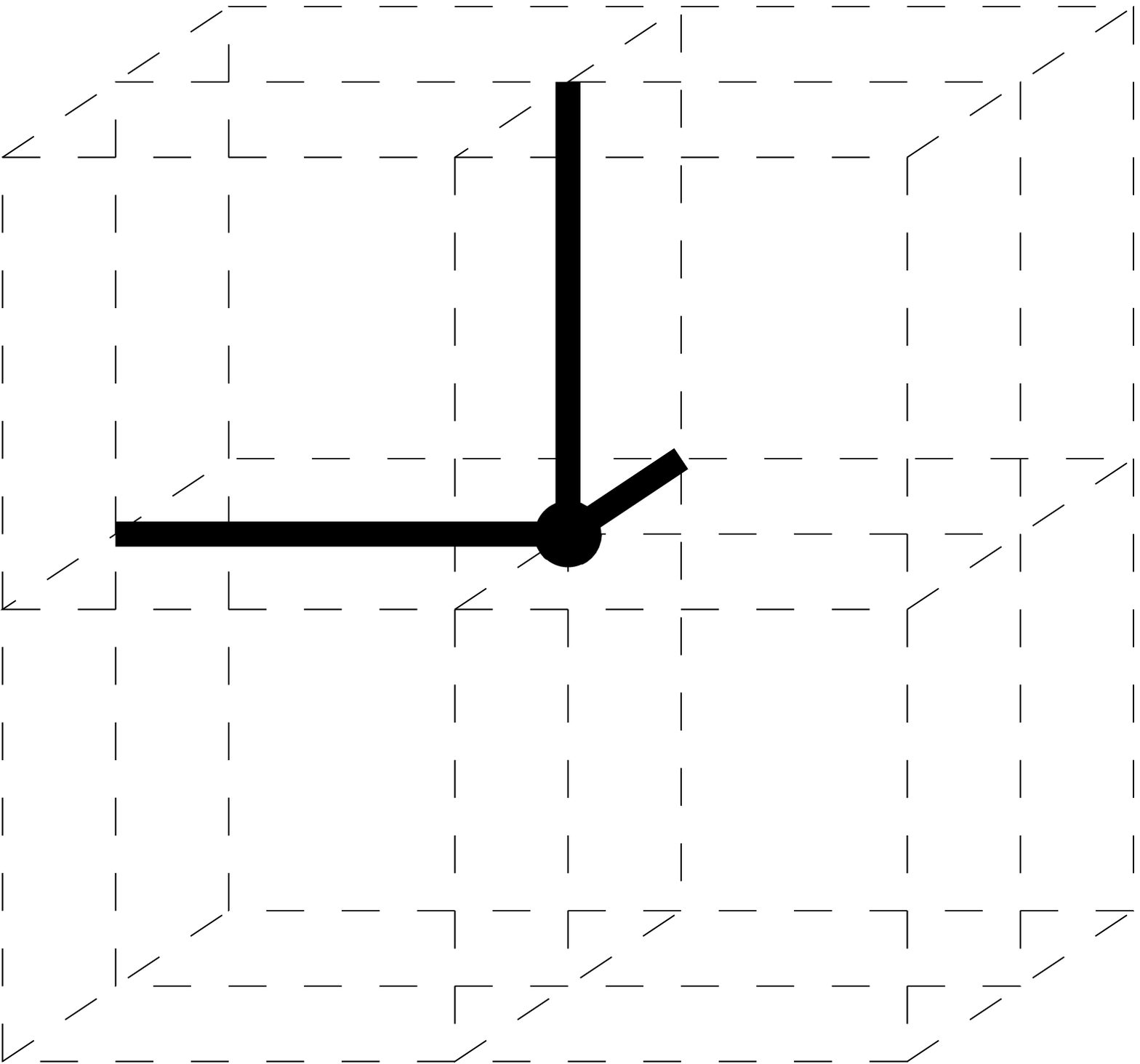}}}
\put(6,0.9){\mbox{$x^3$}}
\epsfxsize=0.7cm
\put(8,0.5){\mbox{\epsfbox{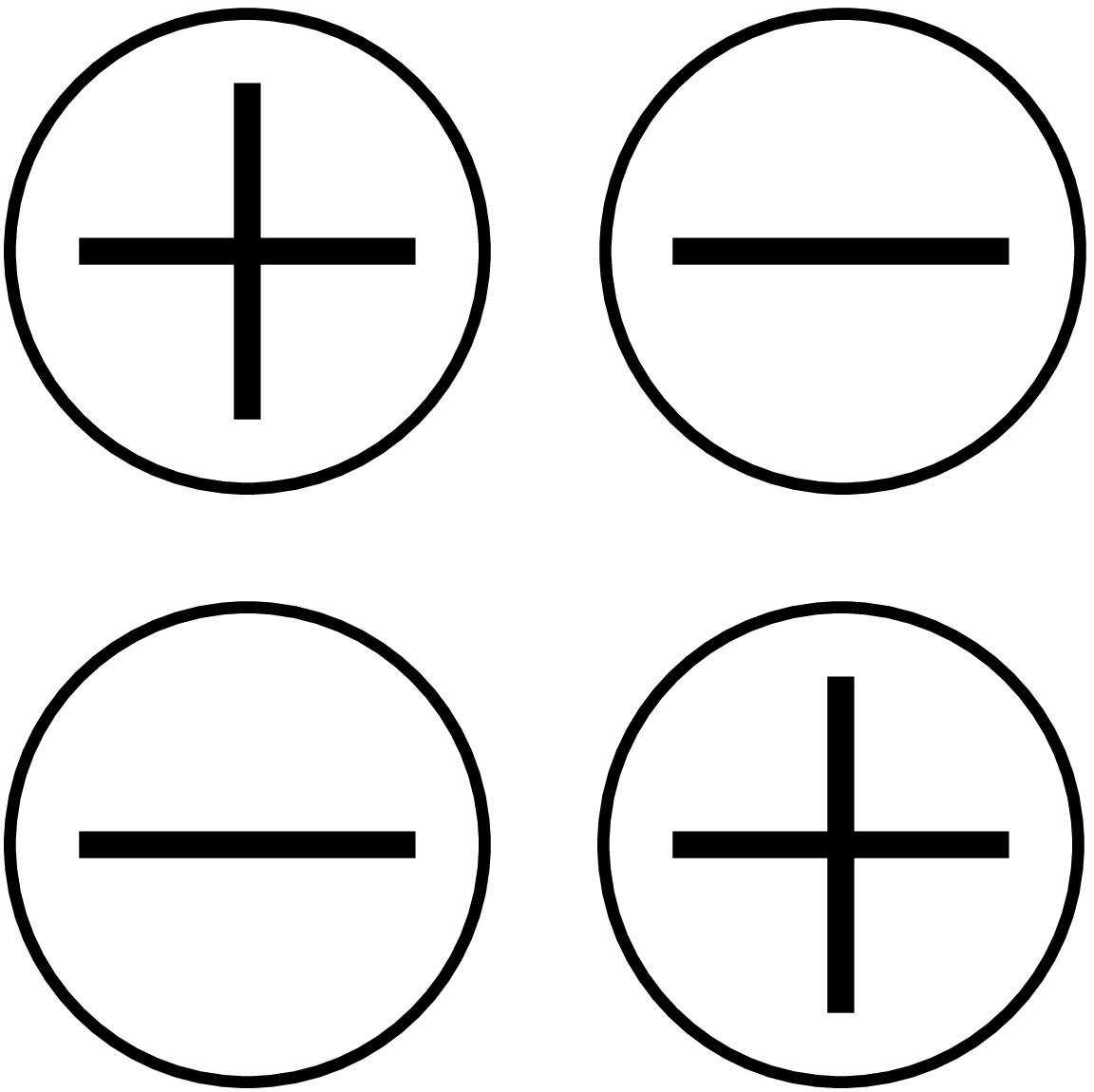}}}
\epsfxsize=0.7cm
\put(9,0.5){\mbox{\epsfbox{spin01.eps}}}
\epsfxsize=1.8cm
\put(10.6,0.1){\mbox{\epsfbox{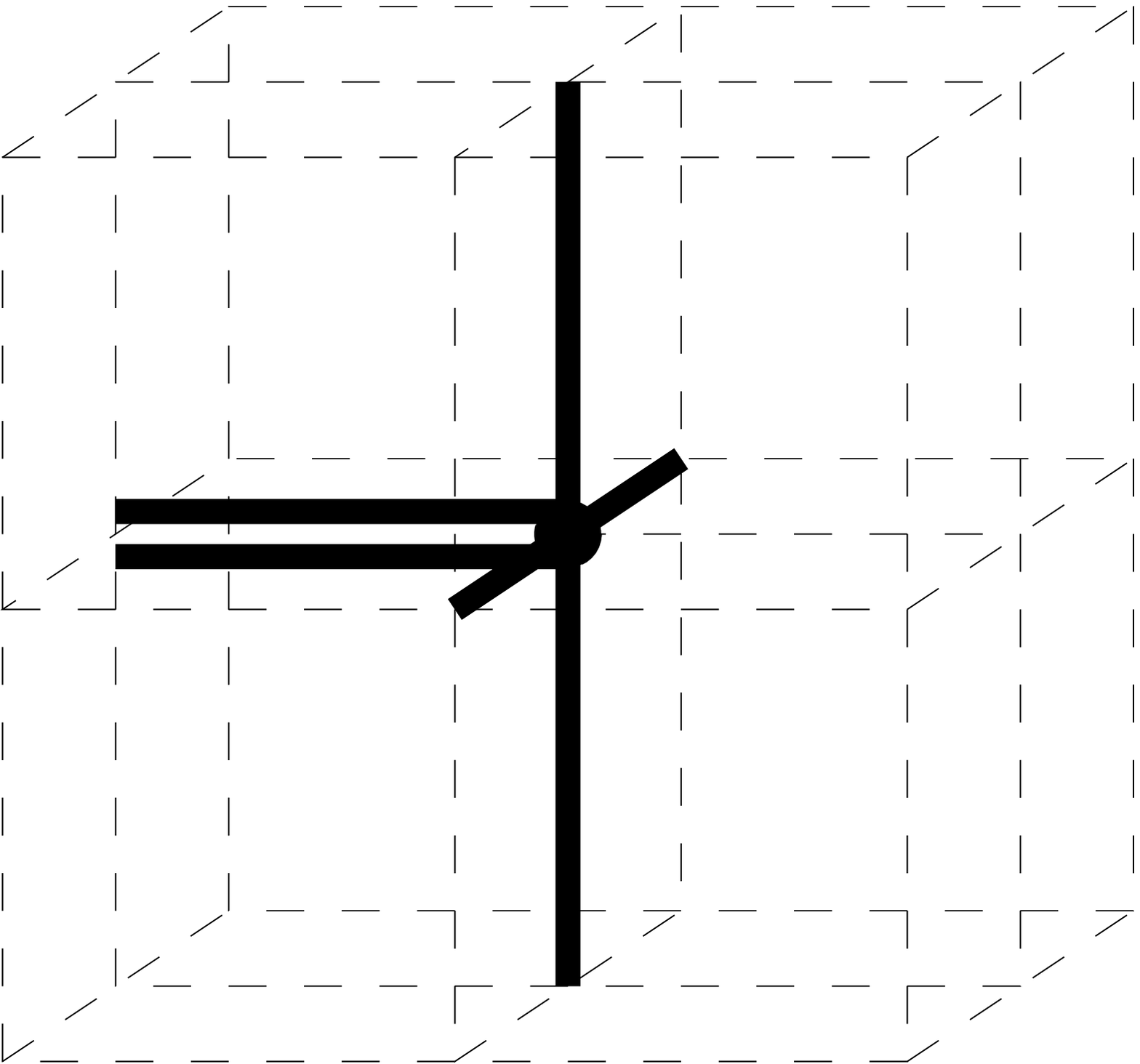}}}
\put(13.5,0.9){\mbox{$x^8$}}
\end{picture}\\
\begin{picture}(15,2)
\put(7.5,0){\line(0,1){2}}
\epsfxsize=0.7cm
\put(0.5,0.5){\mbox{\epsfbox{espin.eps}}}
\epsfxsize=0.7cm
\put(1.5,0.5){\mbox{\epsfbox{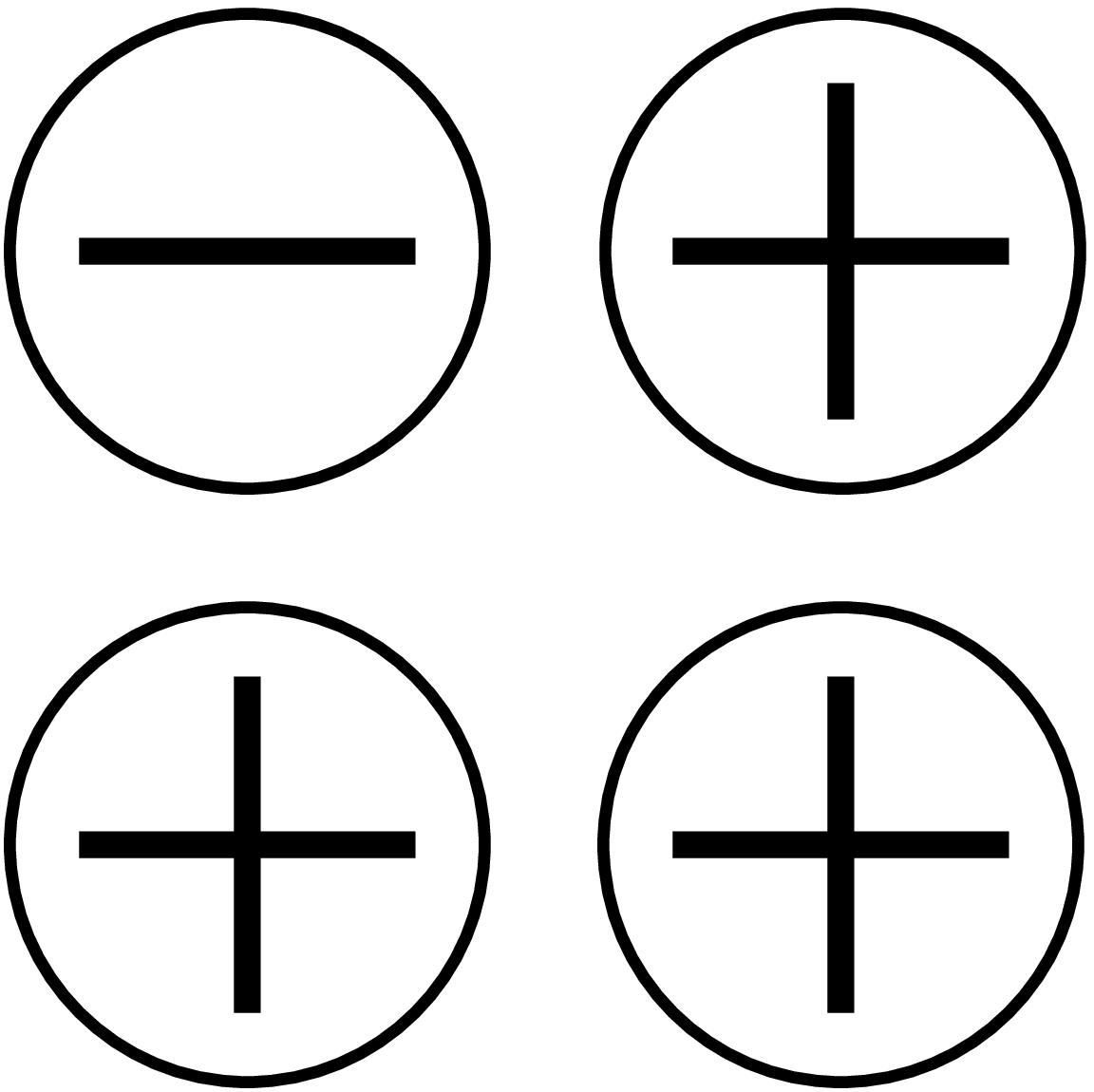}}}
\epsfxsize=1.8cm
\put(3.1,0.1){\mbox{\epsfbox{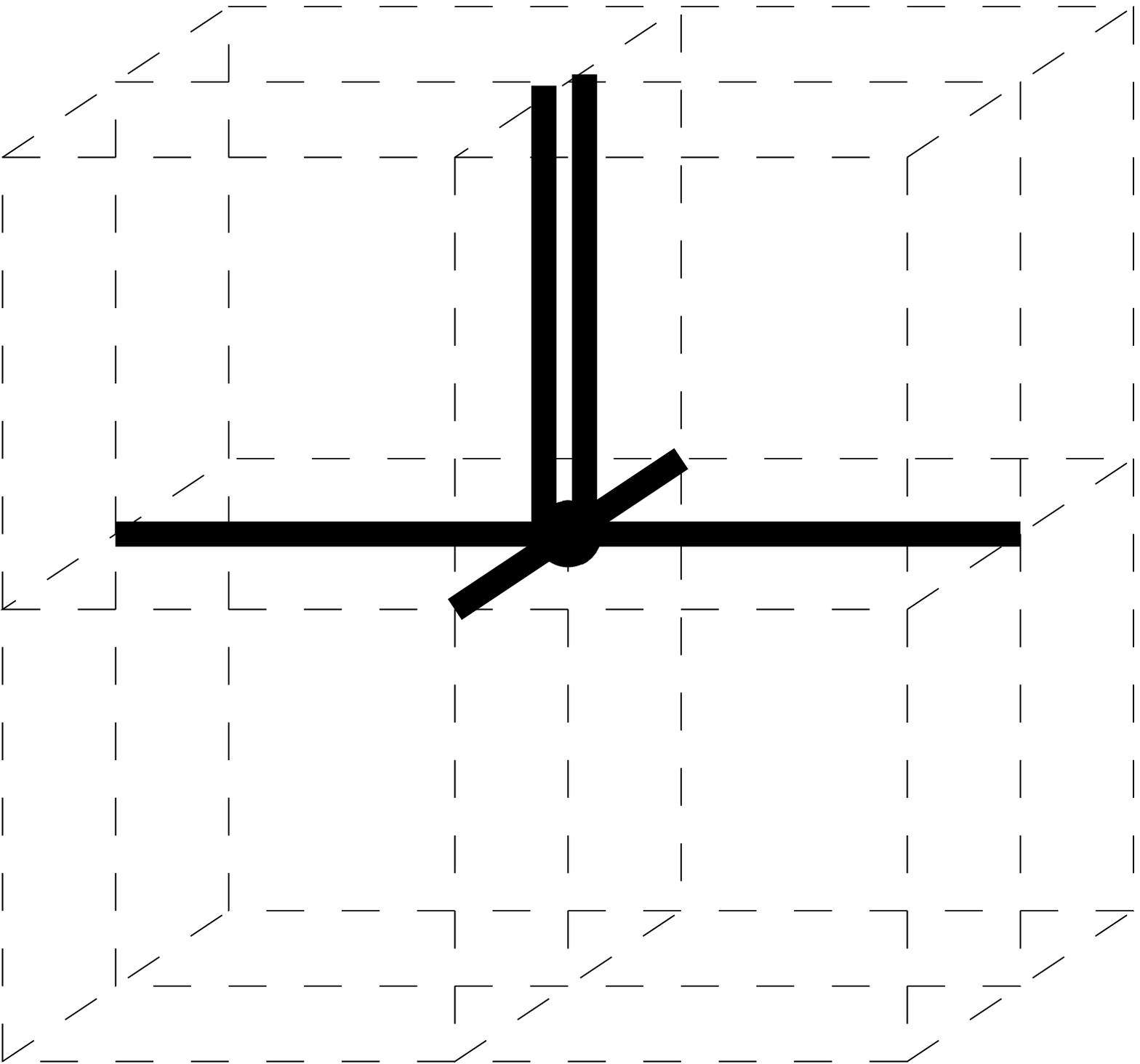}}}
\put(6,0.9){\mbox{$x^8$}}
\epsfxsize=0.7cm
\put(8,0.5){\mbox{\epsfbox{dspin.eps}}}
\epsfxsize=0.7cm
\put(9,0.5){\mbox{\epsfbox{spin02.eps}}}
\epsfxsize=1.8cm
\put(10.6,0.1){\mbox{\epsfbox{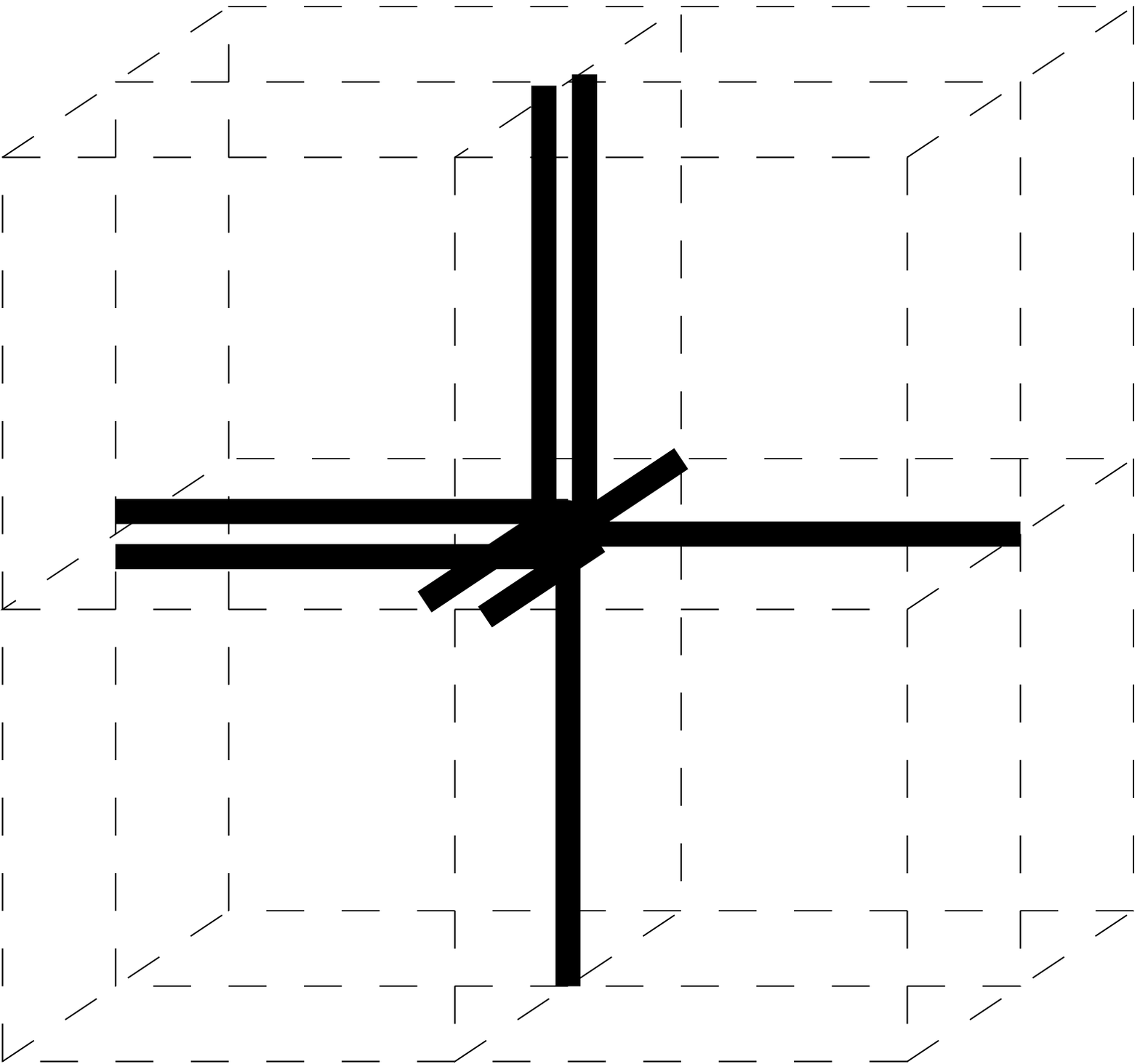}}}
\put(13.5,0.9){\mbox{$x^{15}$}}
\end{picture}\\
\begin{picture}(15,2)
\put(7.5,0){\line(0,1){2}}
\epsfxsize=0.7cm
\put(0.5,0.5){\mbox{\epsfbox{espin.eps}}}
\epsfxsize=0.7cm
\put(1.5,0.5){\mbox{\epsfbox{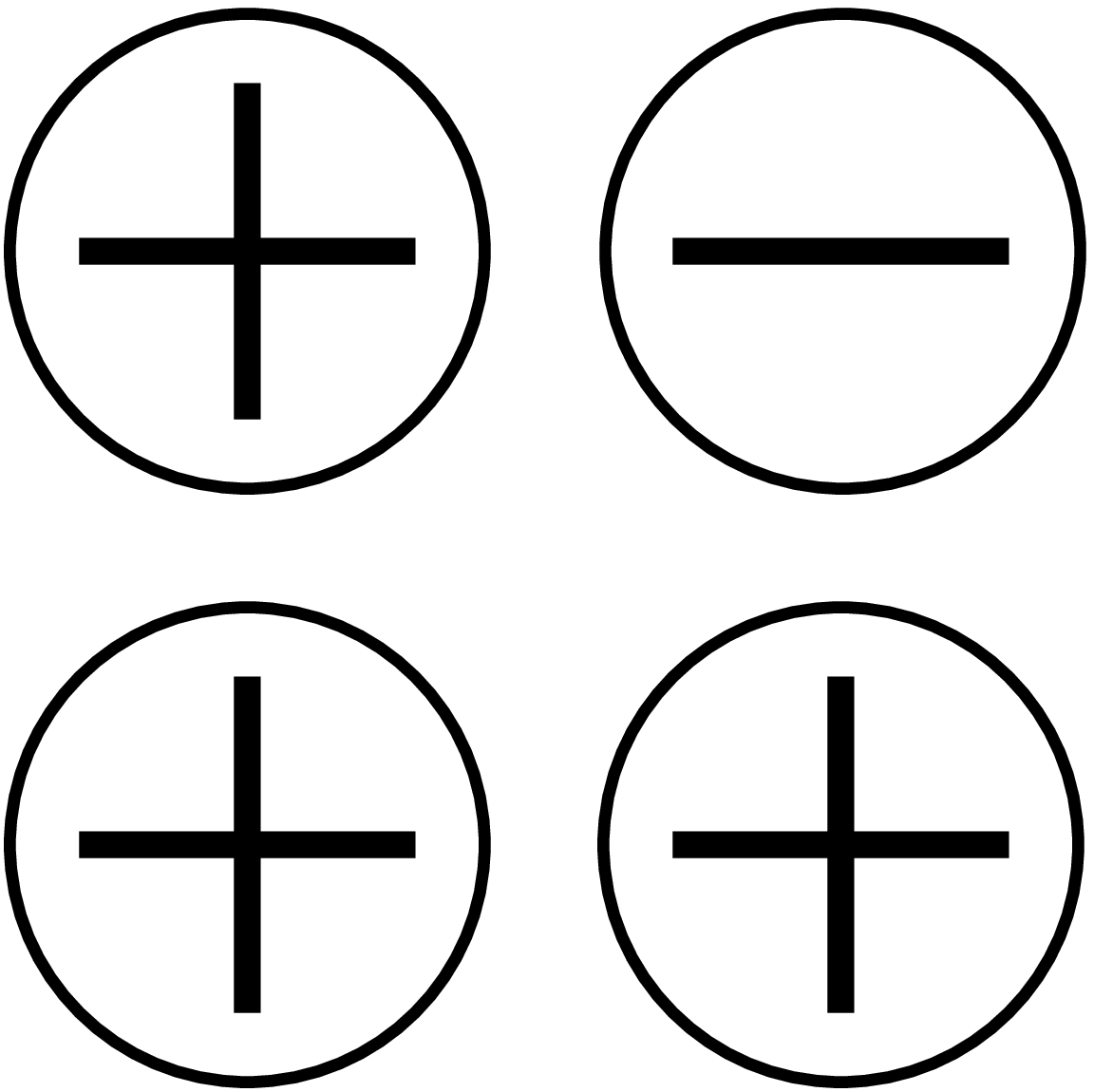}}}
\epsfxsize=1.8cm
\put(3.1,0.1){\mbox{\epsfbox{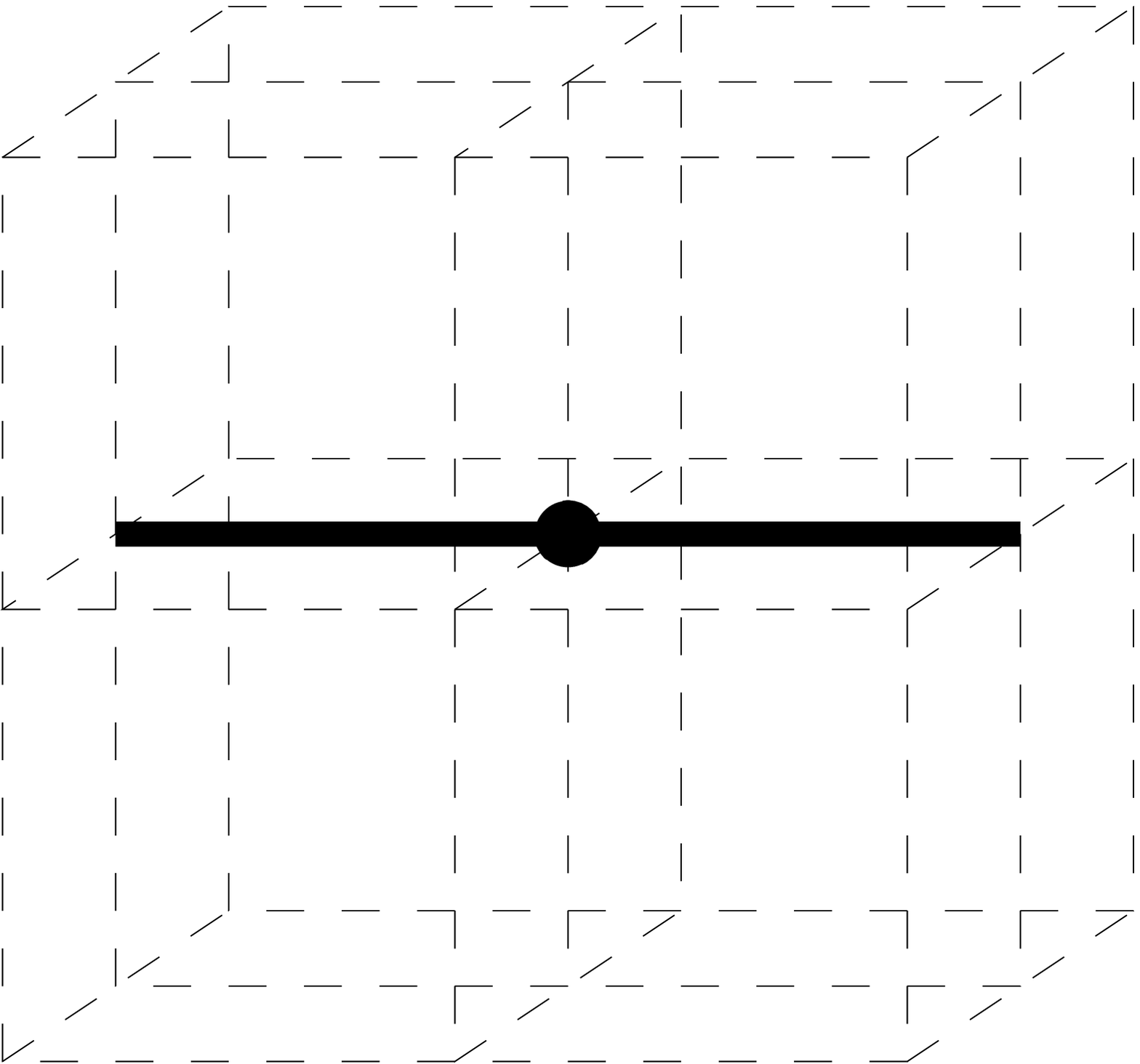}}}
\put(6,0.9){\mbox{$x^2$}}
\epsfxsize=0.7cm
\put(8,0.5){\mbox{\epsfbox{dspin.eps}}}
\epsfxsize=0.7cm
\put(9,0.5){\mbox{\epsfbox{spin03.eps}}}
\epsfxsize=1.8cm
\put(10.6,0.1){\mbox{\epsfbox{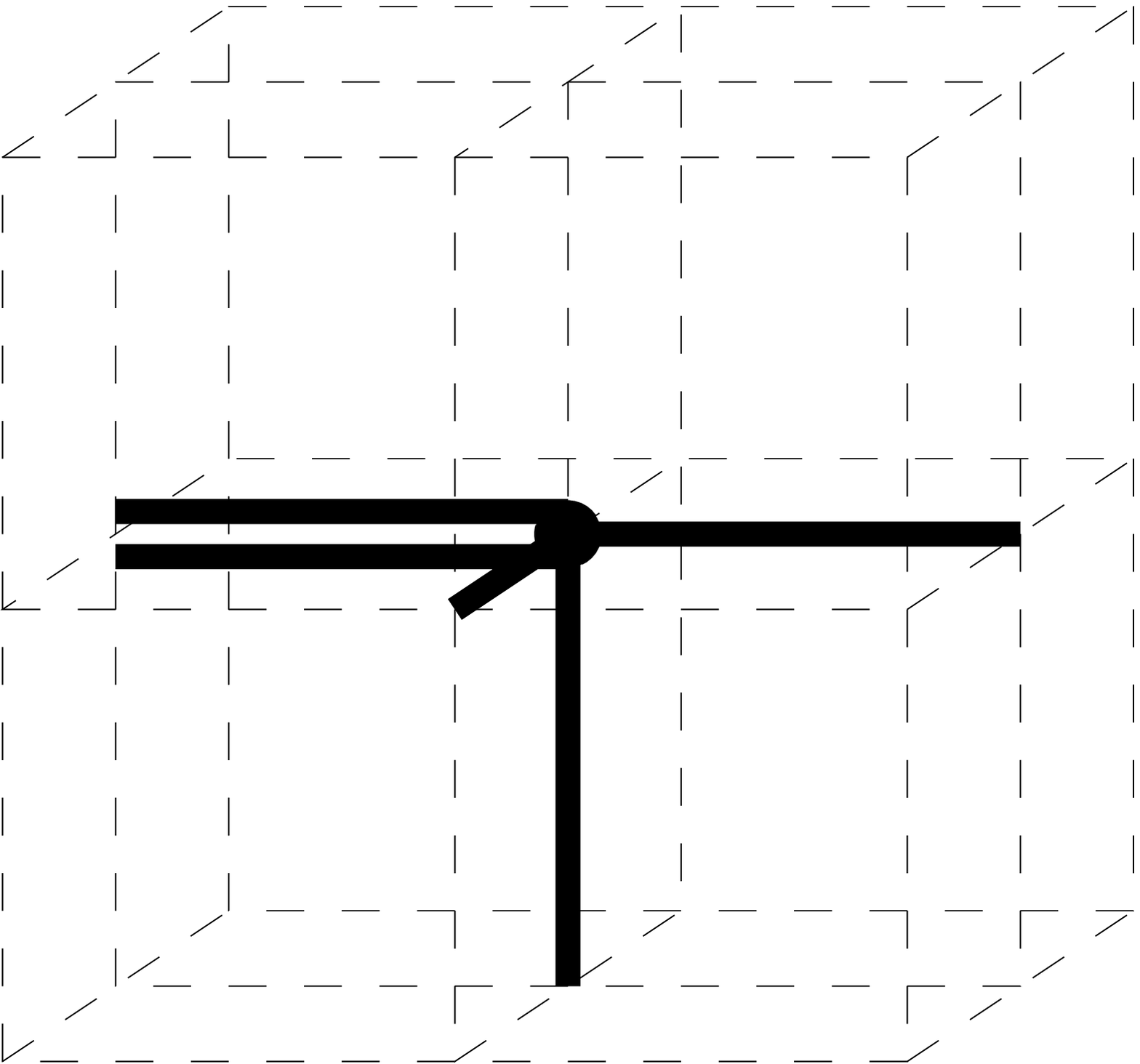}}}
\put(13.5,0.9){\mbox{$x^7$}}
\end{picture}\\
\begin{picture}(15,2)
\put(7.5,0){\line(0,1){2}}
\epsfxsize=0.7cm
\put(0.5,0.5){\mbox{\epsfbox{espin.eps}}}
\epsfxsize=0.7cm
\put(1.5,0.5){\mbox{\epsfbox{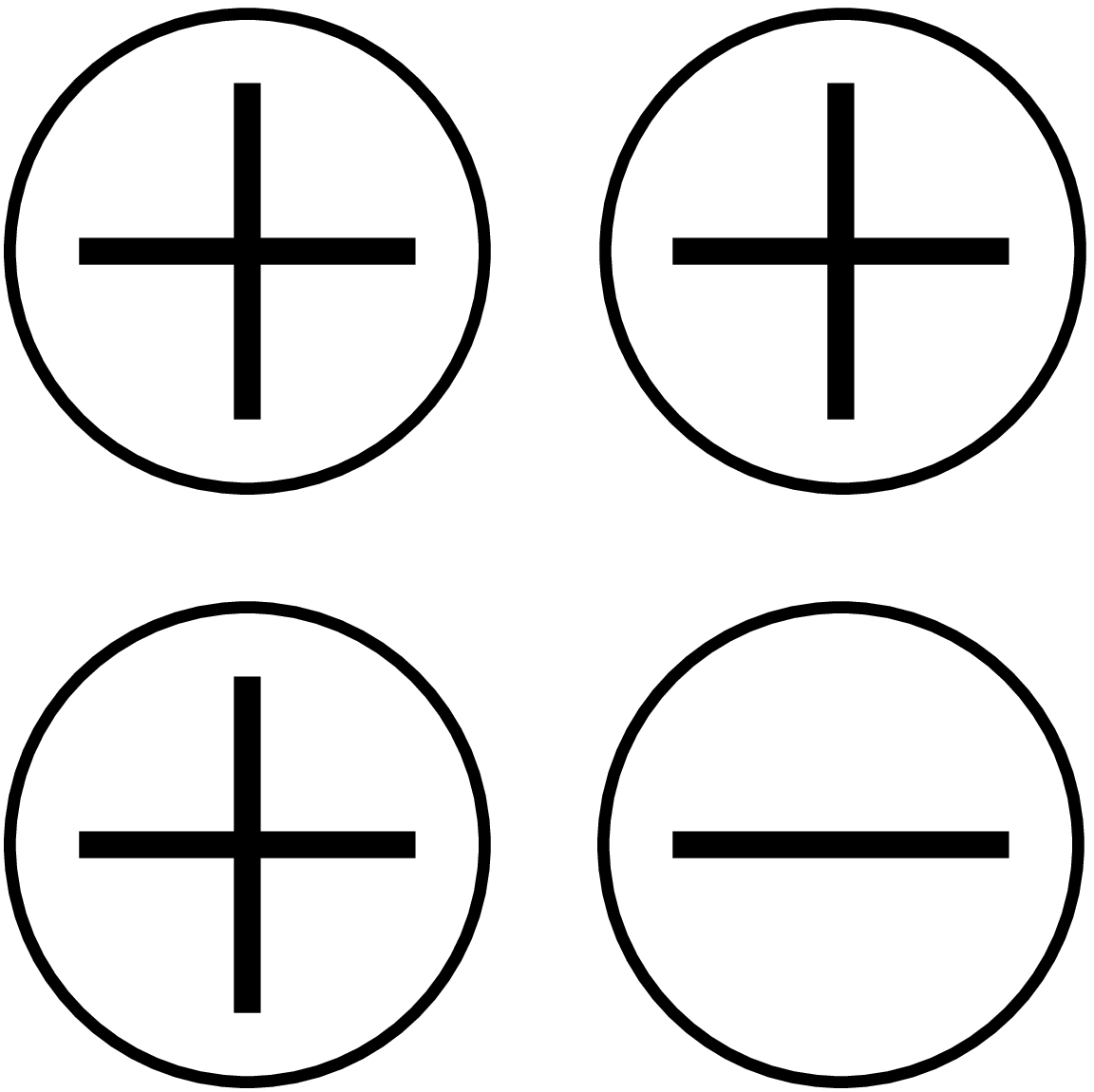}}}
\epsfxsize=1.8cm
\put(3.1,0.1){\mbox{\epsfbox{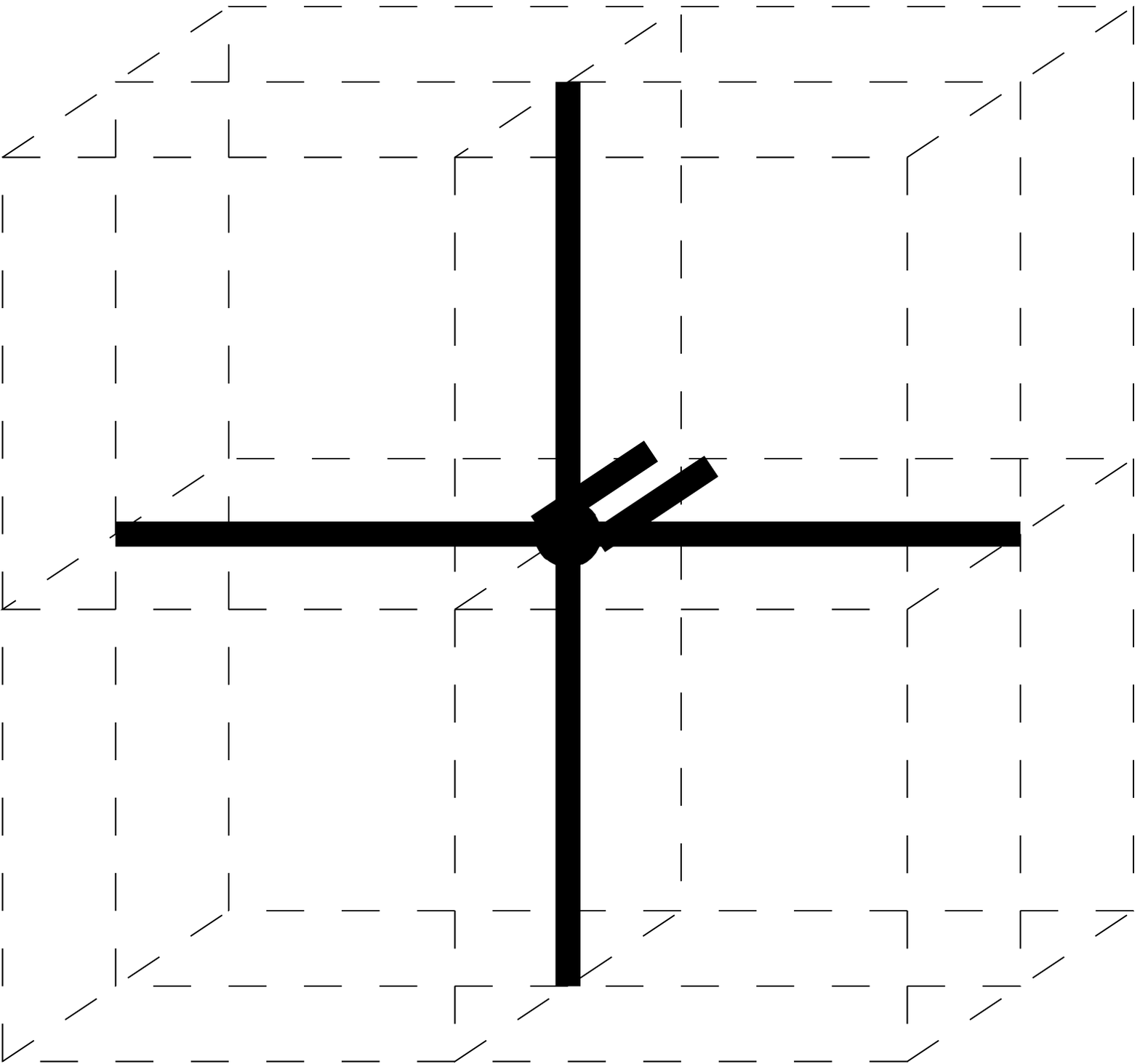}}}
\put(6,0.9){\mbox{$x^8$}}
\epsfxsize=0.7cm
\put(8,0.5){\mbox{\epsfbox{dspin.eps}}}
\epsfxsize=0.7cm
\put(9,0.5){\mbox{\epsfbox{spin04.eps}}}
\epsfxsize=1.8cm
\put(10.6,0.1){\mbox{\epsfbox{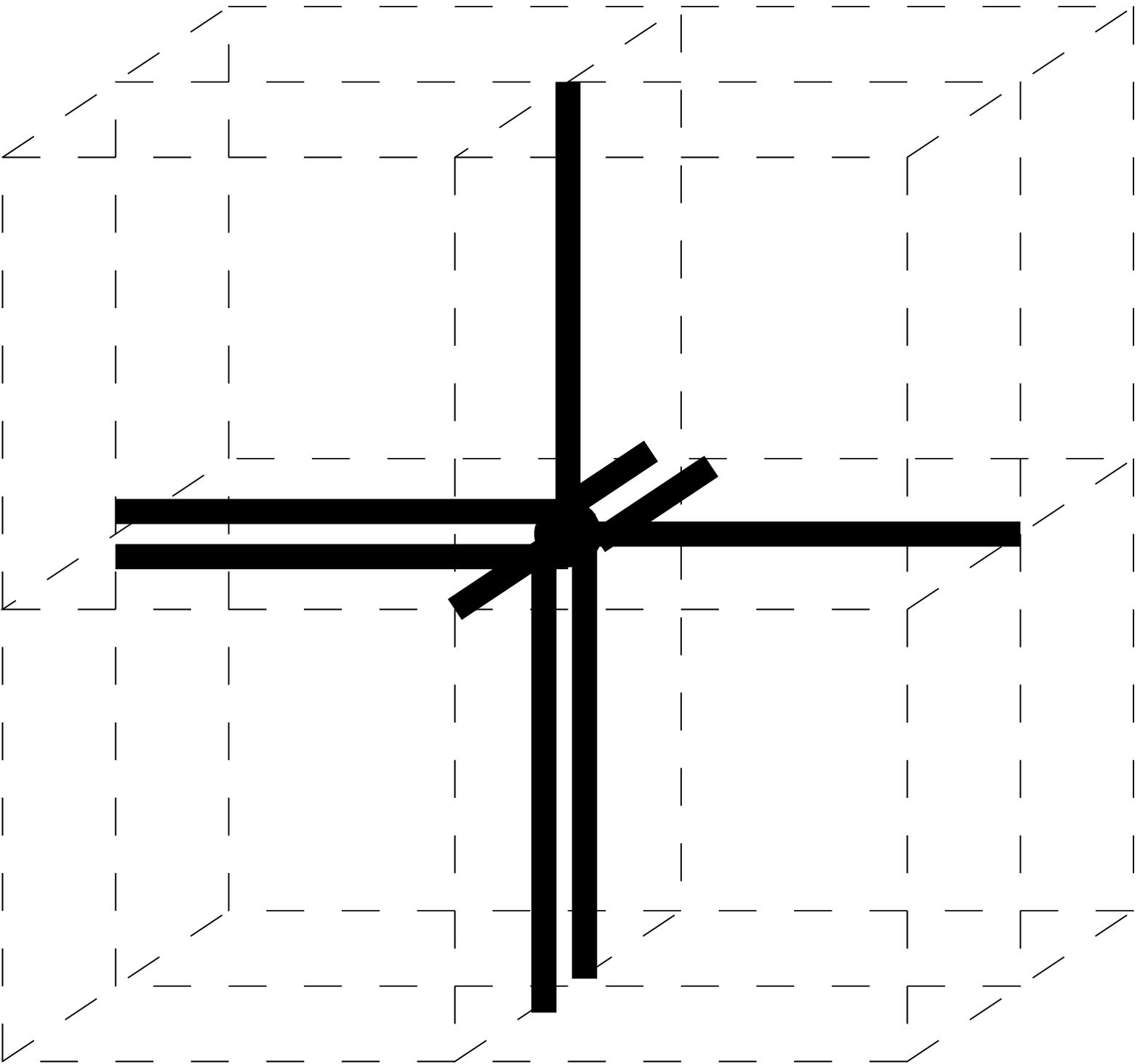}}}
\put(13.5,0.9){\mbox{$x^{15}$}}
\end{picture}\\
\begin{picture}(15,2)
\put(7.5,0){\line(0,1){2}}
\epsfxsize=0.7cm
\put(0.5,0.5){\mbox{\epsfbox{espin.eps}}}
\epsfxsize=0.7cm
\put(1.5,0.5){\mbox{\epsfbox{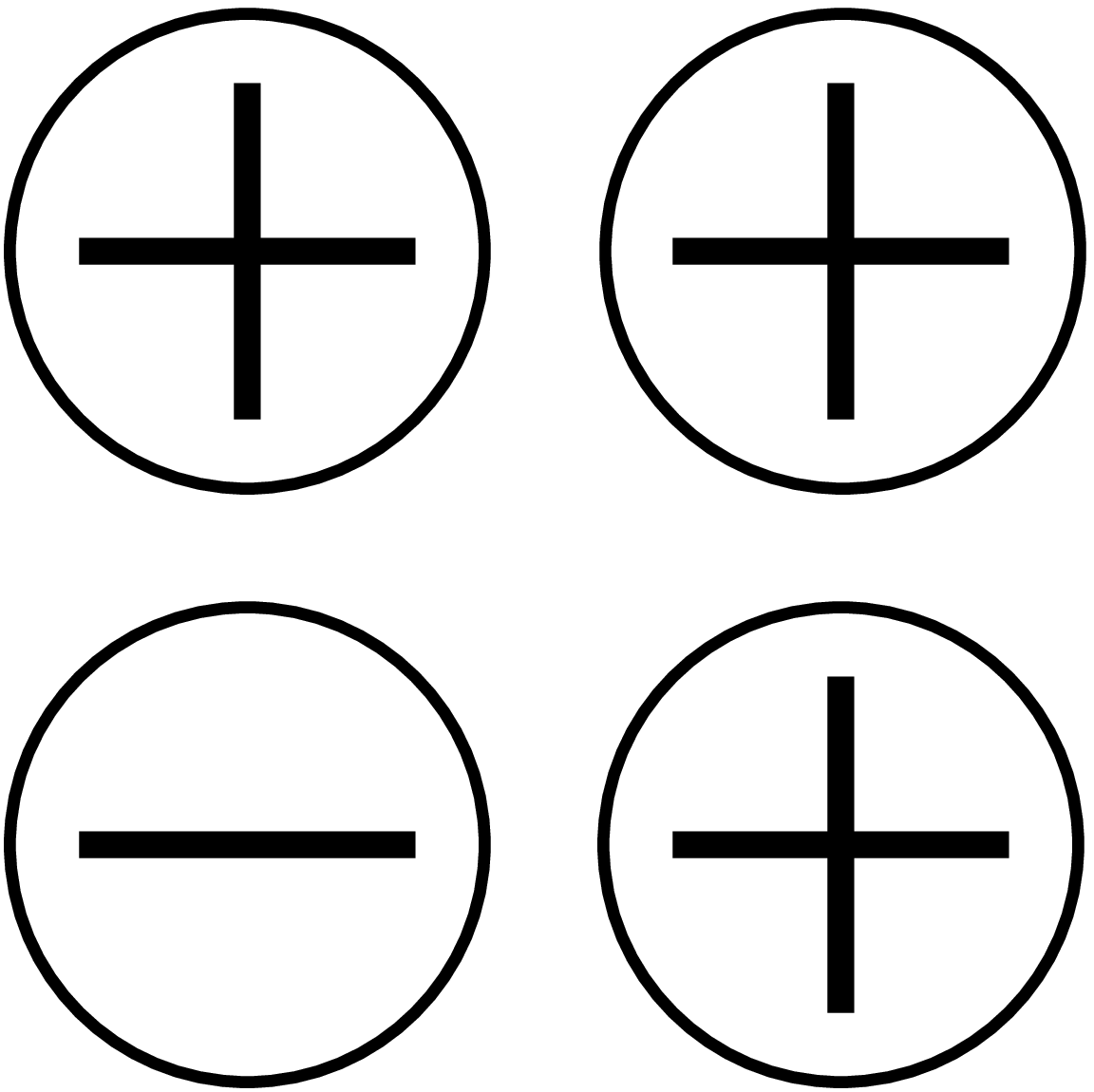}}}
\epsfxsize=1.8cm
\put(3.1,0.1){\mbox{\epsfbox{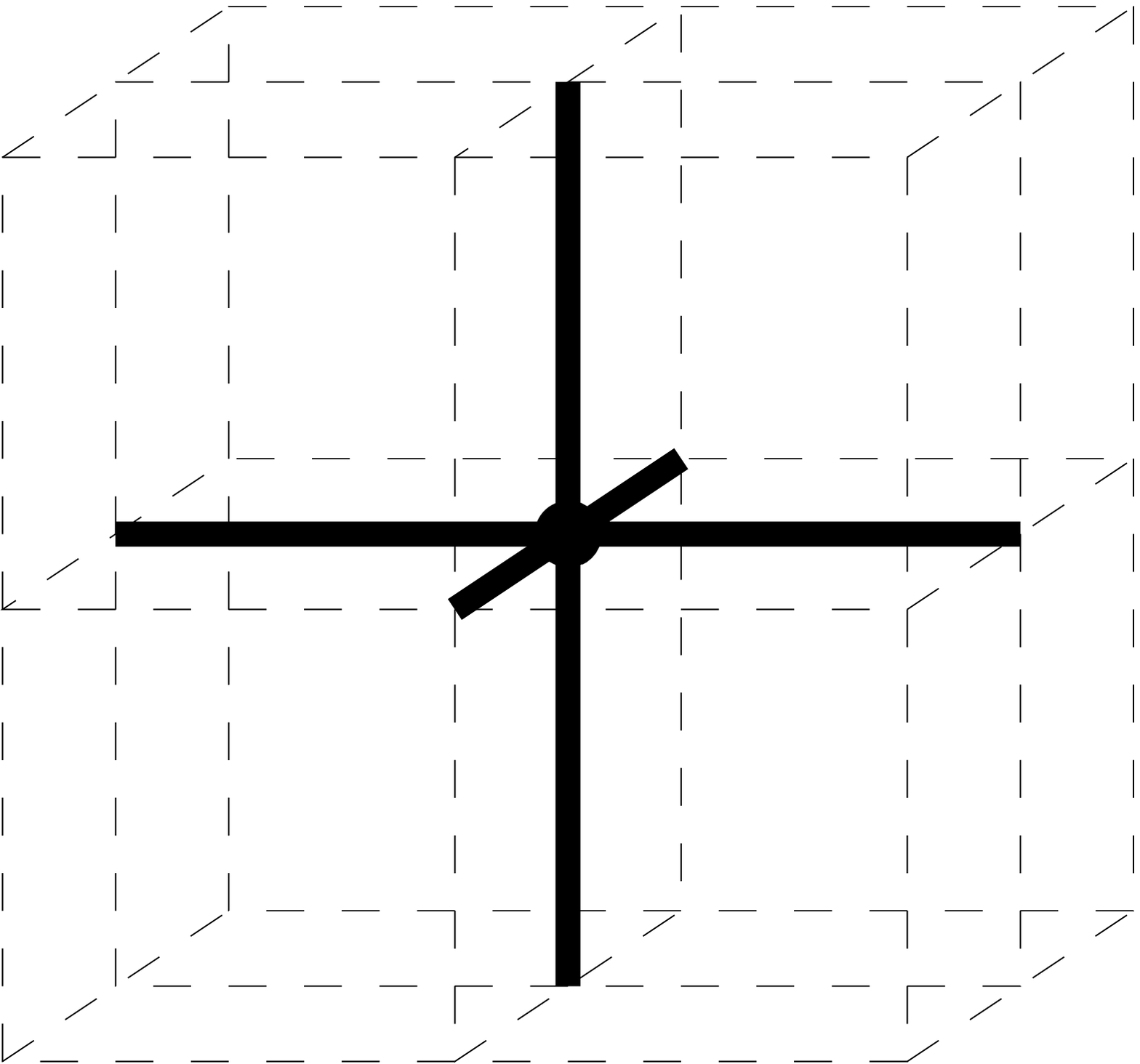}}}
\put(6,0.9){\mbox{$x^6$}}
\epsfxsize=0.7cm
\put(8,0.5){\mbox{\epsfbox{dspin.eps}}}
\epsfxsize=0.7cm
\put(9,0.5){\mbox{\epsfbox{spin05.eps}}}
\epsfxsize=1.8cm
\put(10.6,0.1){\mbox{\epsfbox{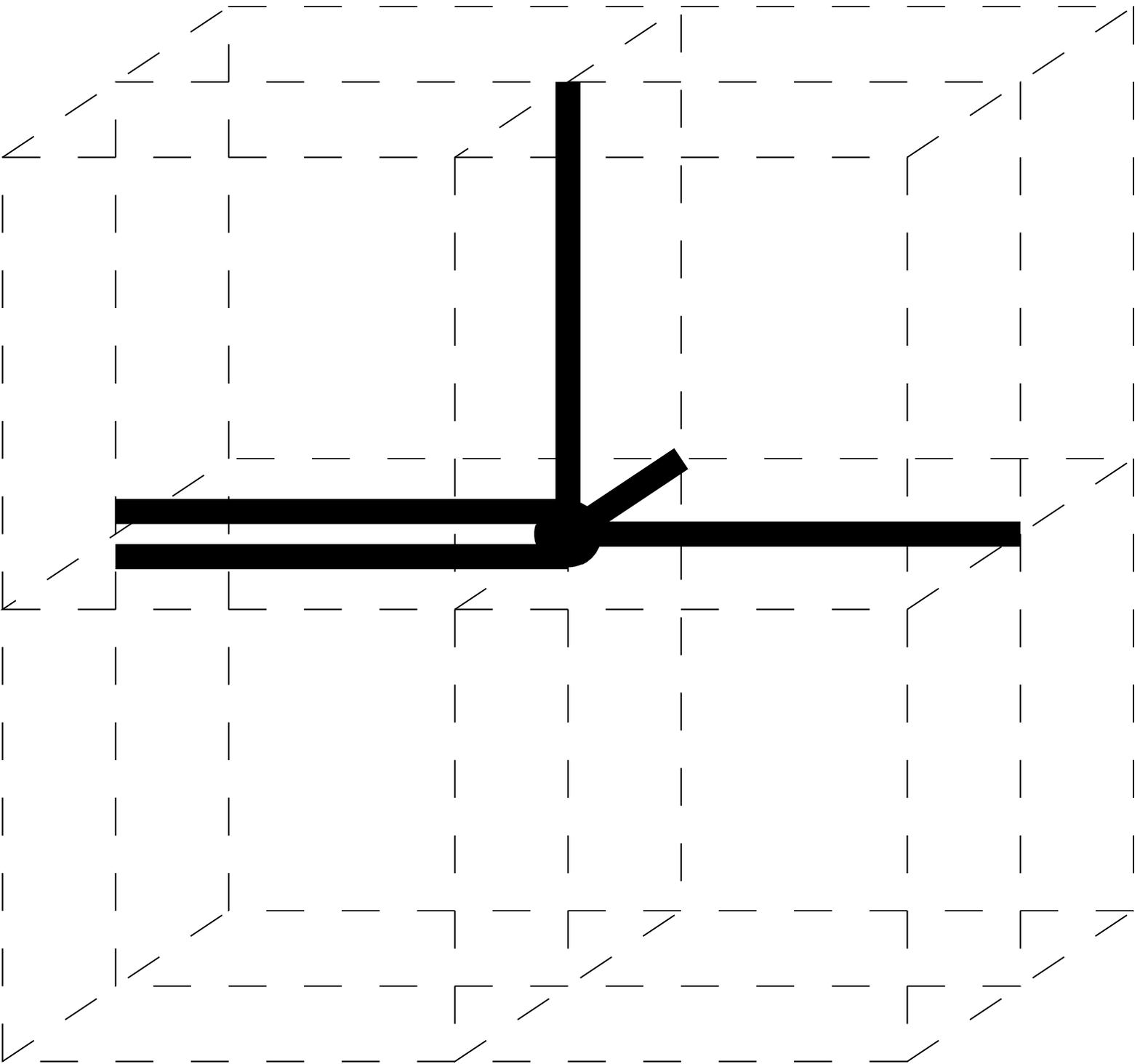}}}
\put(13.5,0.9){\mbox{$x^7$}}
\end{picture}\\
\begin{picture}(15,2)
\put(7.5,0){\line(0,1){2}}
\epsfxsize=0.7cm
\put(0.5,0.5){\mbox{\epsfbox{espin.eps}}}
\epsfxsize=0.7cm
\put(1.5,0.5){\mbox{\epsfbox{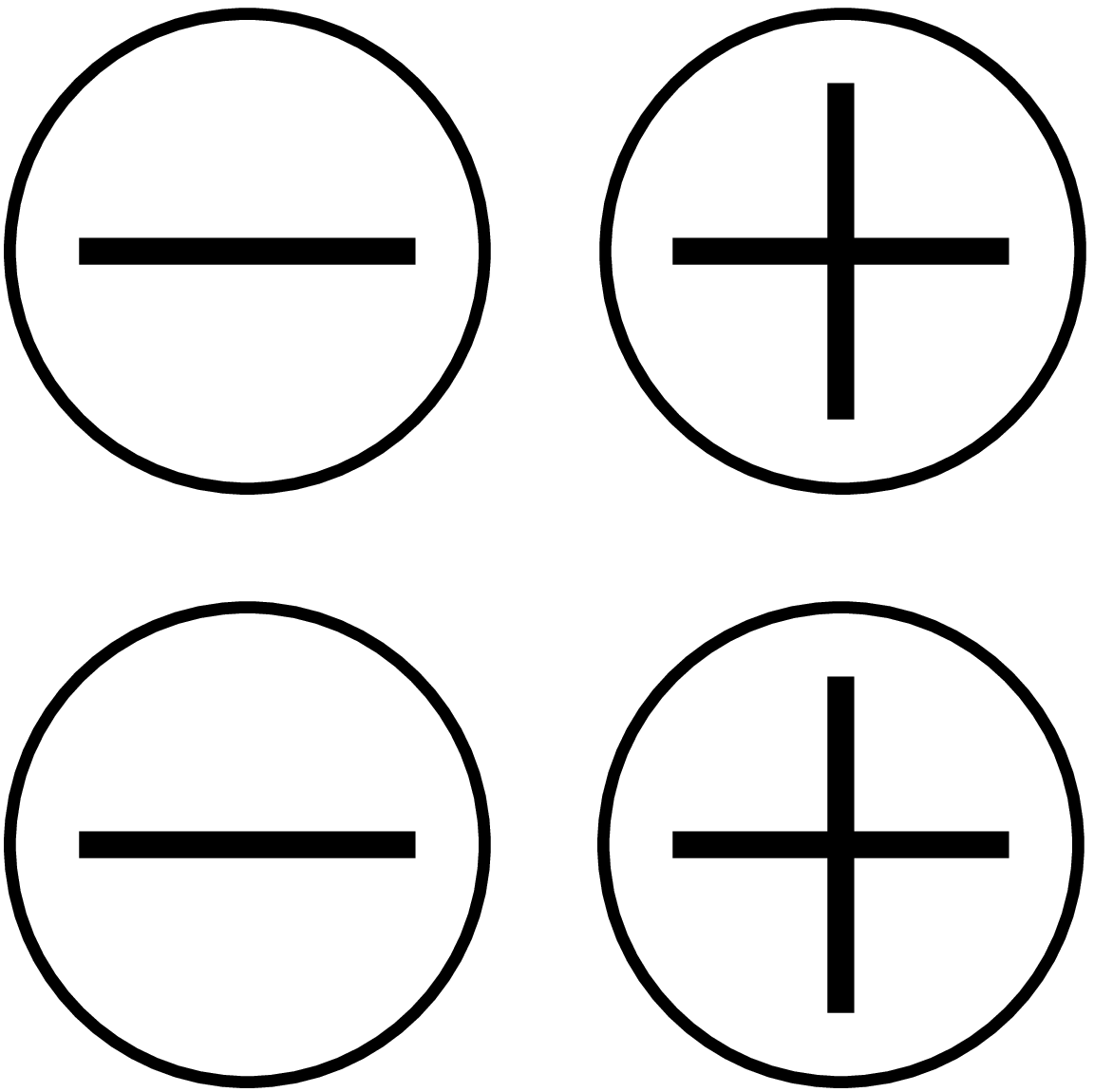}}}
\epsfxsize=1.8cm
\put(3.1,0.1){\mbox{\epsfbox{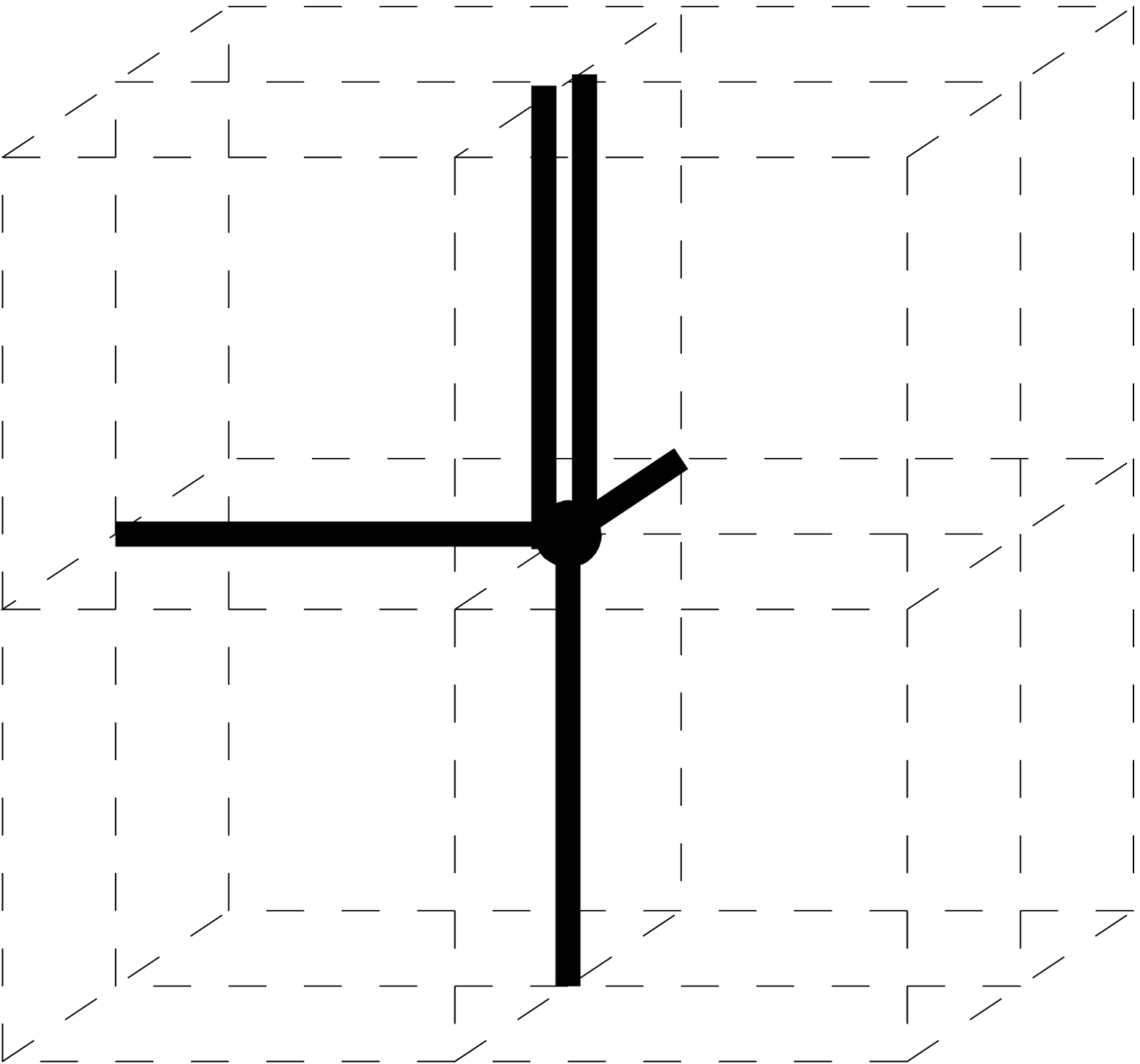}}}
\put(6,0.9){\mbox{$x^7$}}
\epsfxsize=0.7cm
\put(8,0.5){\mbox{\epsfbox{dspin.eps}}}
\epsfxsize=0.7cm
\put(9,0.5){\mbox{\epsfbox{spin06.eps}}}
\epsfxsize=1.8cm
\put(10.6,0.1){\mbox{\epsfbox{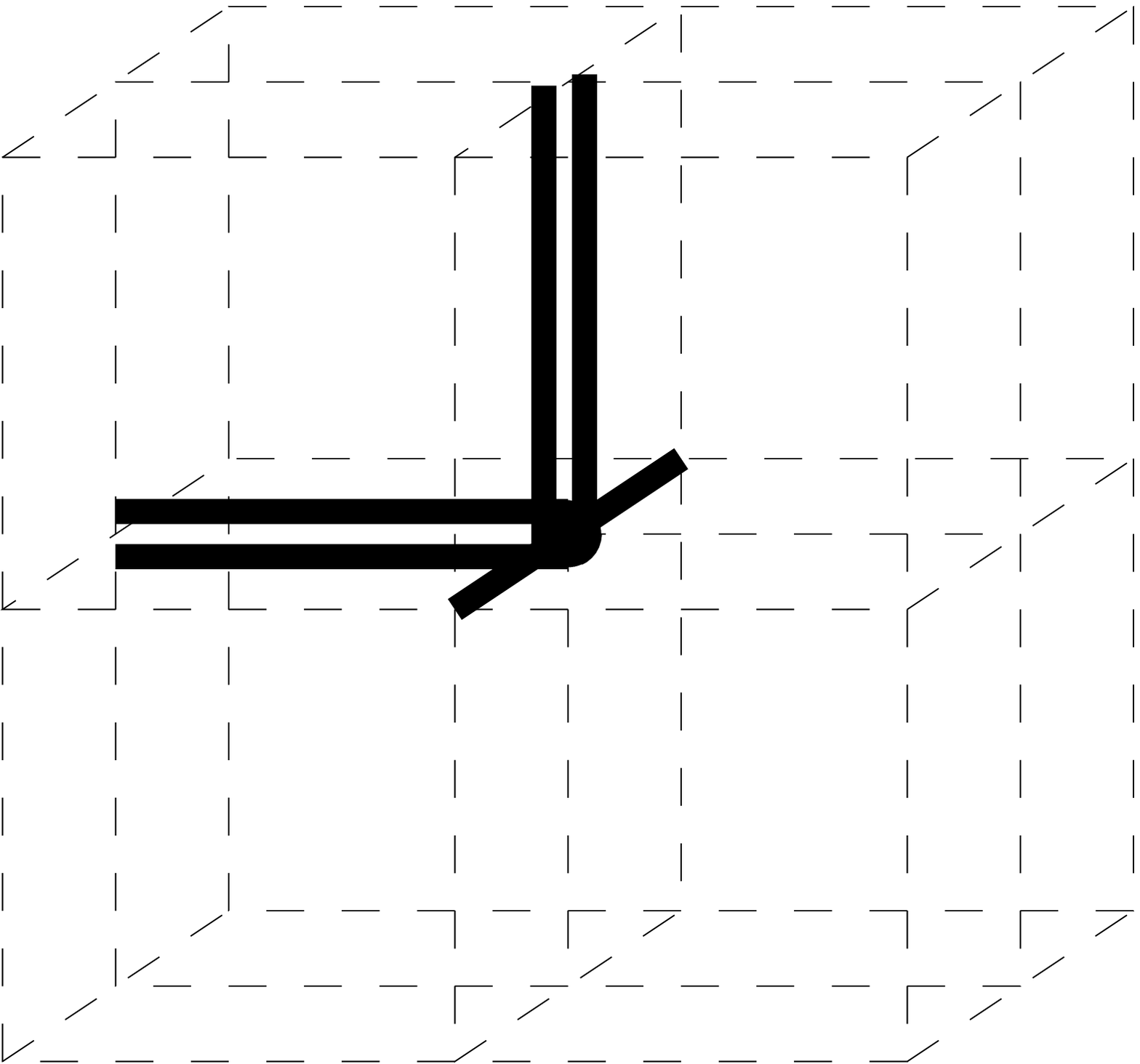}}}
\put(13.5,0.9){\mbox{$x^{10}$}}
\end{picture}\\
\begin{picture}(15,2)
\put(7.5,0){\line(0,1){2}}
\epsfxsize=0.7cm
\put(0.5,0.5){\mbox{\epsfbox{espin.eps}}}
\epsfxsize=0.7cm
\put(1.5,0.5){\mbox{\epsfbox{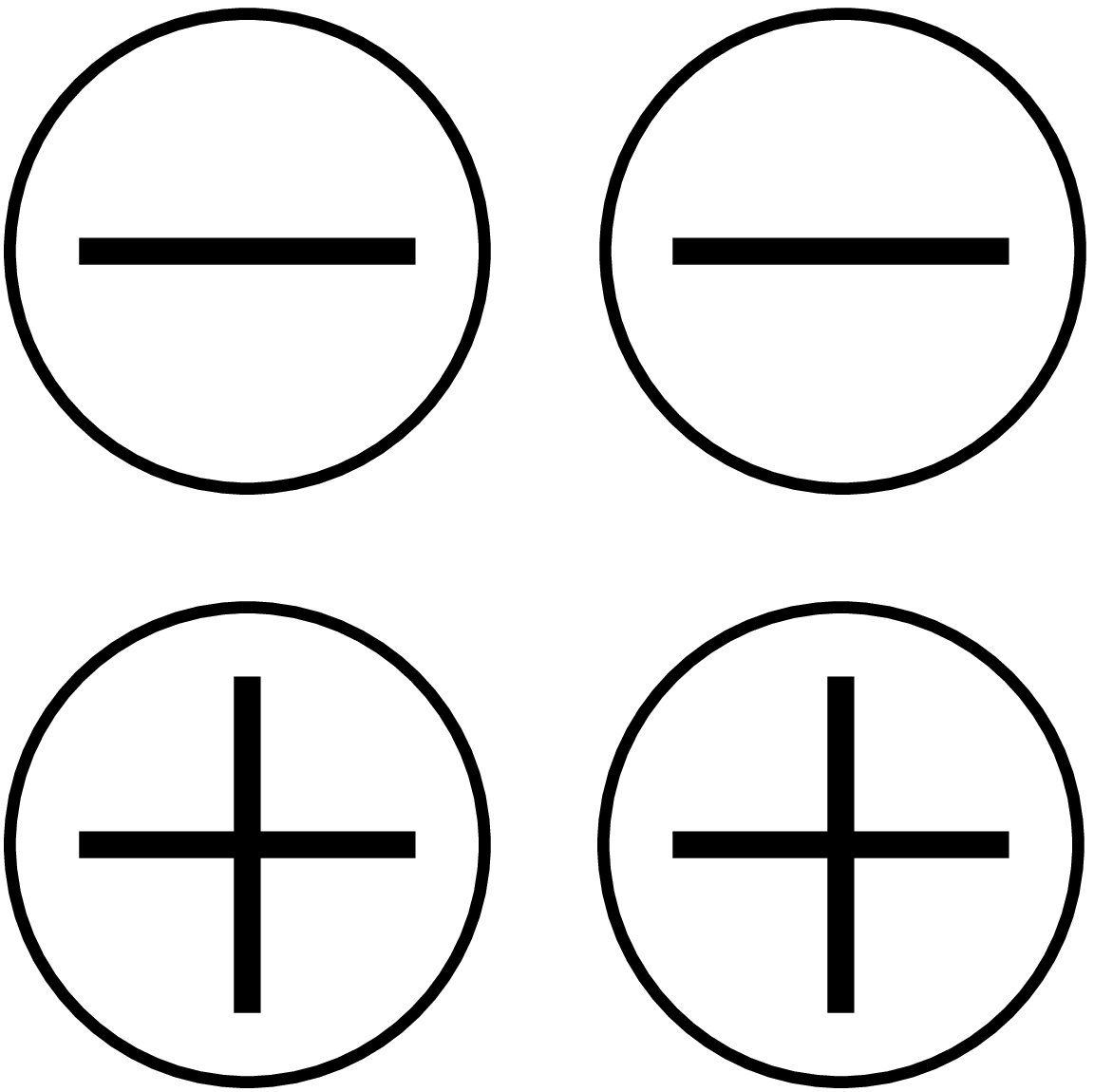}}}
\epsfxsize=1.8cm
\put(3.1,0.1){\mbox{\epsfbox{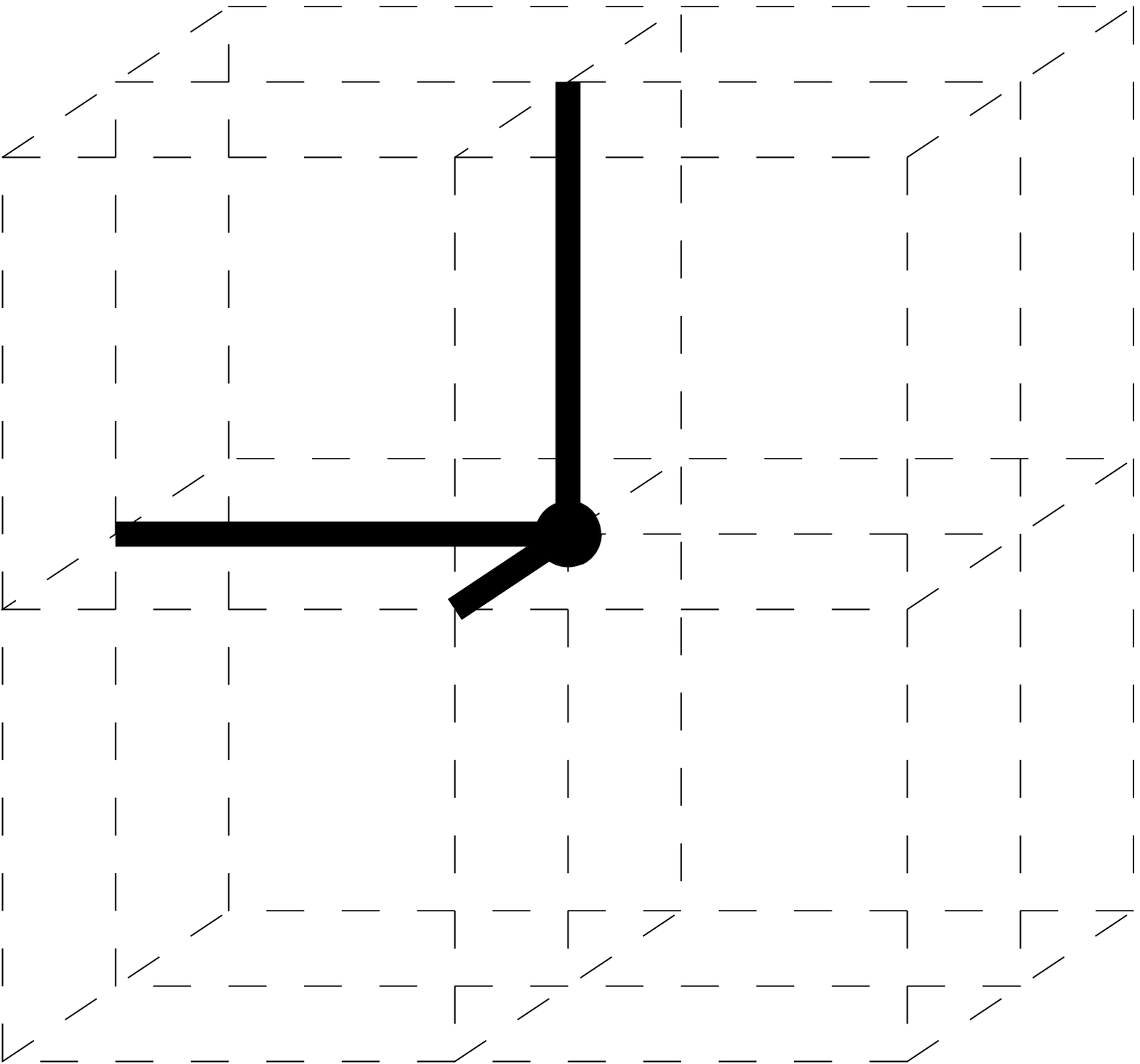}}}
\put(6,0.9){\mbox{$x^3$}}
\epsfxsize=0.7cm
\put(8,0.5){\mbox{\epsfbox{dspin.eps}}}
\epsfxsize=0.7cm
\put(9,0.5){\mbox{\epsfbox{spin07.eps}}}
\epsfxsize=1.8cm
\put(10.6,0.1){\mbox{\epsfbox{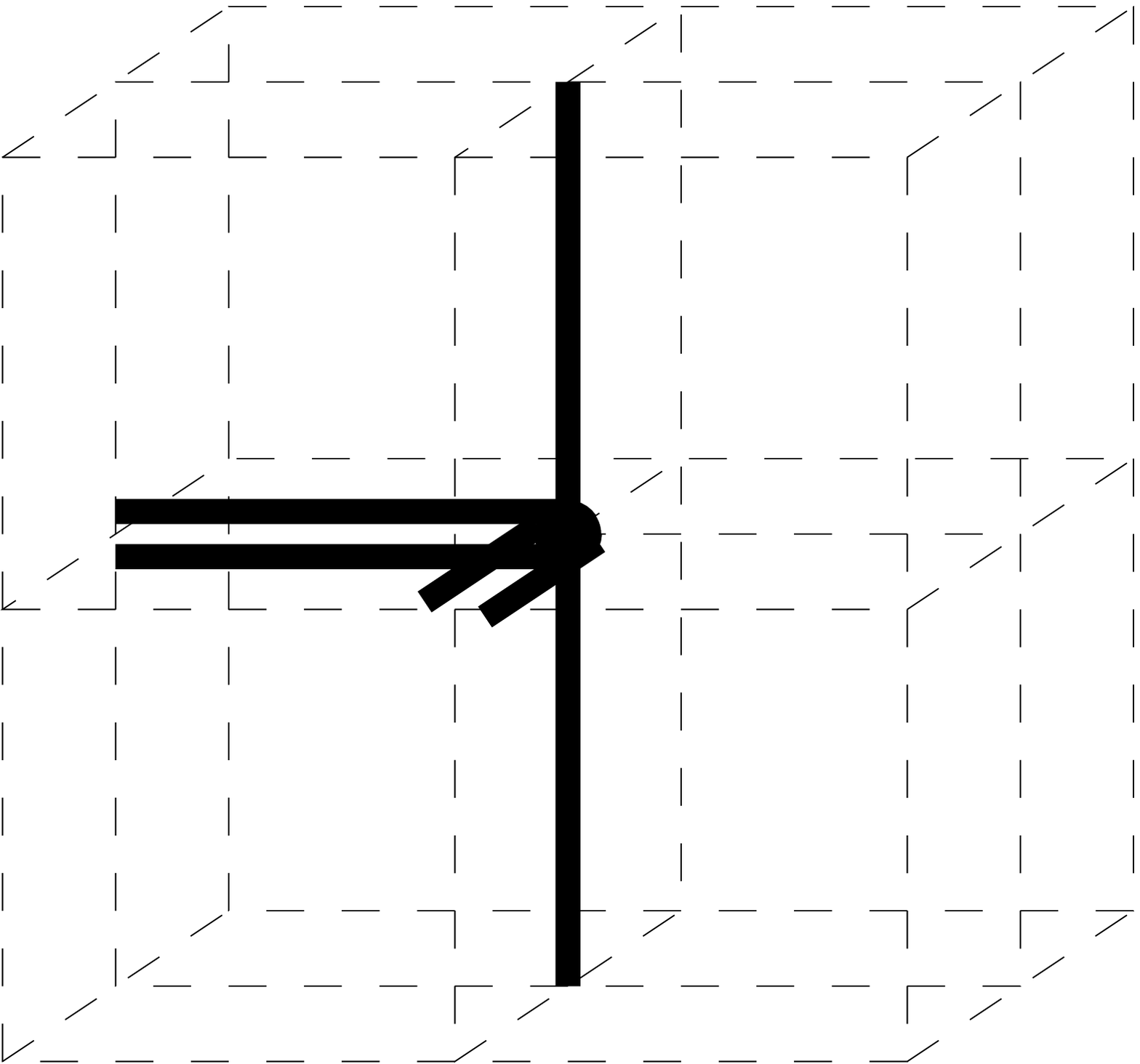}}}
\put(13.5,0.9){\mbox{$x^{10}$}}
\end{picture}\\
\begin{picture}(15,2)
\put(7.5,0){\line(0,1){2}}
\epsfxsize=0.7cm
\put(0.5,0.5){\mbox{\epsfbox{espin.eps}}}
\epsfxsize=0.7cm
\put(1.5,0.5){\mbox{\epsfbox{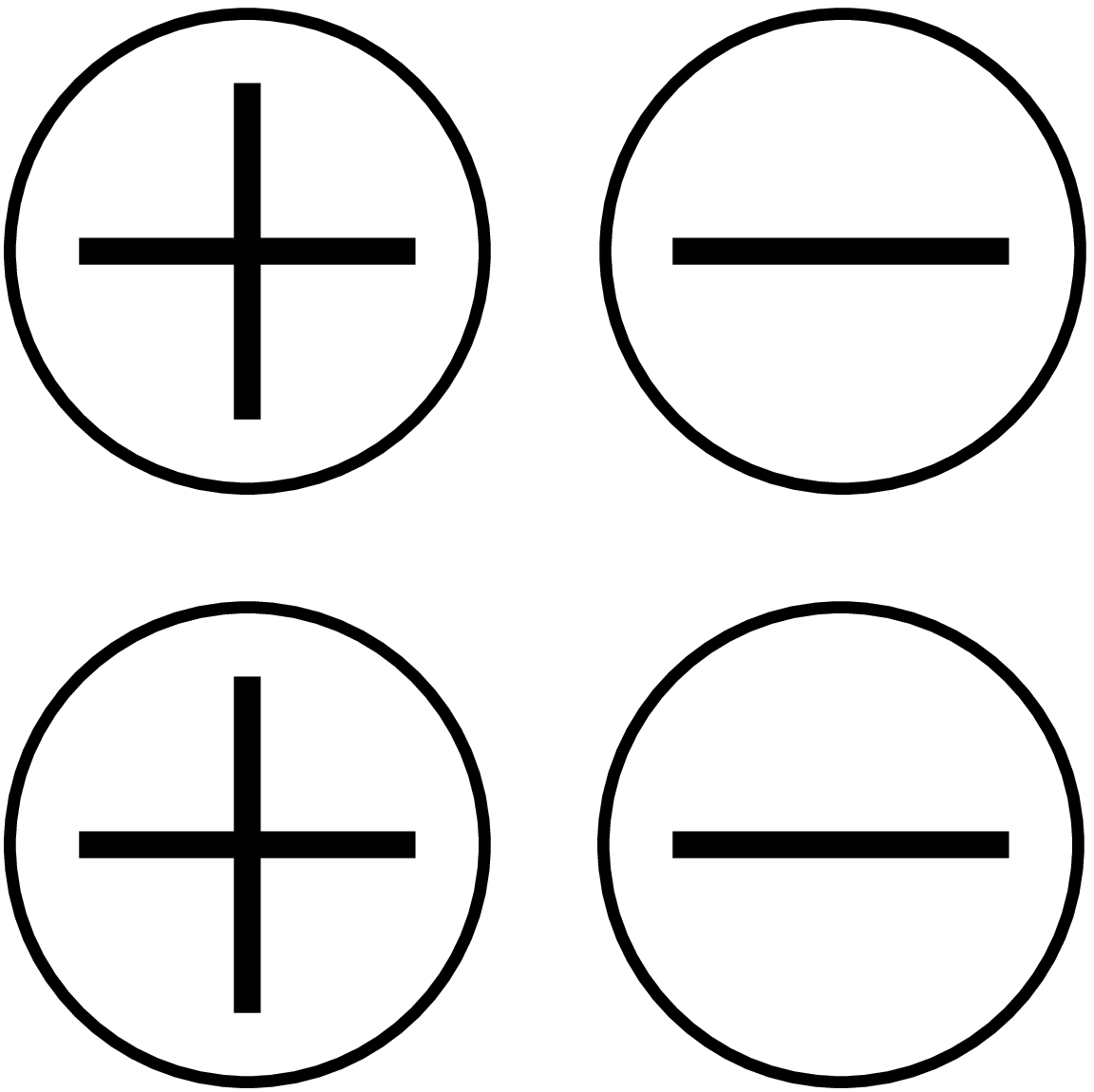}}}
\epsfxsize=1.8cm
\put(3.1,0.1){\mbox{\epsfbox{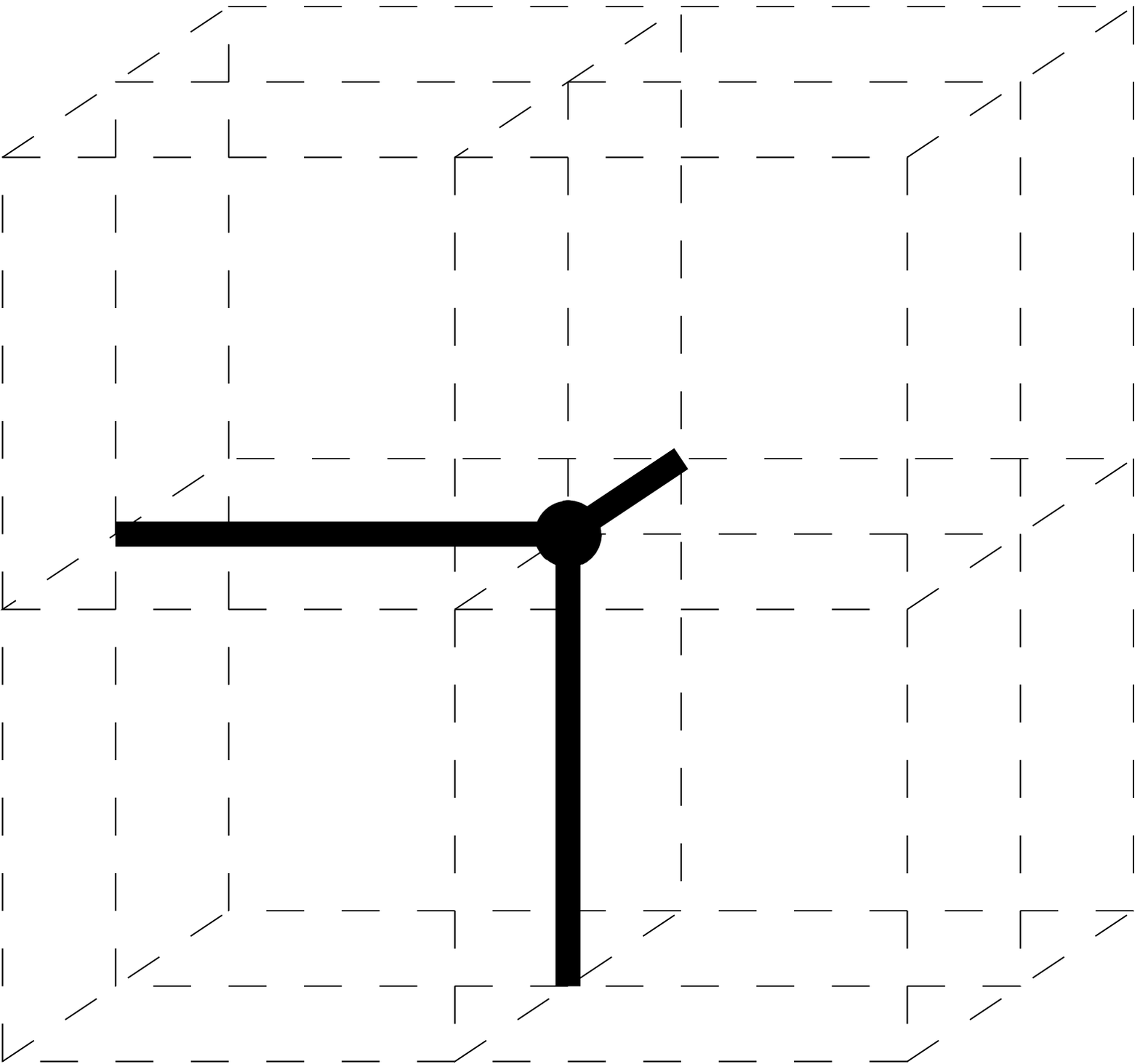}}}
\put(6,0.9){\mbox{$x^3$}}
\epsfxsize=0.7cm
\put(8,0.5){\mbox{\epsfbox{dspin.eps}}}
\epsfxsize=0.7cm
\put(9,0.5){\mbox{\epsfbox{spin08.eps}}}
\epsfxsize=1.8cm
\put(10.6,0.1){\mbox{\epsfbox{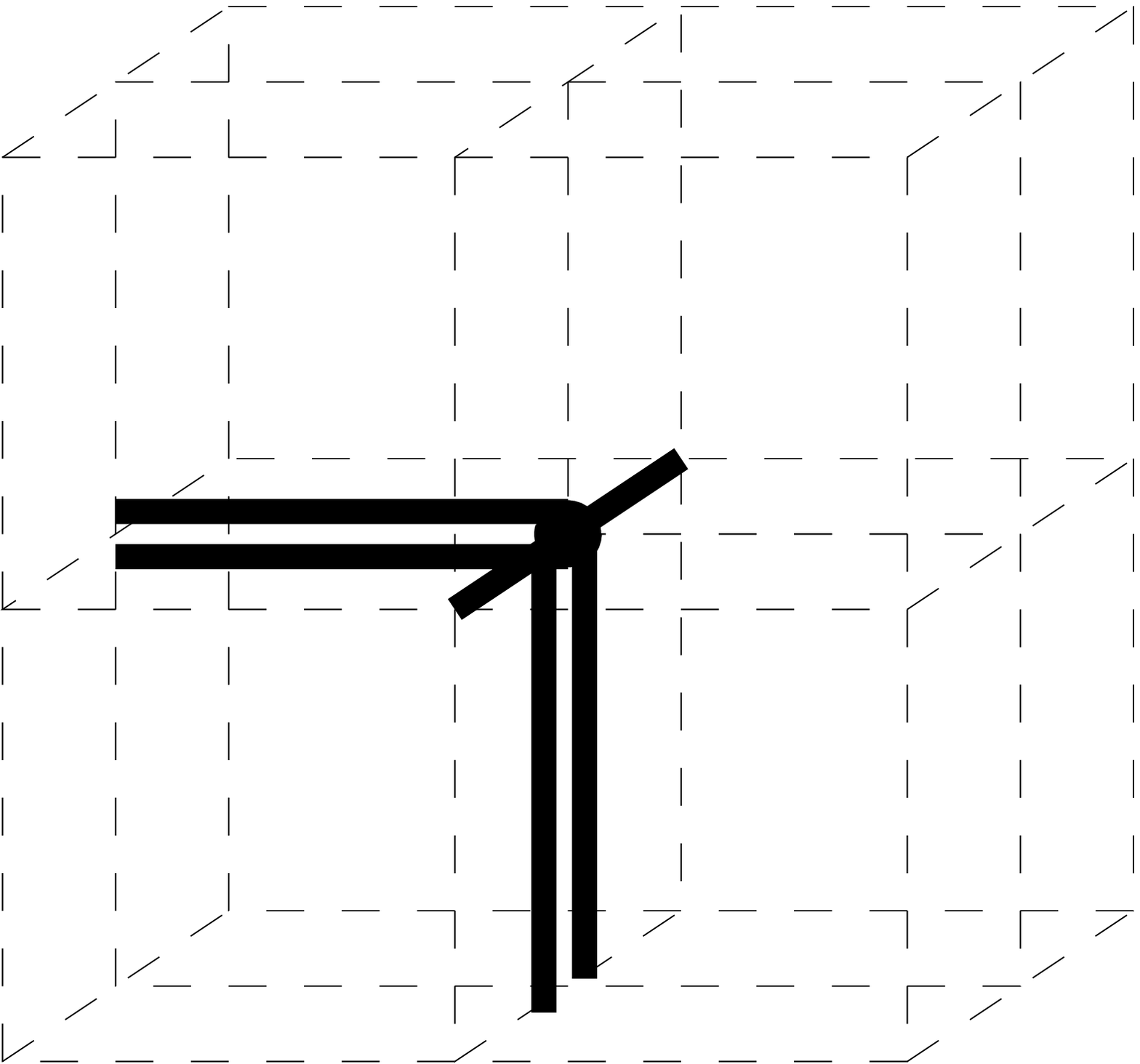}}}
\put(13.5,0.9){\mbox{$x^{10}$}}
\end{picture}\\
\begin{picture}(15,2)
\put(7.5,0){\line(0,1){2}}
\epsfxsize=0.7cm
\put(0.5,0.5){\mbox{\epsfbox{espin.eps}}}
\epsfxsize=0.7cm
\put(1.5,0.5){\mbox{\epsfbox{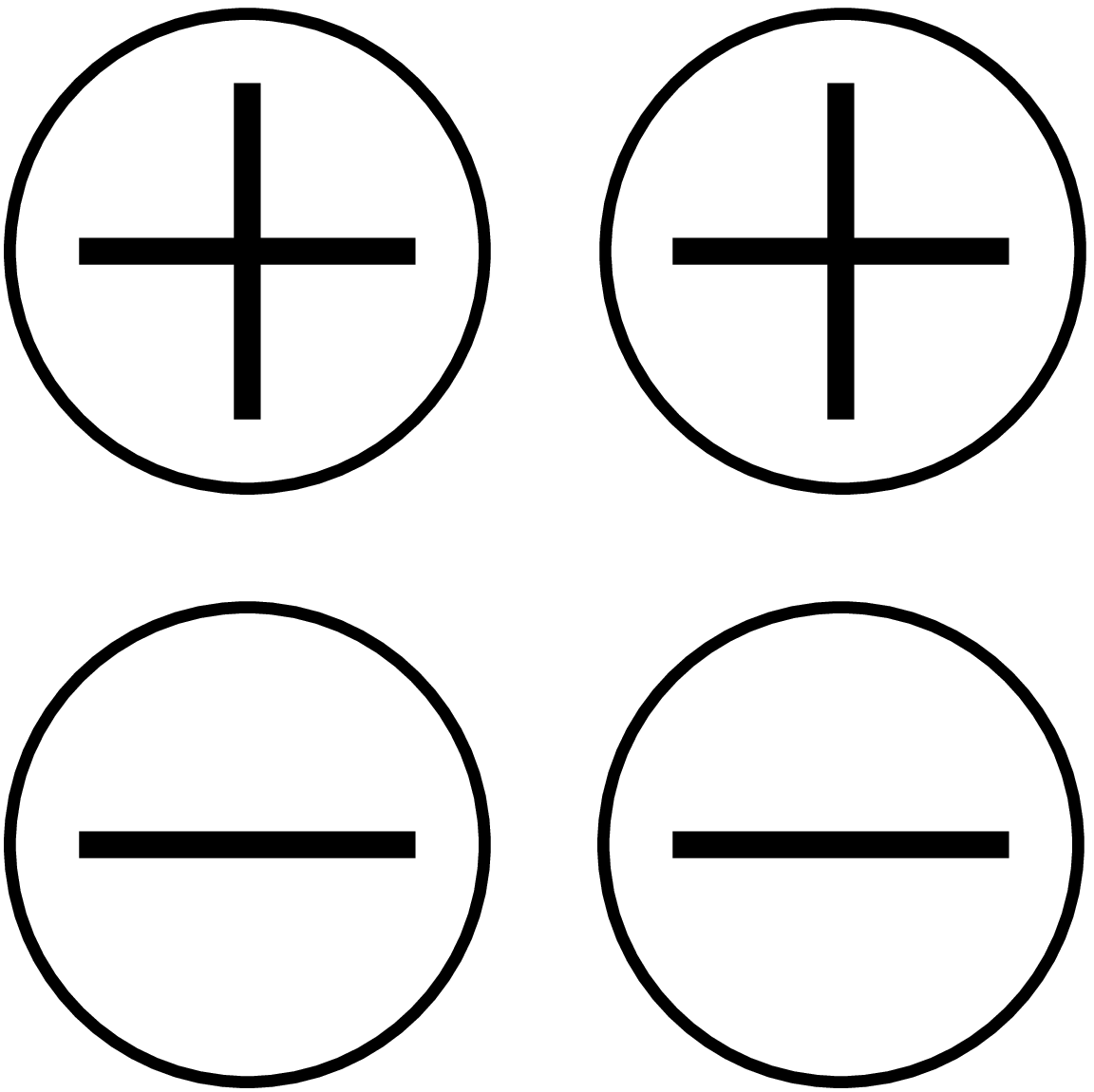}}}
\epsfxsize=1.8cm
\put(3.1,0.1){\mbox{\epsfbox{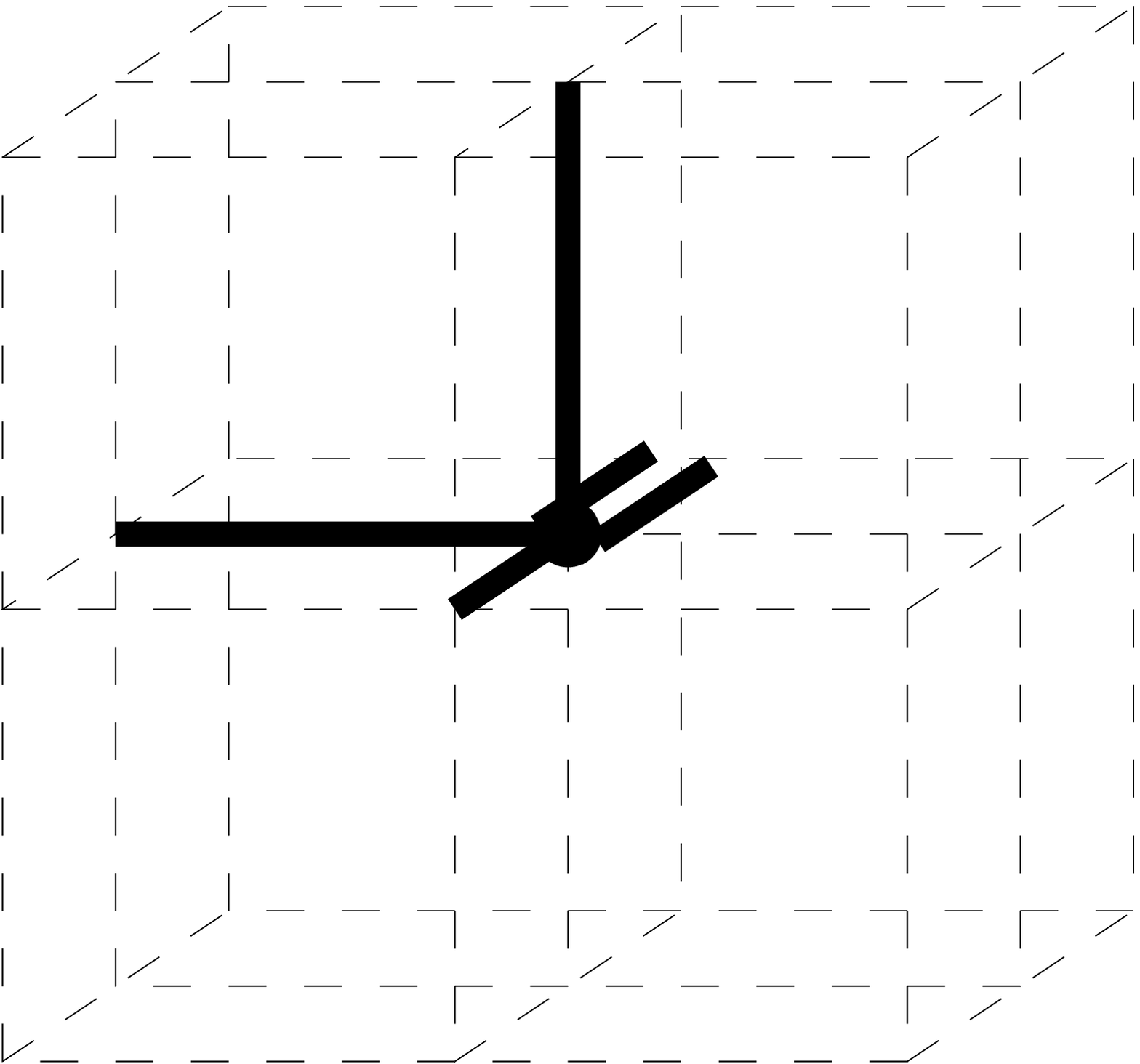}}}
\put(6,0.9){\mbox{$x^7$}}
\epsfxsize=0.7cm
\put(8,0.5){\mbox{\epsfbox{dspin.eps}}}
\epsfxsize=0.7cm
\put(9,0.5){\mbox{\epsfbox{spin09.eps}}}
\epsfxsize=1.8cm
\put(10.6,0.1){\mbox{\epsfbox{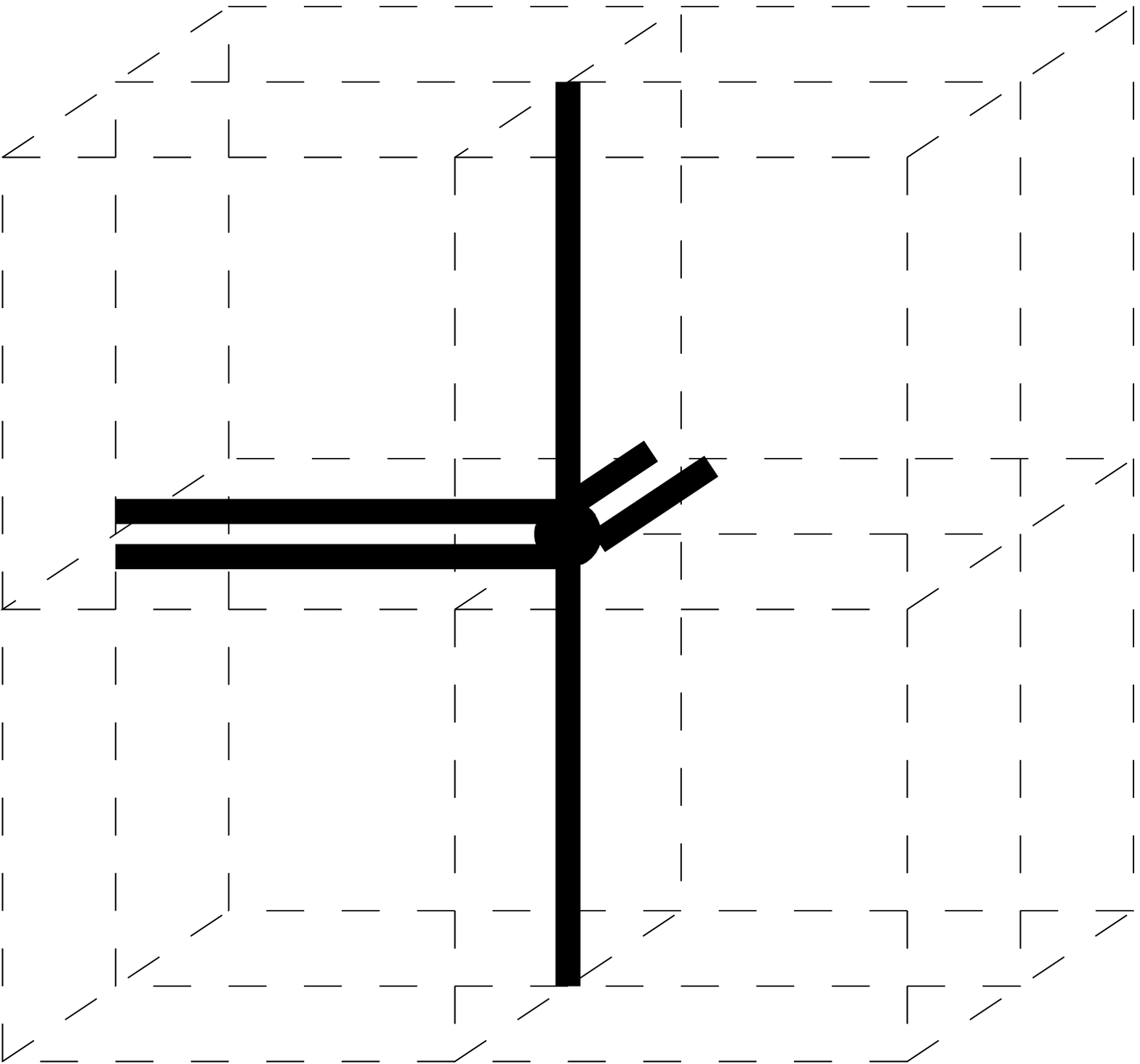}}}
\put(13.5,0.9){\mbox{$x^{10}$}}
\end{picture}\\
\begin{picture}(15,2)
\put(7.5,0){\line(0,1){2}}
\epsfxsize=0.7cm
\put(0.5,0.5){\mbox{\epsfbox{espin.eps}}}
\epsfxsize=0.7cm
\put(1.5,0.5){\mbox{\epsfbox{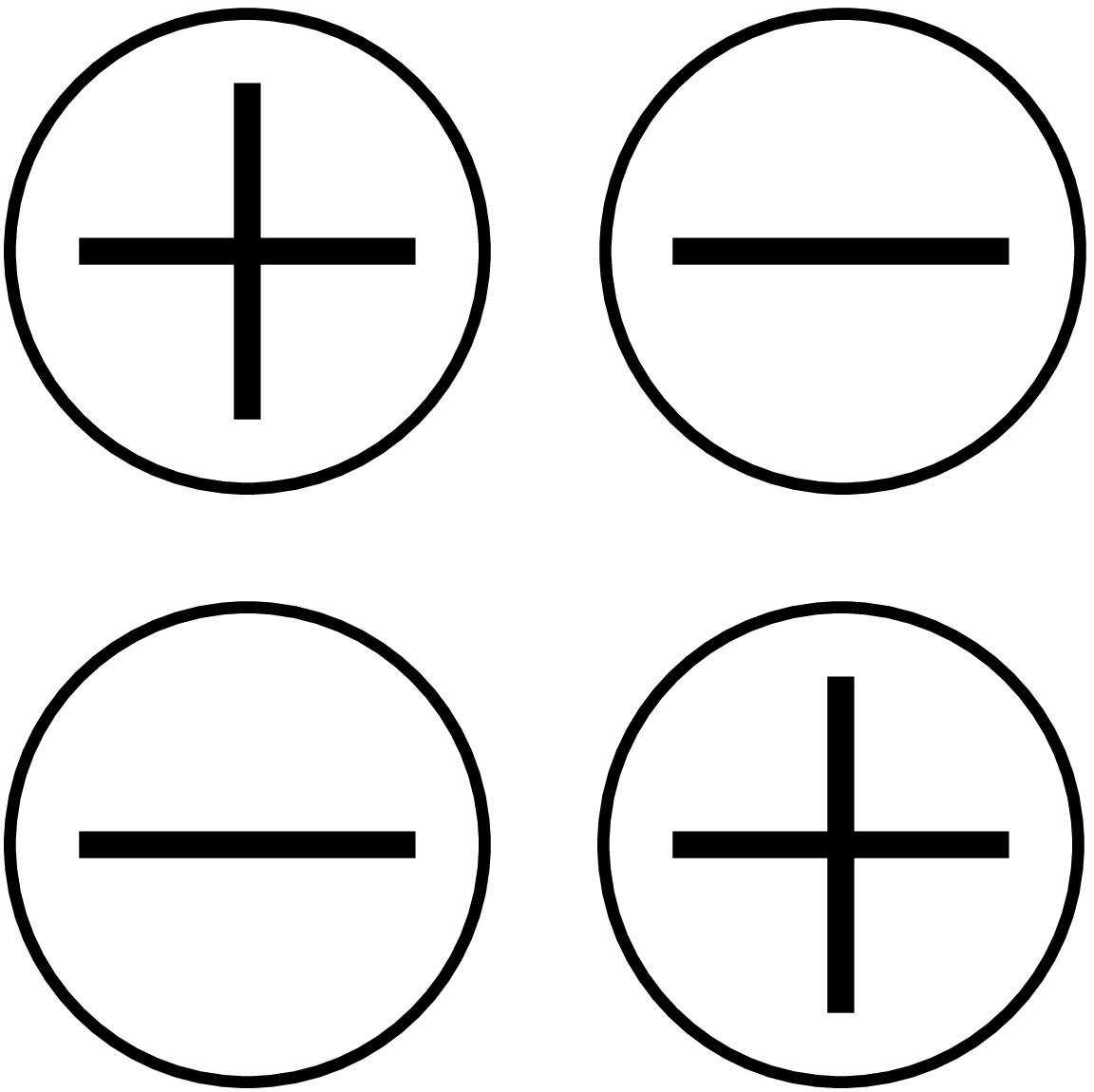}}}
\epsfxsize=1.8cm
\put(3.1,0.1){\mbox{\epsfbox{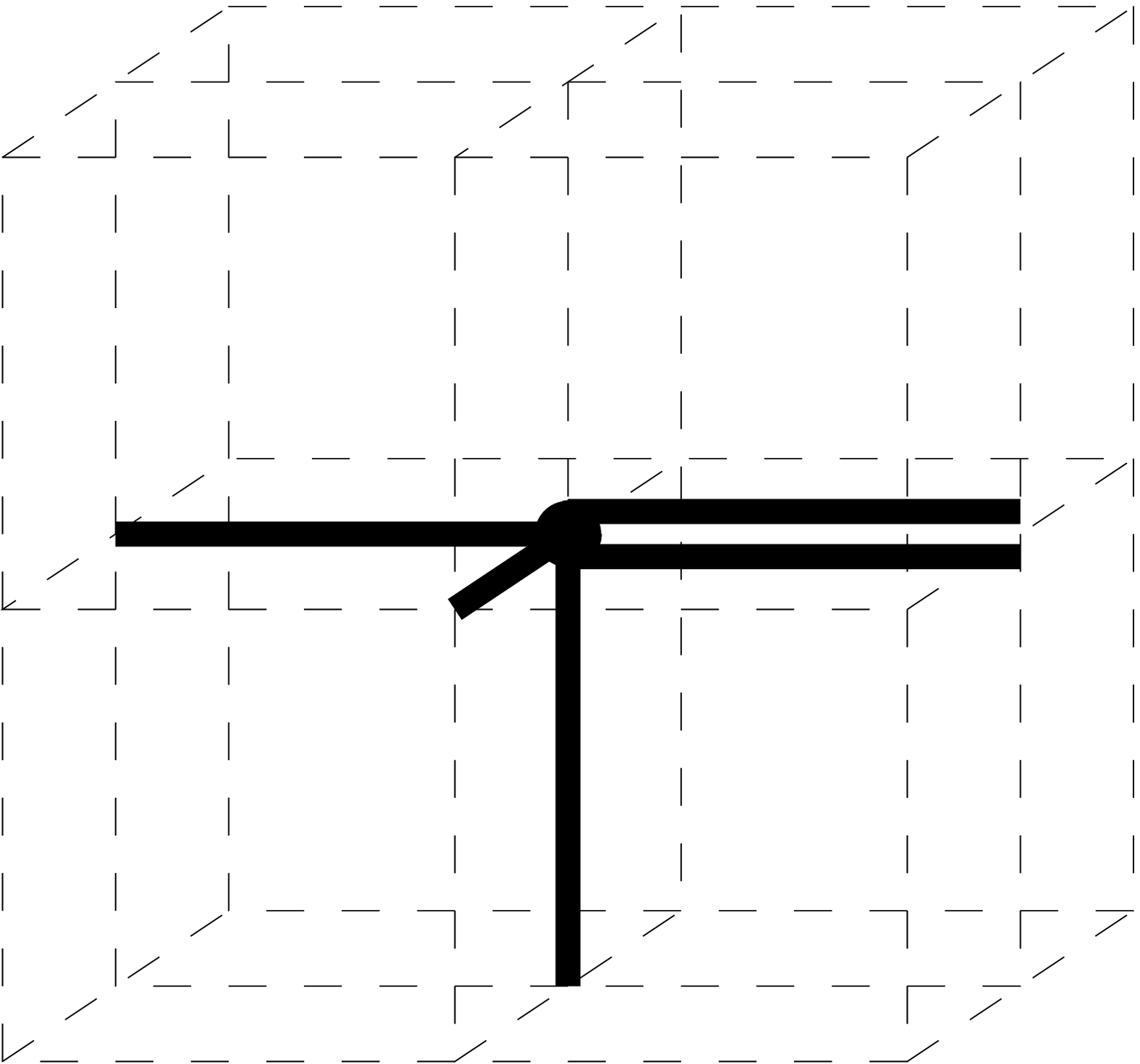}}}
\put(6,0.9){\mbox{$x^7$}}
\epsfxsize=0.7cm
\put(8,0.5){\mbox{\epsfbox{dspin.eps}}}
\epsfxsize=0.7cm
\put(9,0.5){\mbox{\epsfbox{spin10.eps}}}
\epsfxsize=1.8cm
\put(10.6,0.1){\mbox{\epsfbox{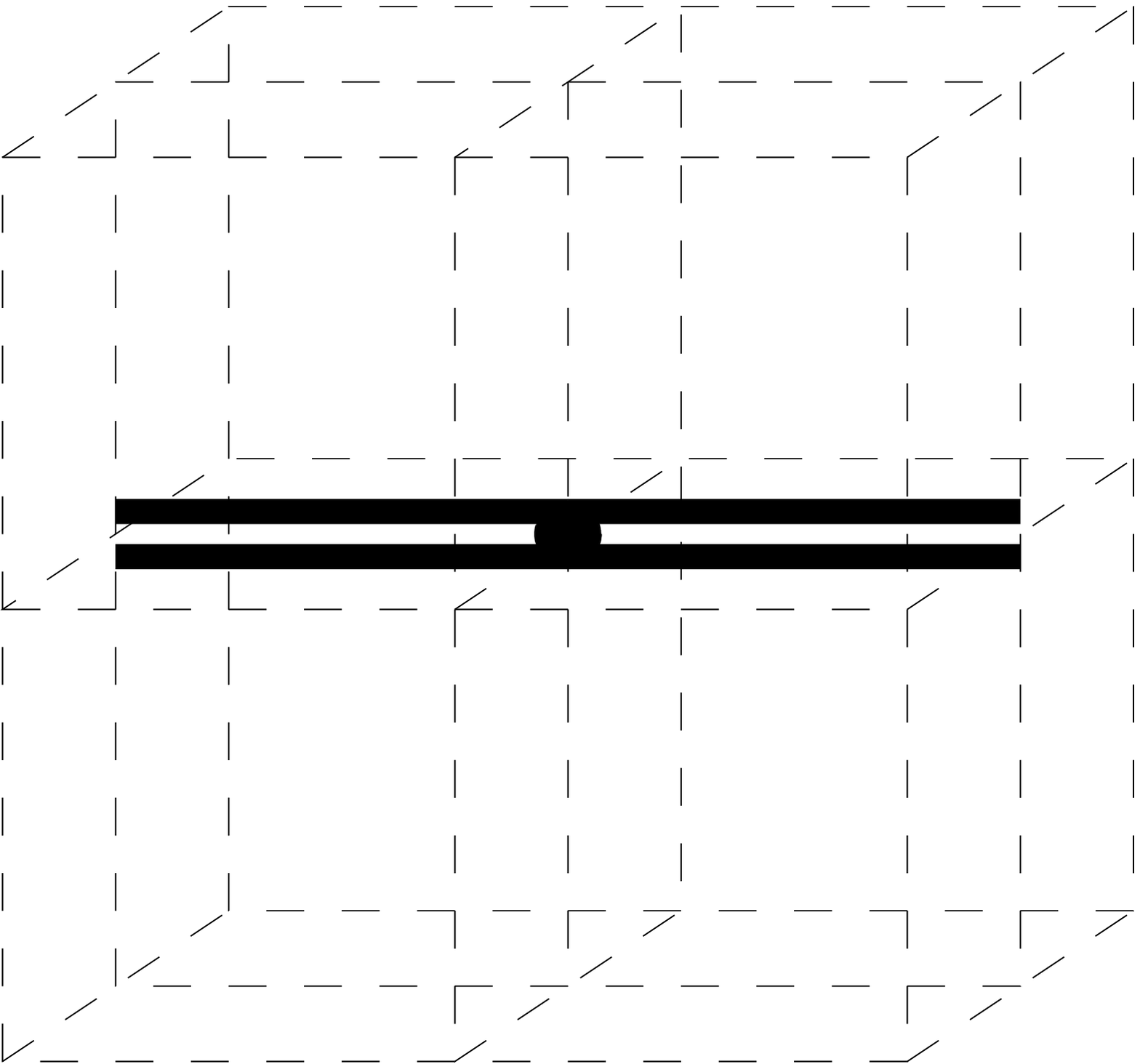}}}
\put(13.5,0.9){\mbox{$x^8$}}
\end{picture}\\
\begin{picture}(15,2)
\put(7.5,0){\line(0,1){2}}
\epsfxsize=0.7cm
\put(0.5,0.5){\mbox{\epsfbox{espin.eps}}}
\epsfxsize=0.7cm
\put(1.5,0.5){\mbox{\epsfbox{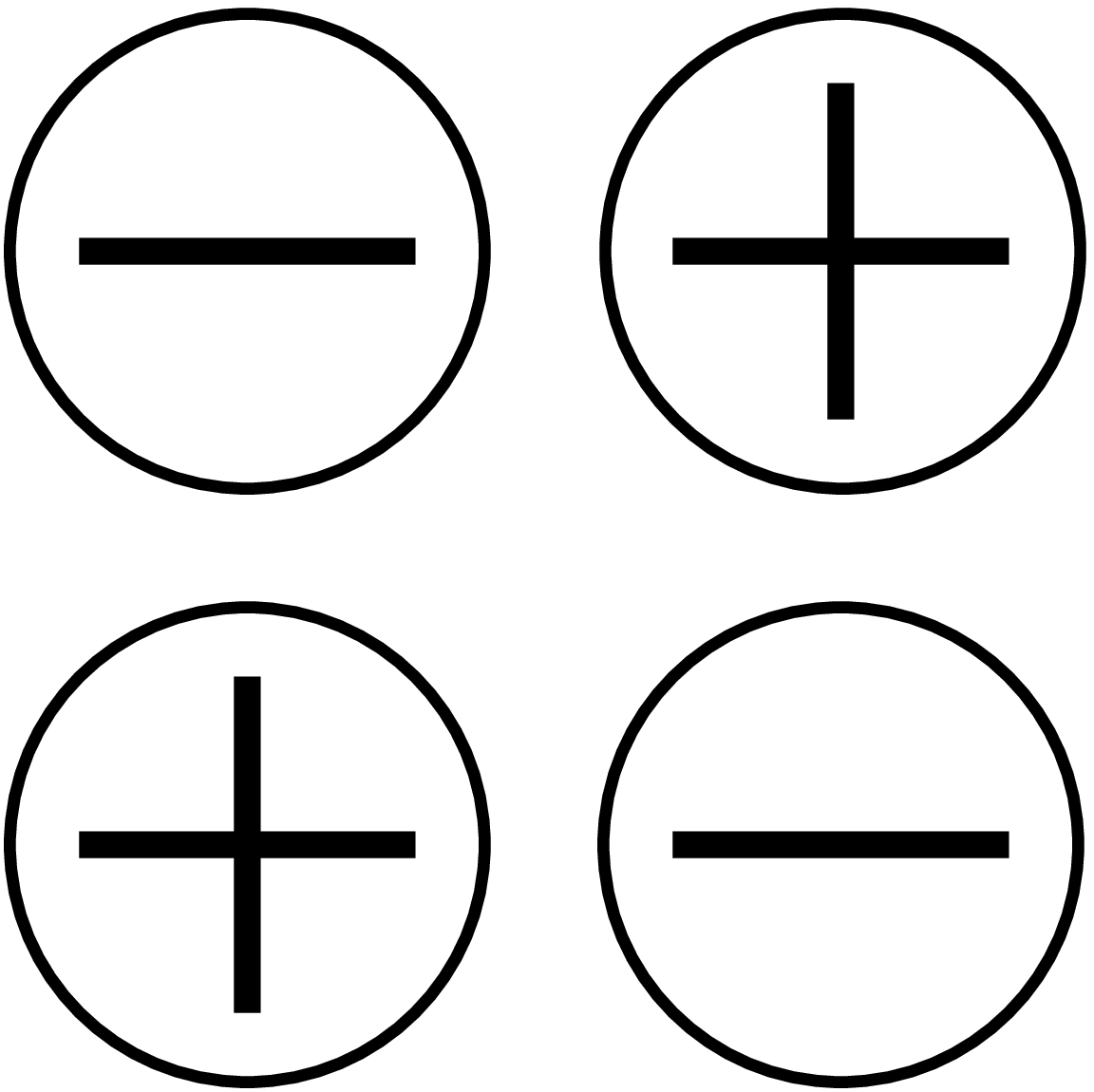}}}
\epsfxsize=1.8cm
\put(3.1,0.1){\mbox{\epsfbox{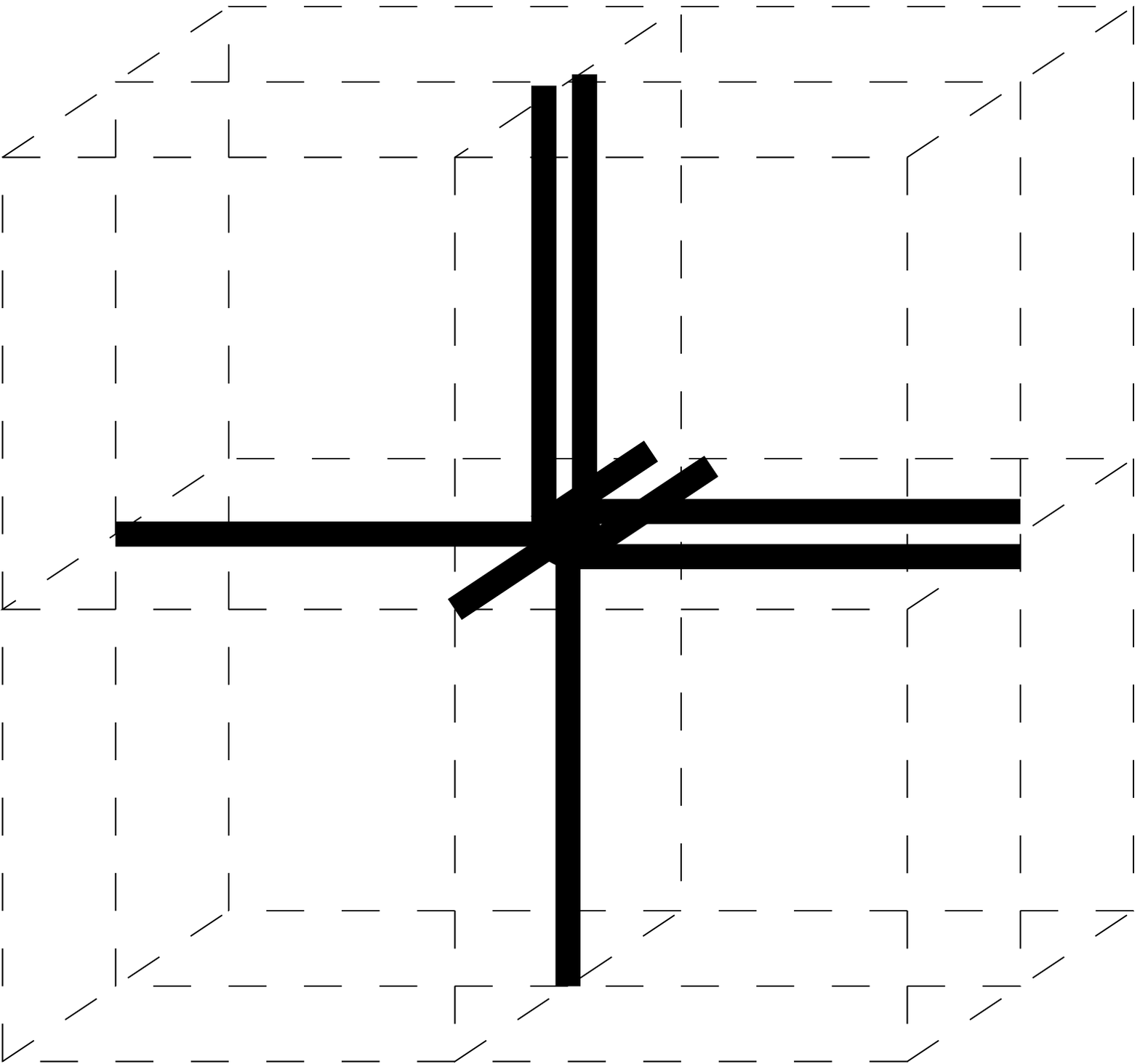}}}
\put(6,0.9){\mbox{$x^{15}$}}
\epsfxsize=0.7cm
\put(8,0.5){\mbox{\epsfbox{dspin.eps}}}
\epsfxsize=0.7cm
\put(9,0.5){\mbox{\epsfbox{spin11.eps}}}
\epsfxsize=1.8cm
\put(10.6,0.1){\mbox{\epsfbox{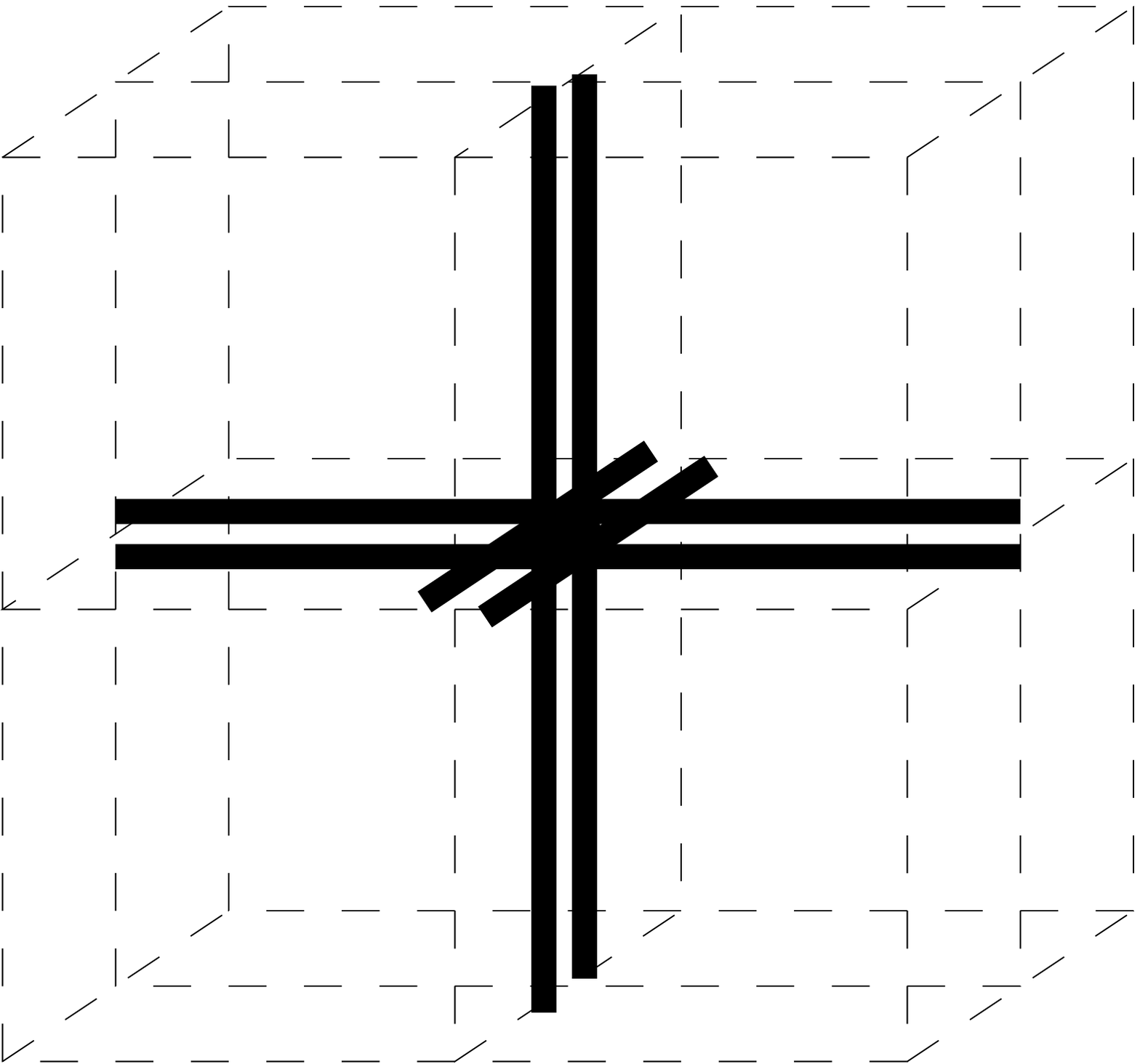}}}
\put(13.5,0.9){\mbox{$x^{24}$}}
\end{picture}\\
\begin{picture}(15,2)
\put(7.5,0){\line(0,1){2}}
\epsfxsize=0.7cm
\put(0.5,0.5){\mbox{\epsfbox{espin.eps}}}
\epsfxsize=0.7cm
\put(1.5,0.5){\mbox{\epsfbox{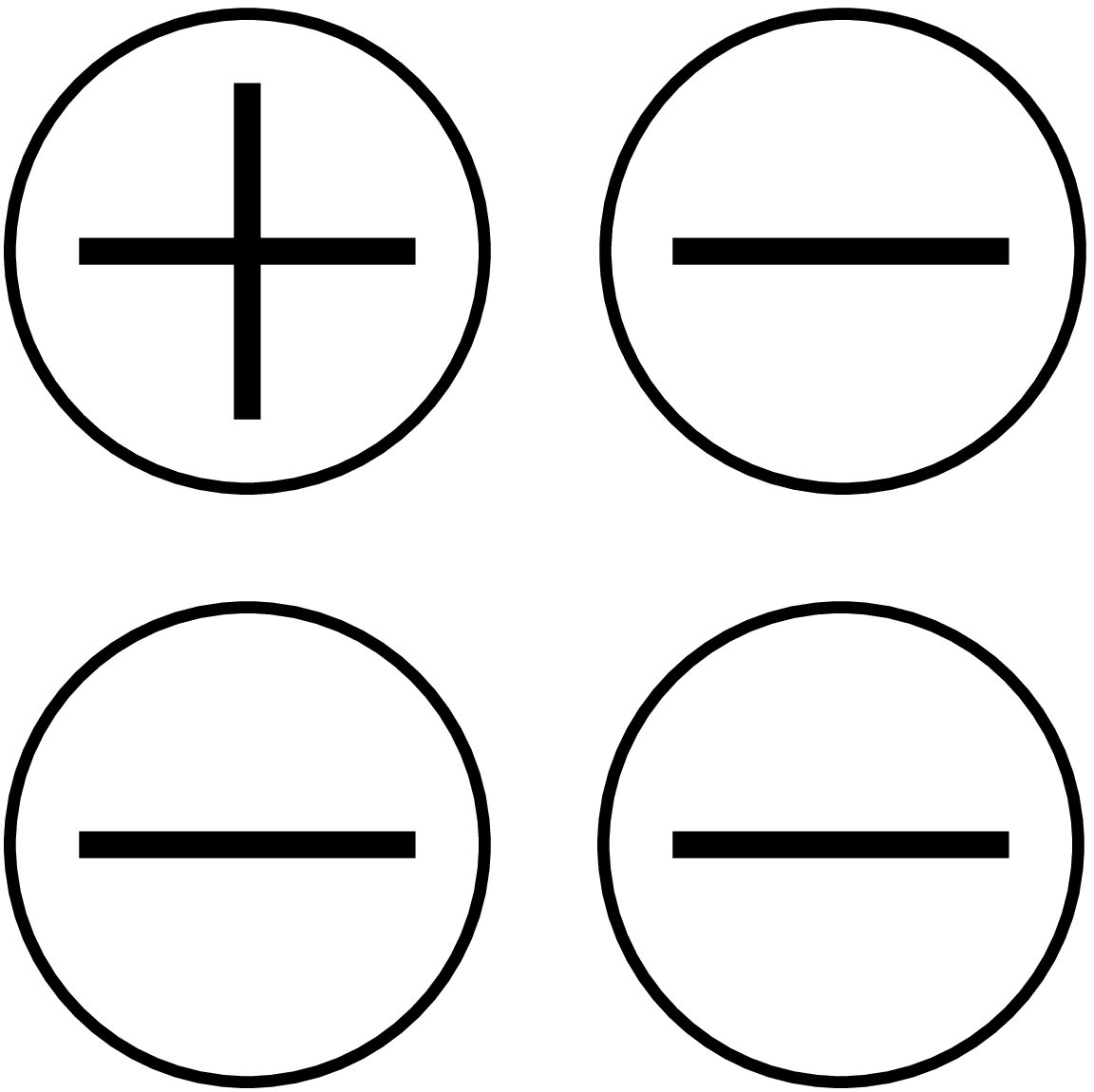}}}
\epsfxsize=1.8cm
\put(3.1,0.1){\mbox{\epsfbox{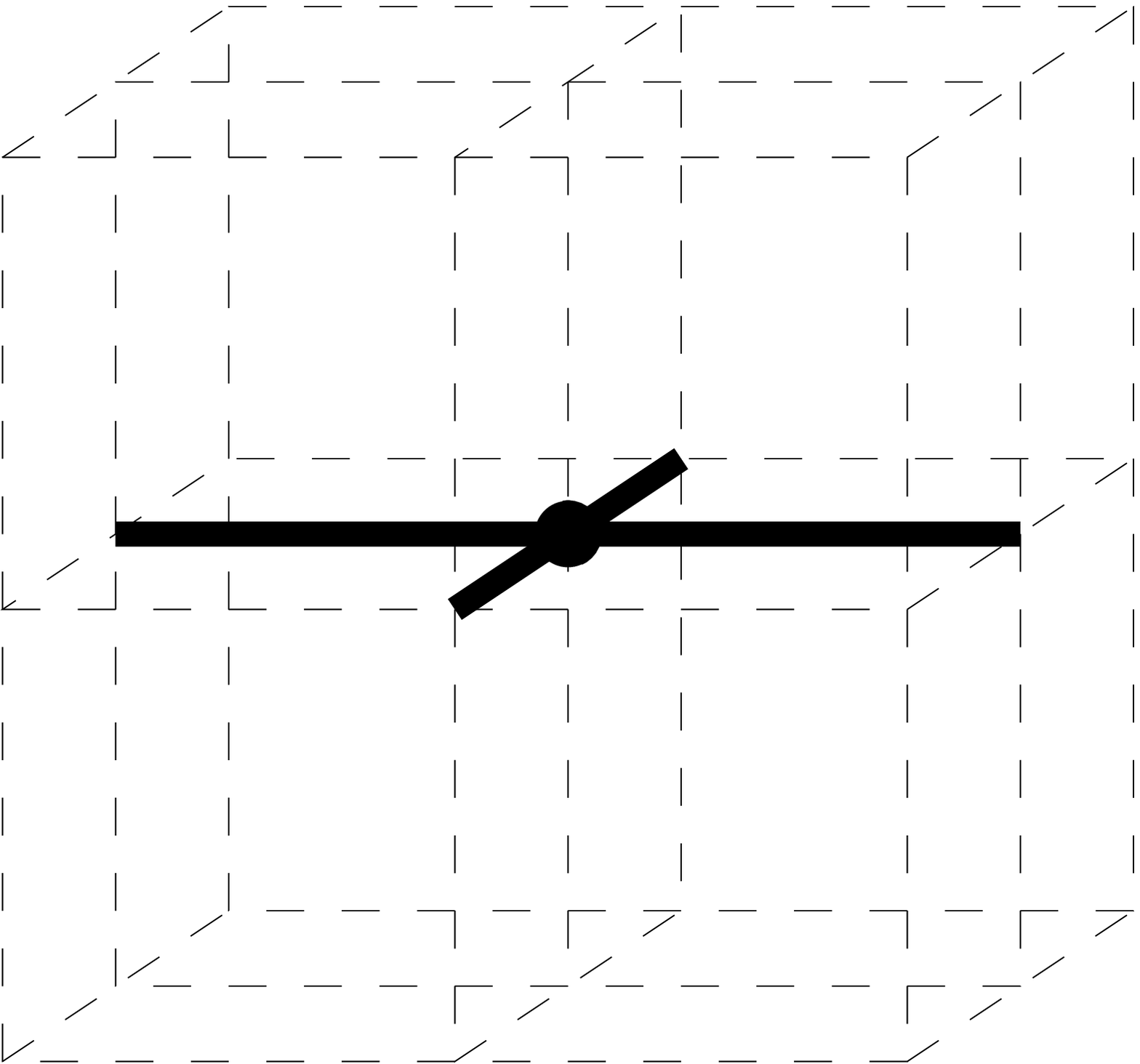}}}
\put(6,0.9){\mbox{$x^4$}}
\epsfxsize=0.7cm
\put(8,0.5){\mbox{\epsfbox{dspin.eps}}}
\epsfxsize=0.7cm
\put(9,0.5){\mbox{\epsfbox{spin12.eps}}}
\epsfxsize=1.8cm
\put(10.6,0.1){\mbox{\epsfbox{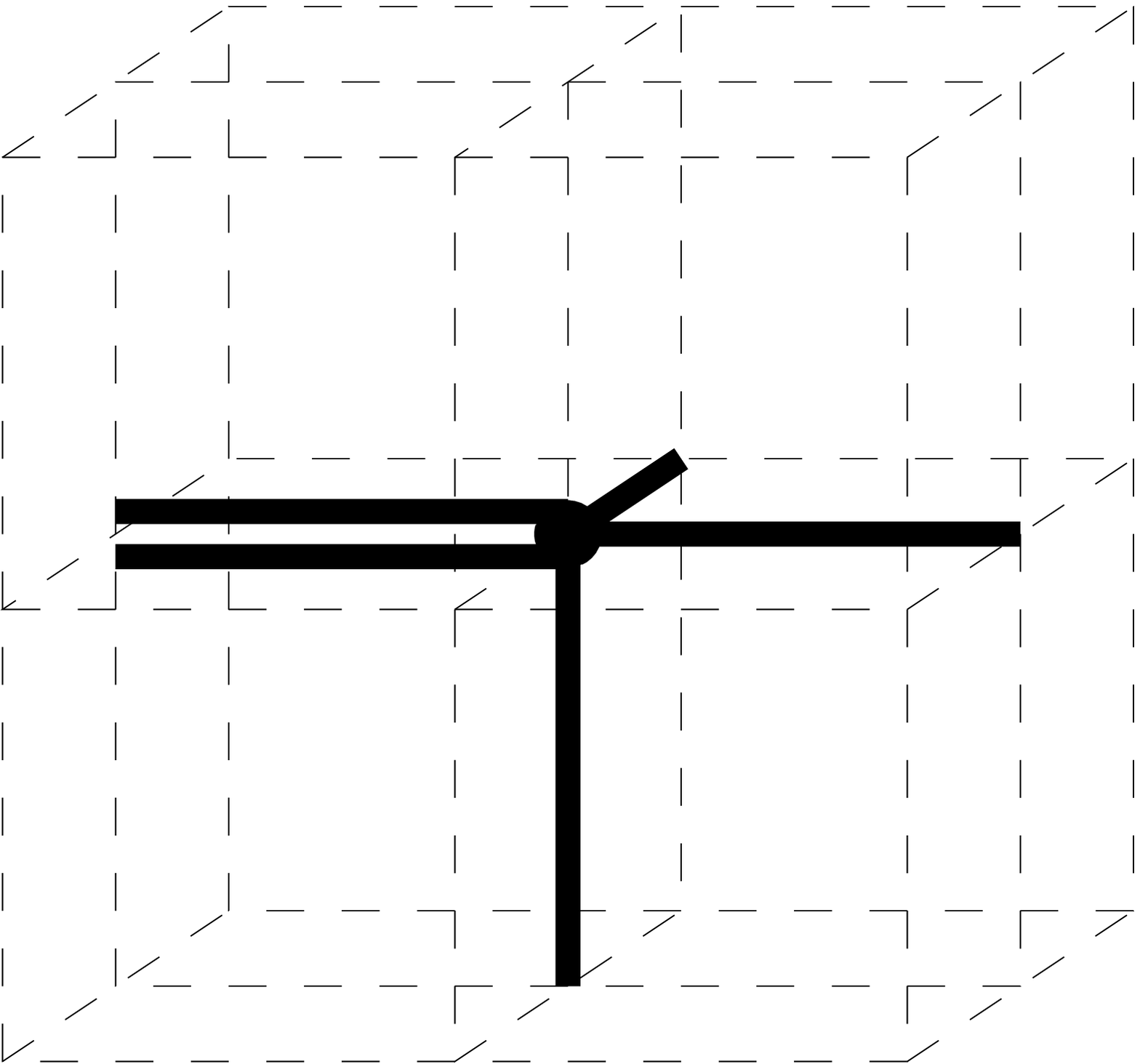}}}
\put(13.5,0.9){\mbox{$x^7$}}
\end{picture}\\
\begin{picture}(15,2)
\put(7.5,0){\line(0,1){2}}
\epsfxsize=0.7cm
\put(0.5,0.5){\mbox{\epsfbox{espin.eps}}}
\epsfxsize=0.7cm
\put(1.5,0.5){\mbox{\epsfbox{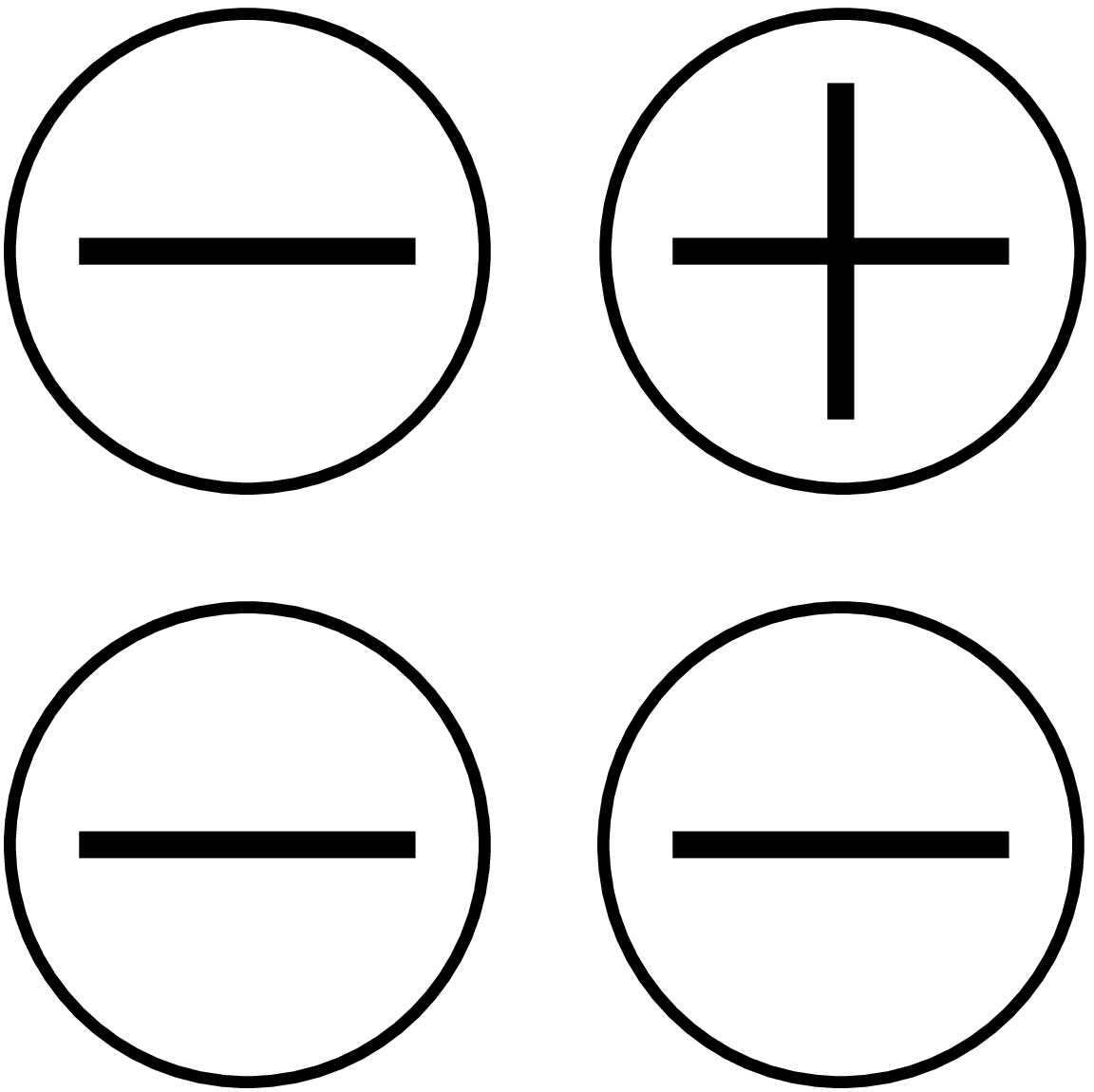}}}
\epsfxsize=1.8cm
\put(3.1,0.1){\mbox{\epsfbox{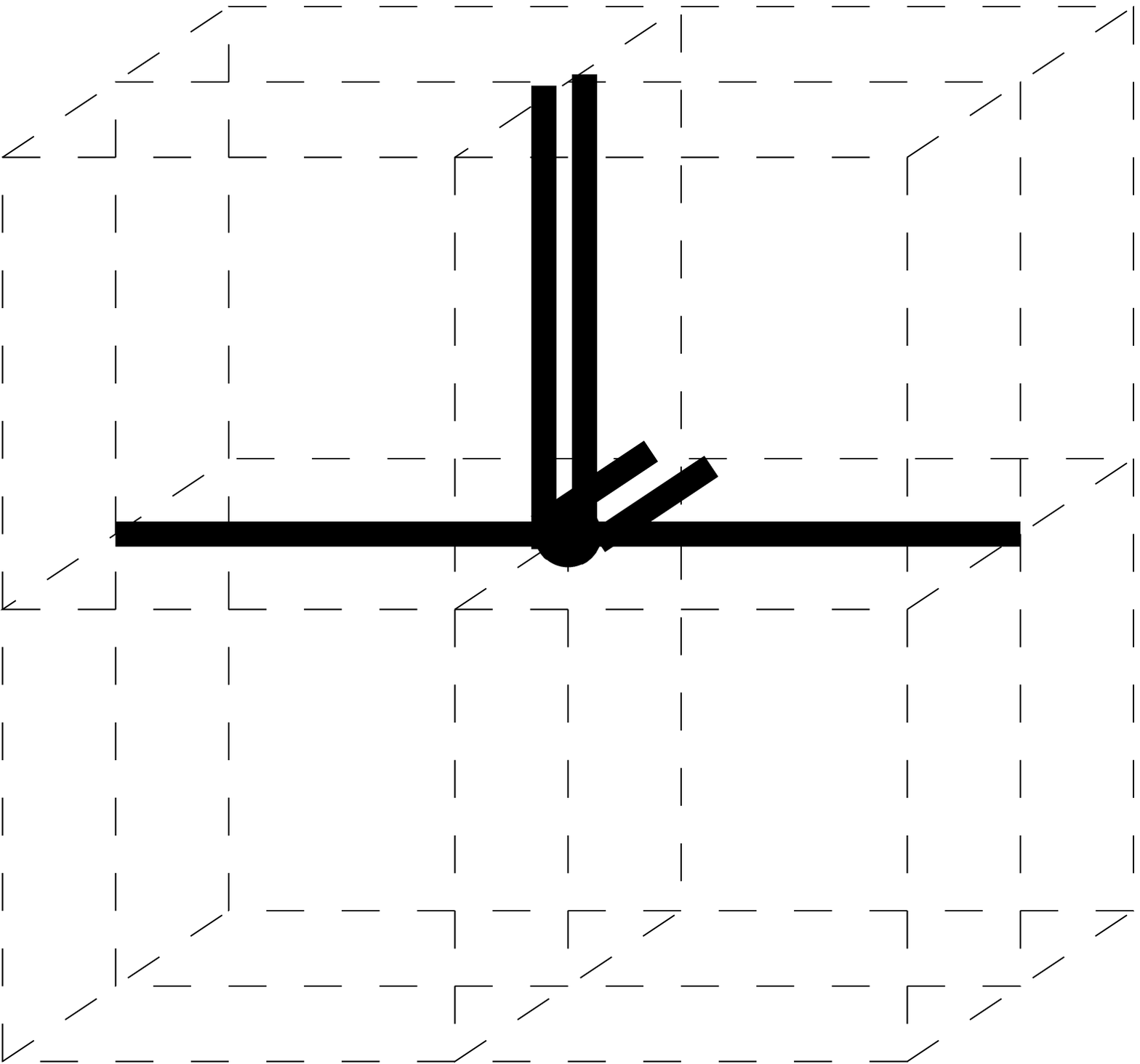}}}
\put(6,0.9){\mbox{$x^{10}$}}
\epsfxsize=0.7cm
\put(8,0.5){\mbox{\epsfbox{dspin.eps}}}
\epsfxsize=0.7cm
\put(9,0.5){\mbox{\epsfbox{spin13.eps}}}
\epsfxsize=1.8cm
\put(10.6,0.1){\mbox{\epsfbox{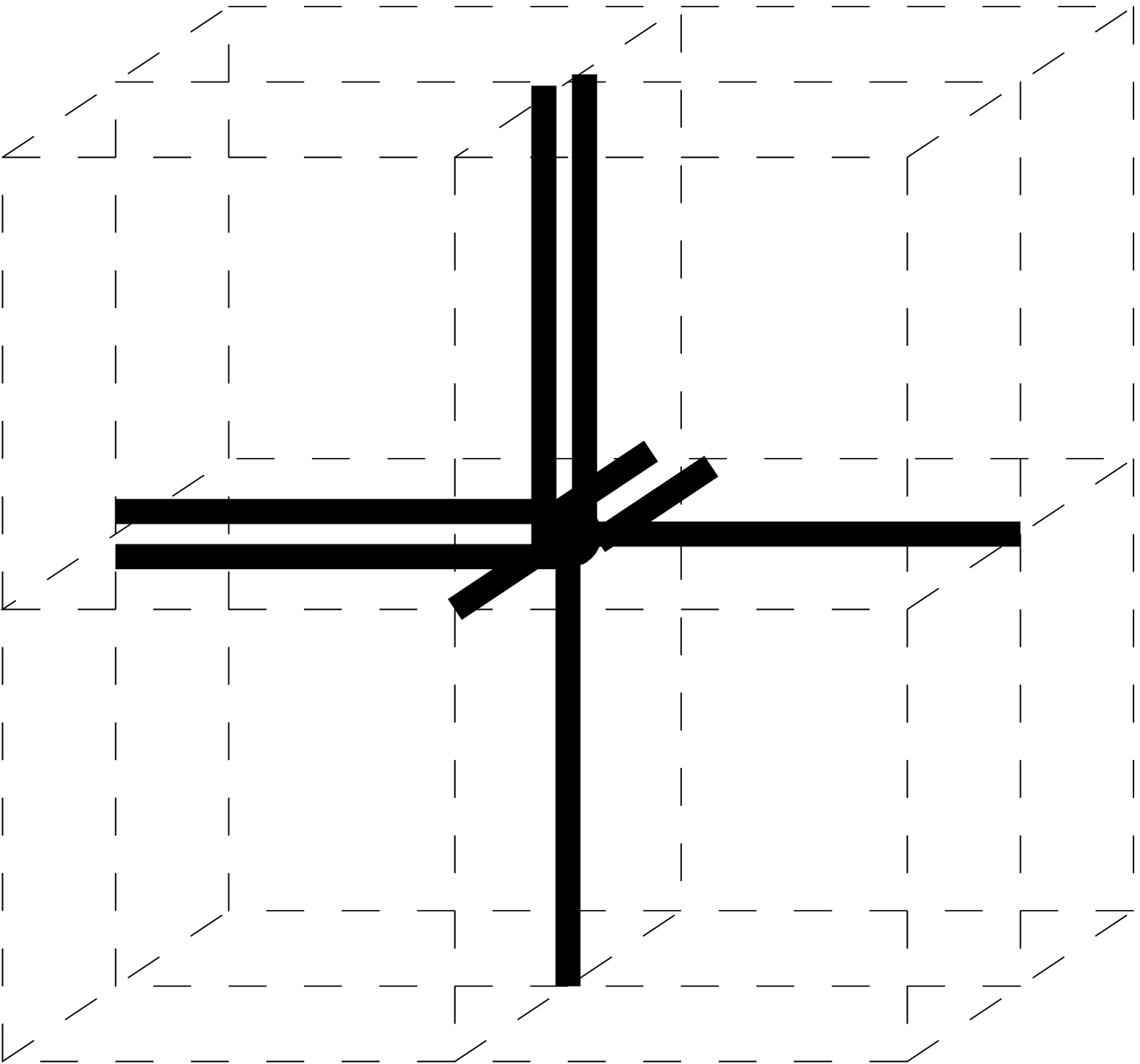}}}
\put(13.5,0.9){\mbox{$x^{15}$}}
\end{picture}\\
\begin{picture}(15,2)
\put(7.5,0){\line(0,1){2}}
\epsfxsize=0.7cm
\put(0.5,0.5){\mbox{\epsfbox{espin.eps}}}
\epsfxsize=0.7cm
\put(1.5,0.5){\mbox{\epsfbox{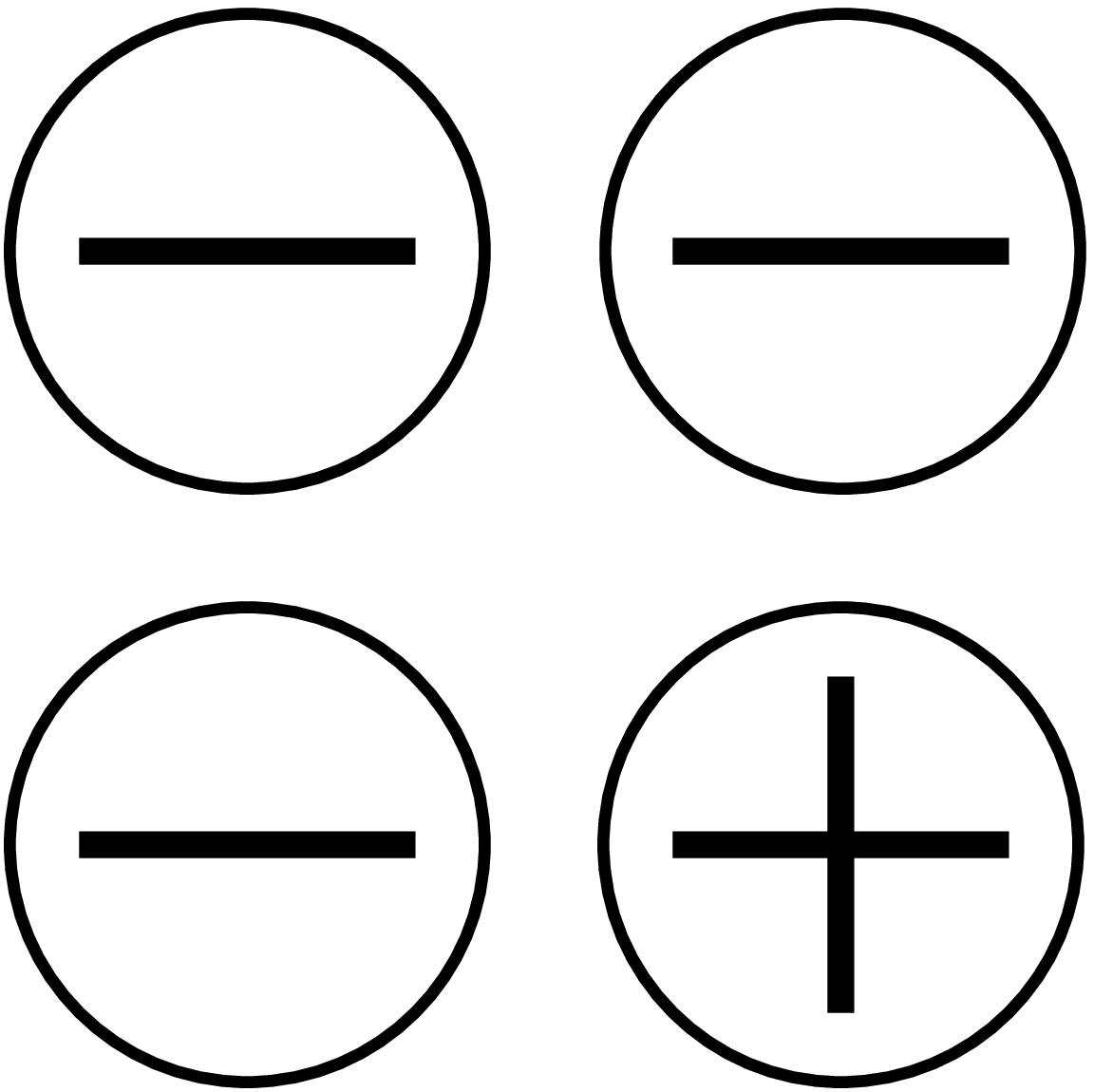}}}
\epsfxsize=1.8cm
\put(3.1,0.1){\mbox{\epsfbox{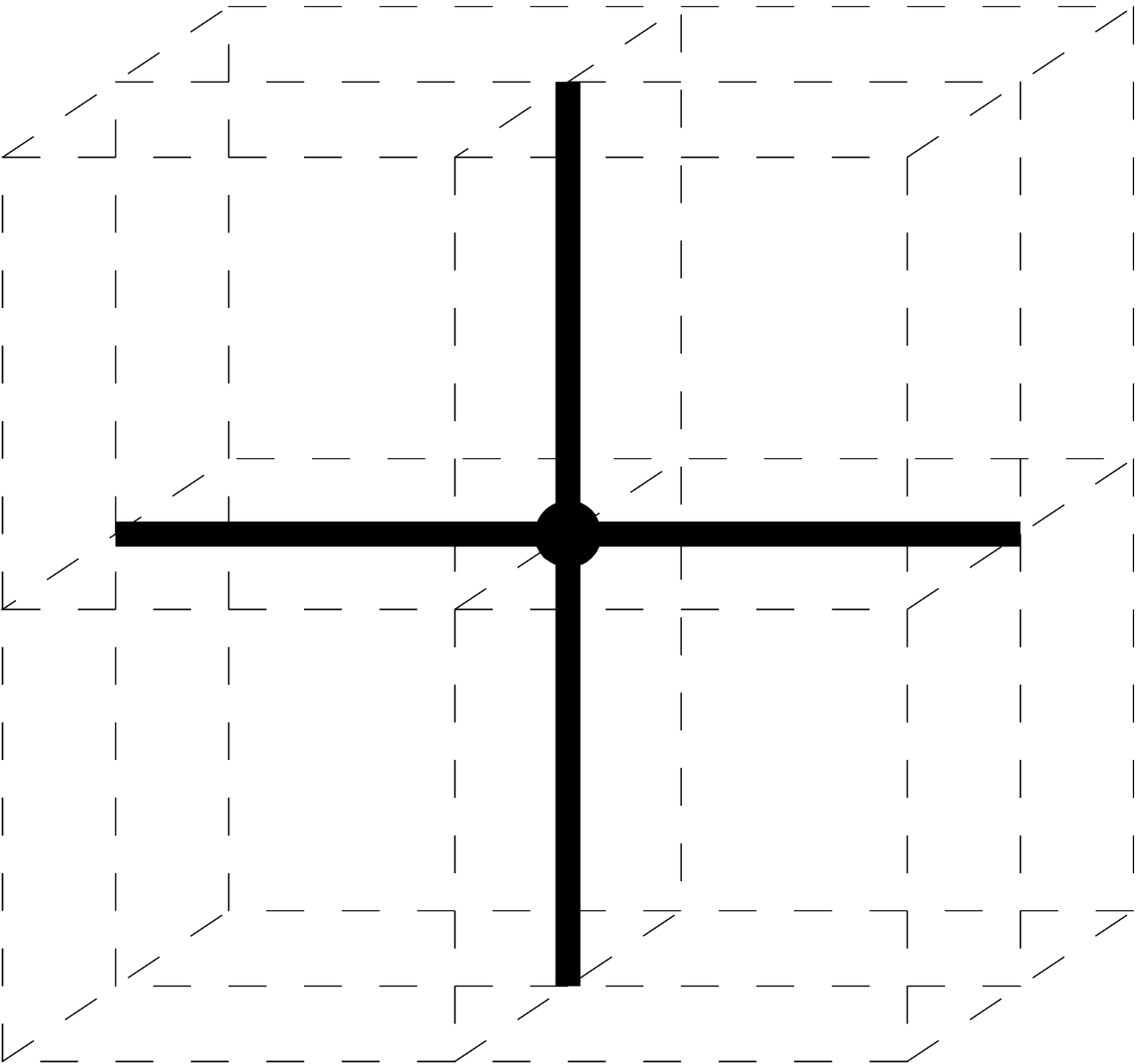}}}
\put(6,0.9){\mbox{$x^4$}}
\epsfxsize=0.7cm
\put(8,0.5){\mbox{\epsfbox{dspin.eps}}}
\epsfxsize=0.7cm
\put(9,0.5){\mbox{\epsfbox{spin14.eps}}}
\epsfxsize=1.8cm
\put(10.6,0.1){\mbox{\epsfbox{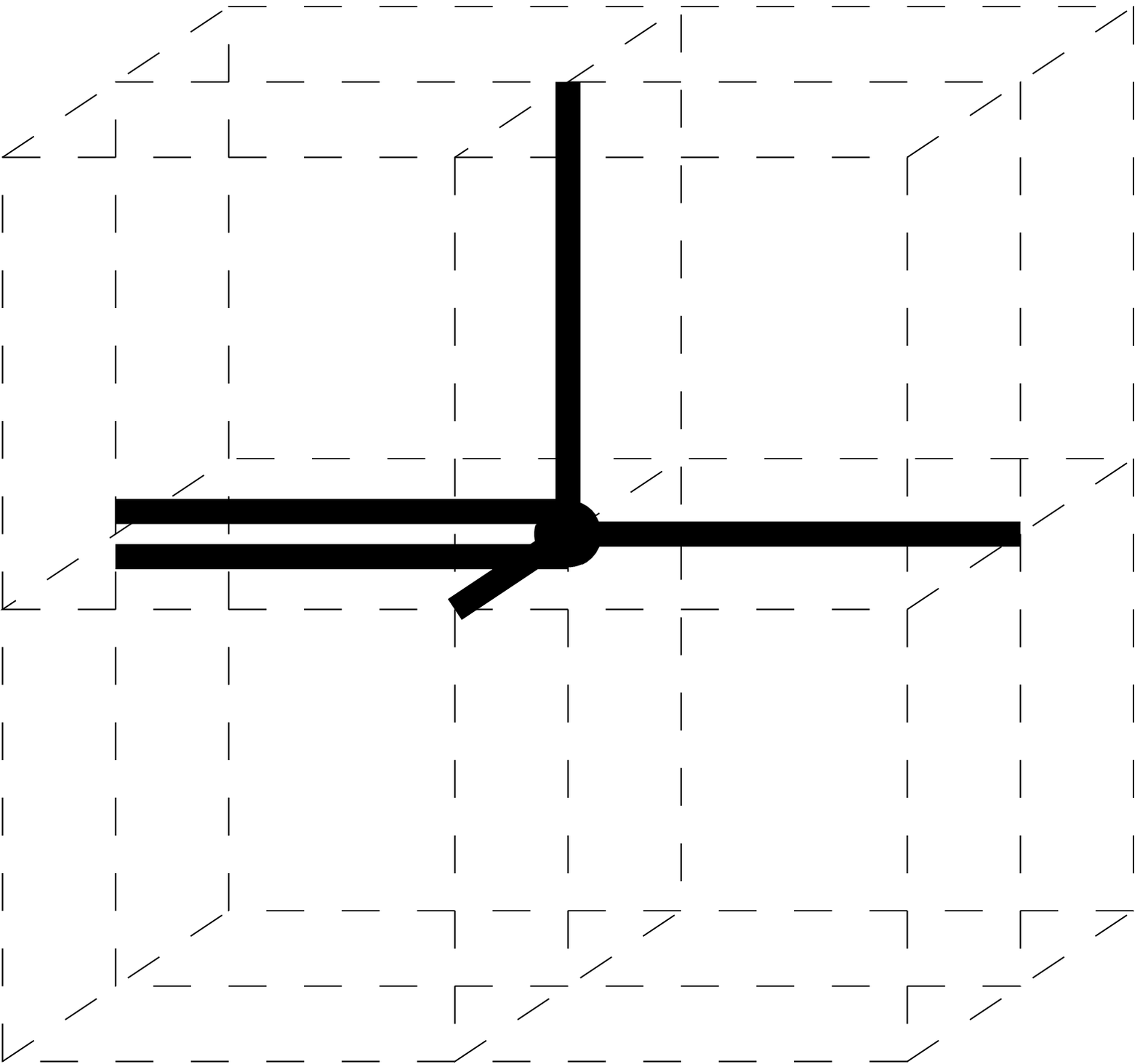}}}
\put(13.5,0.9){\mbox{$x^7$}}
\end{picture}\\
\begin{picture}(15,2)
\put(7.5,0){\line(0,1){2}}
\epsfxsize=0.7cm
\put(0.5,0.5){\mbox{\epsfbox{espin.eps}}}
\epsfxsize=0.7cm
\put(1.5,0.5){\mbox{\epsfbox{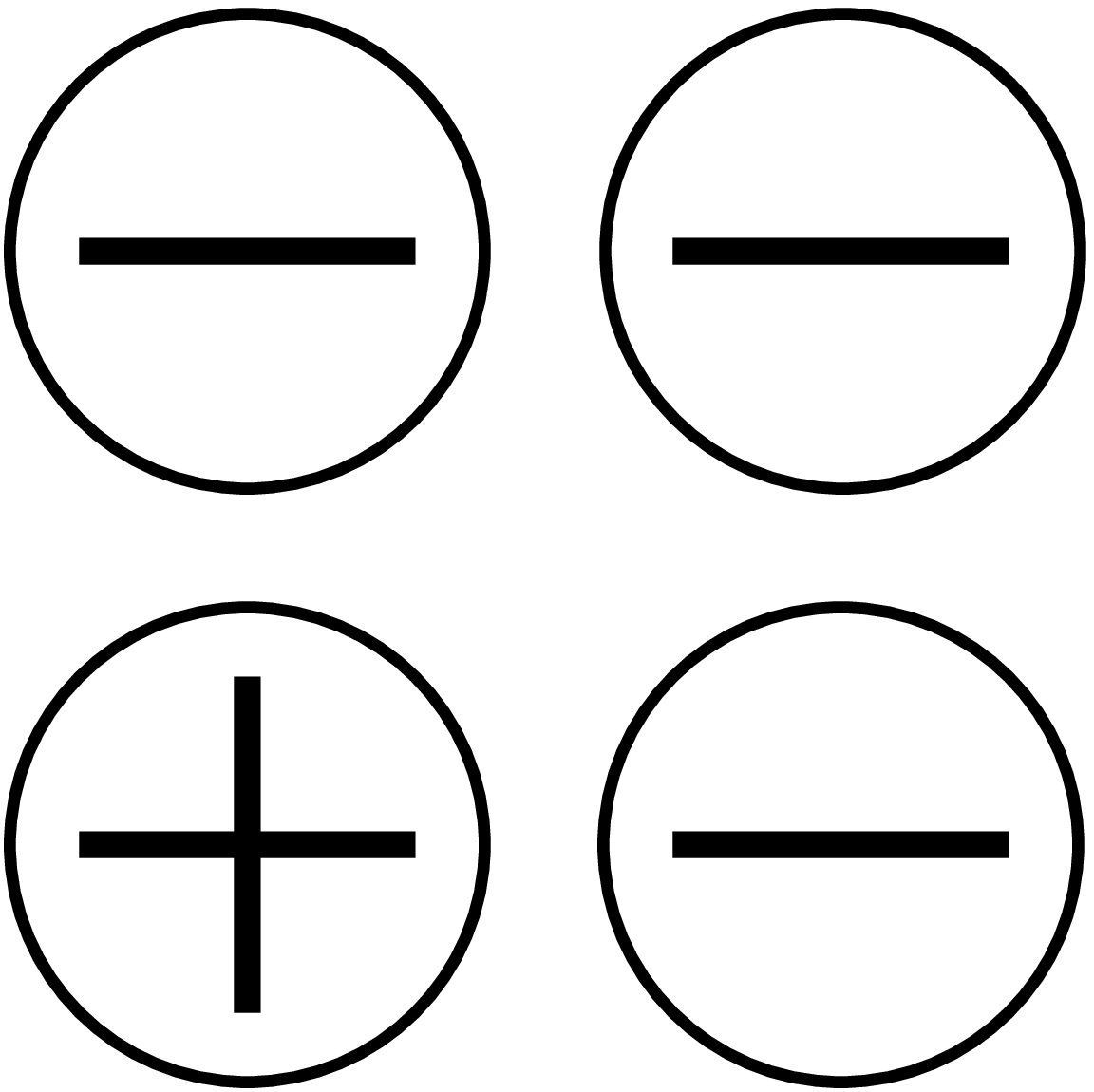}}}
\epsfxsize=1.8cm
\put(3.1,0.1){\mbox{\epsfbox{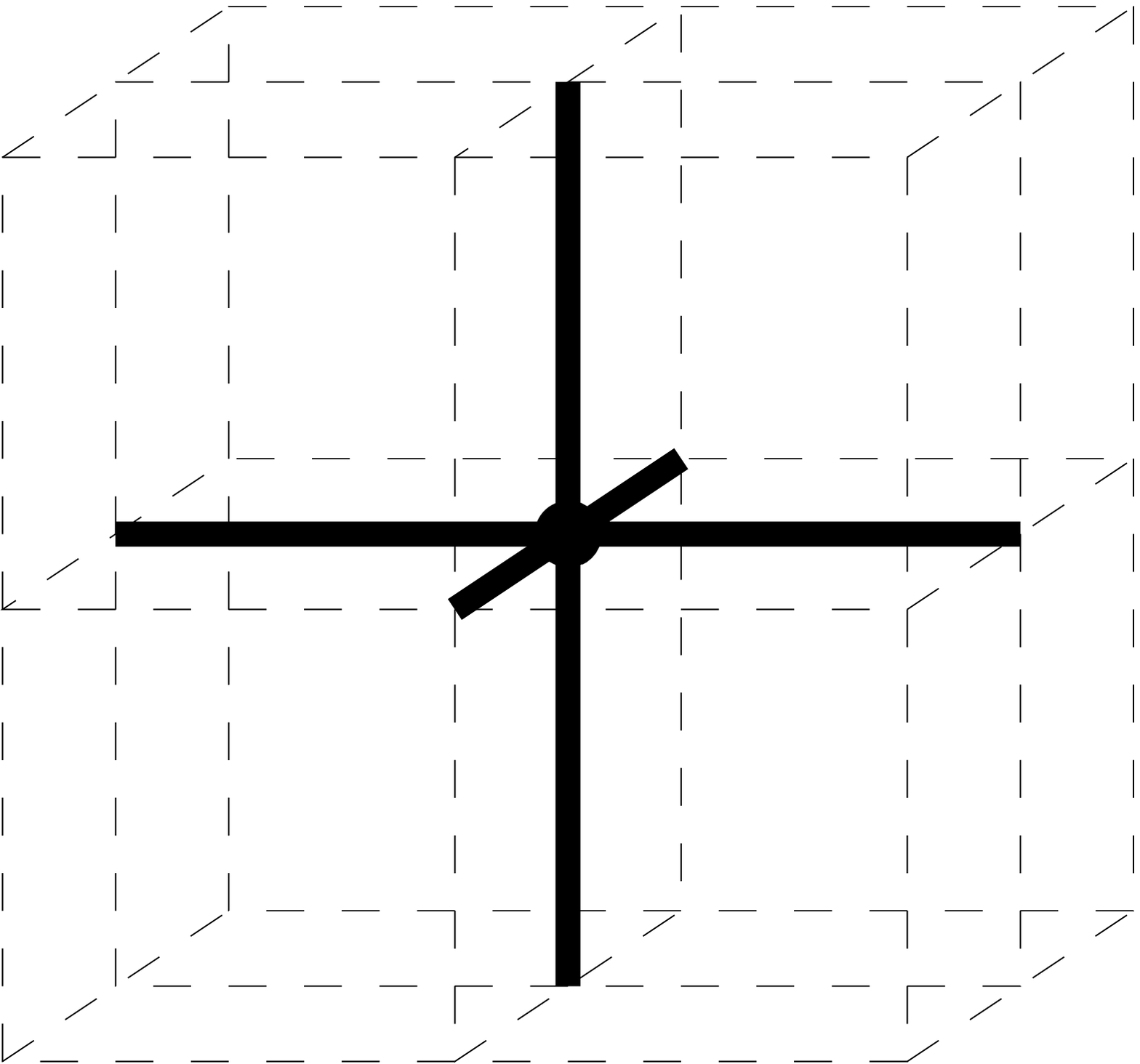}}}
\put(6,0.9){\mbox{$x^6$}}
\epsfxsize=0.7cm
\put(8,0.5){\mbox{\epsfbox{dspin.eps}}}
\epsfxsize=0.7cm
\put(9,0.5){\mbox{\epsfbox{spin15.eps}}}
\epsfxsize=1.8cm
\put(10.6,0.1){\mbox{\epsfbox{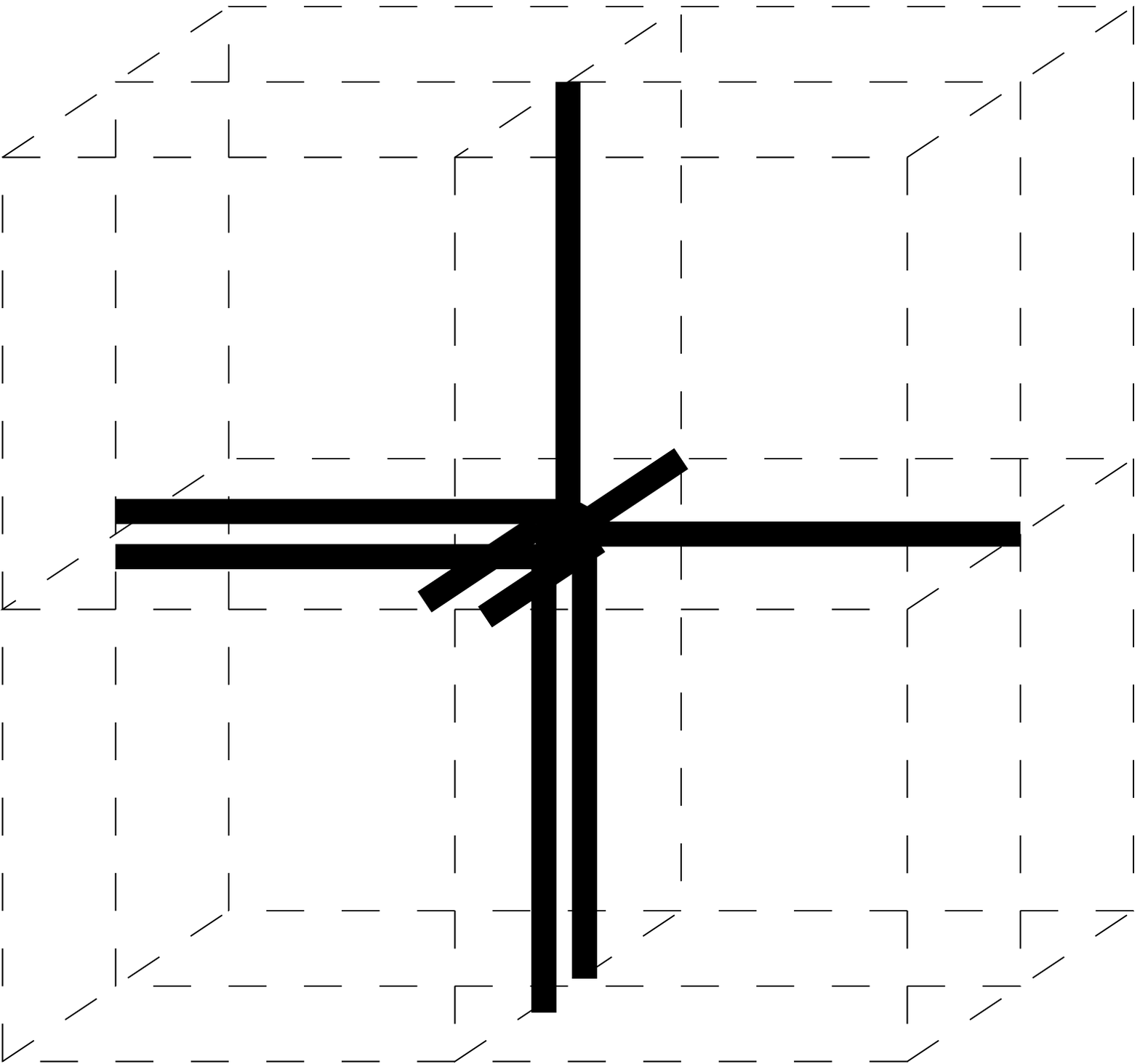}}}
\put(13.5,0.9){\mbox{$x^{15}$}}
\end{picture}\\
\begin{picture}(15,2)
\put(7.5,0){\line(0,1){2}}
\epsfxsize=0.7cm
\put(0.5,0.5){\mbox{\epsfbox{espin.eps}}}
\epsfxsize=0.7cm
\put(1.5,0.5){\mbox{\epsfbox{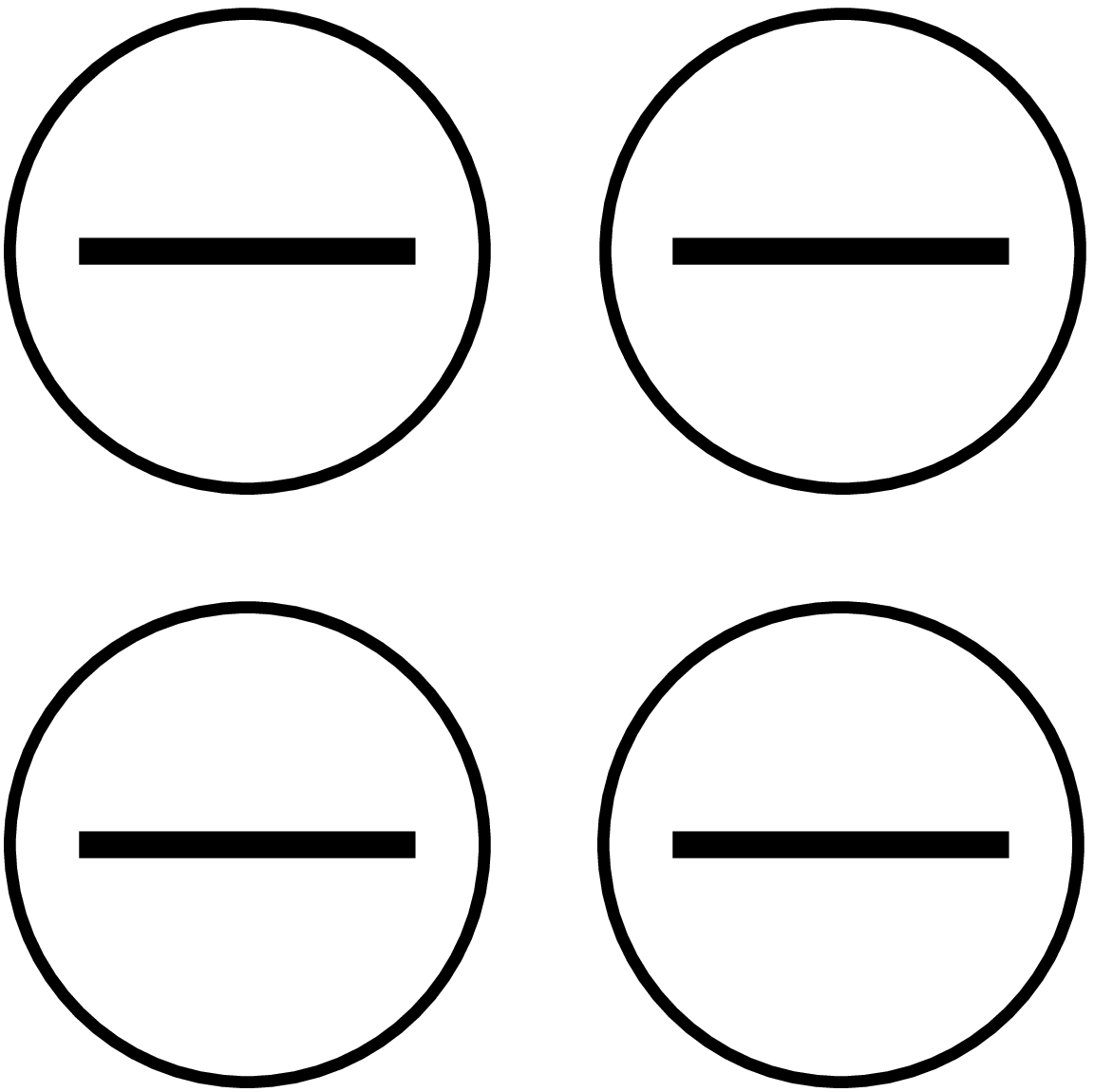}}}
\epsfxsize=1.8cm
\put(3.1,0.1){\mbox{\epsfbox{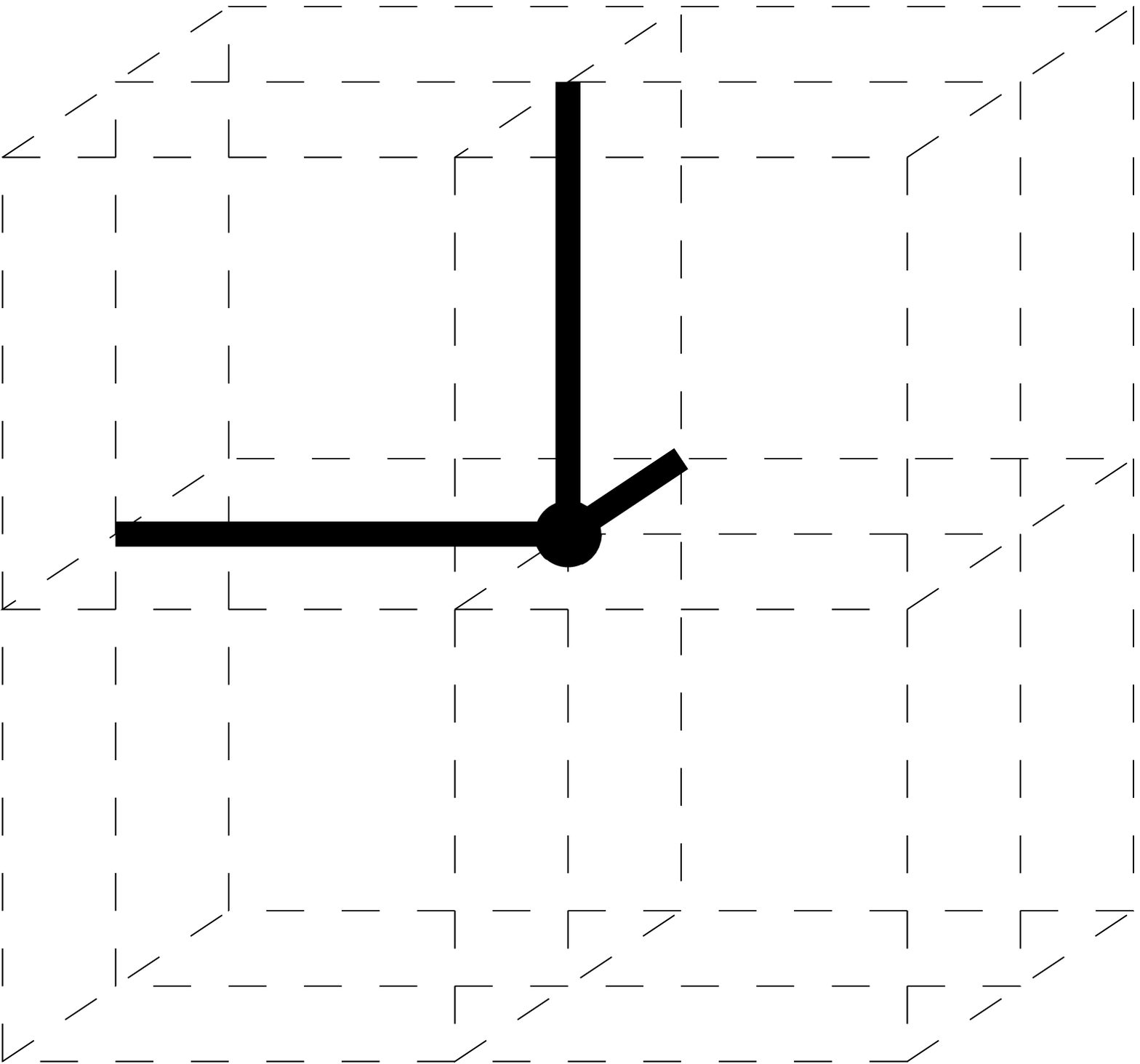}}}
\put(6,0.9){\mbox{$x^3$}}
\epsfxsize=0.7cm
\put(8,0.5){\mbox{\epsfbox{dspin.eps}}}
\epsfxsize=0.7cm
\put(9,0.5){\mbox{\epsfbox{spin16.eps}}}
\epsfxsize=1.8cm
\put(10.6,0.1){\mbox{\epsfbox{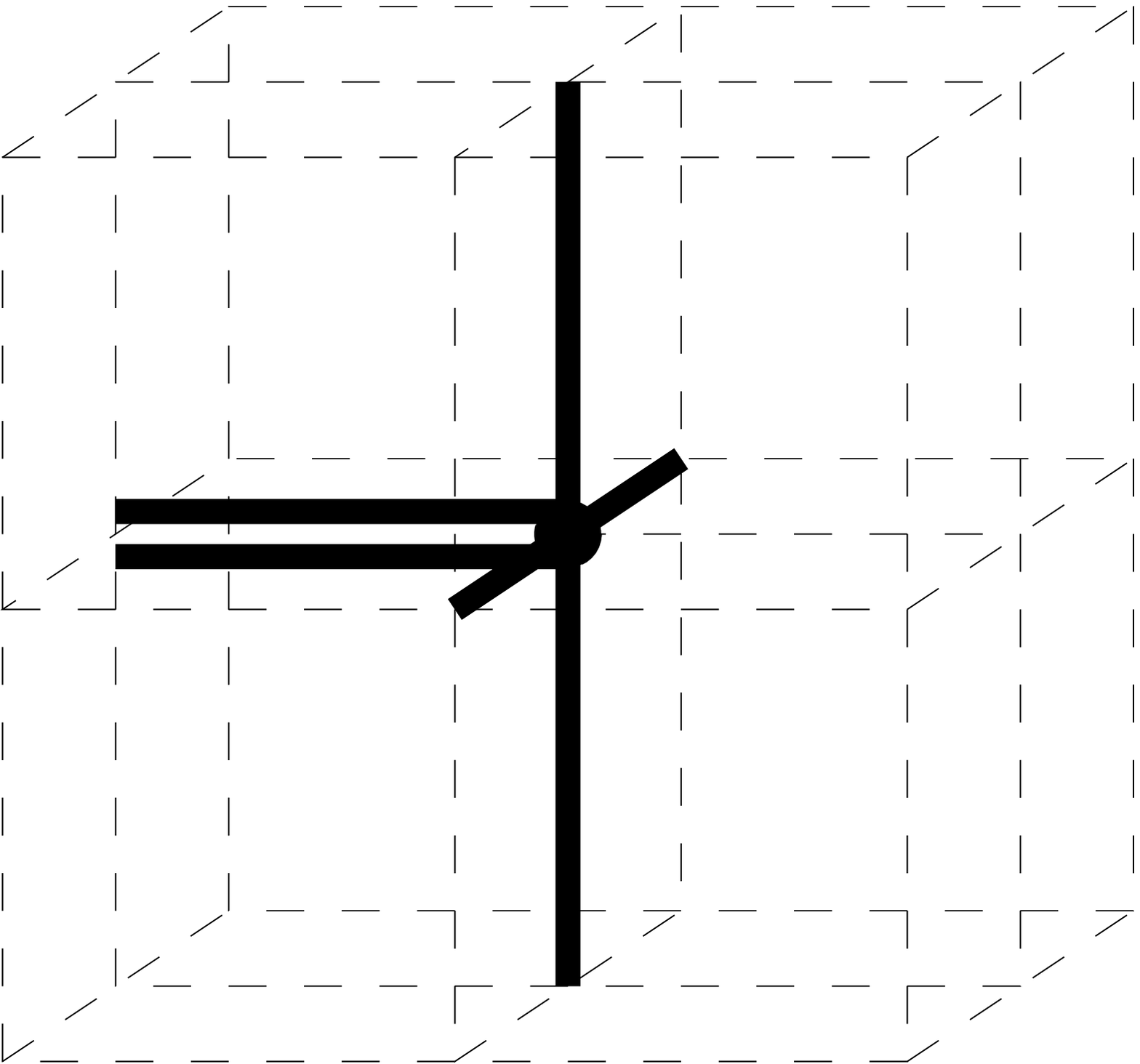}}}
\put(13.5,0.9){\mbox{$x^8$}}
\end{picture}\\
\center{figure 1}
\end{document}